\documentclass[fleqn,usenatbib]{mnras}

\usepackage{latexsym,mathrsfs,amssymb,bm}
\usepackage[tbtags]{amsmath}
\usepackage{comment}
\usepackage[T1]{fontenc}
\usepackage{ae,aecompl,times}
\usepackage{graphicx}	
\usepackage{amsmath}	
\usepackage{amssymb}	
\usepackage{multicol}
\usepackage[normalem]{ulem}
\usepackage{dsfont}
\usepackage{mathptm}

\title[PRFM-vol cosmological]{Learning the Universe with PRFM-vol: Introducing a new subgrid model for star formation in cosmological simulations}

\graphicspath{{./}{figures/}}

\author[J.~D.~Burger et al.]{%
\parbox{0.99\textwidth}
{%
Jan D.~Burger$^{1}$\thanks{E-mail: burger@mpa-garching.mpg.de},
Volker Springel$^{1}$, 
Eve C.~Ostriker$^{2, 3}$,
Chang-Goo Kim$^{2}$,
Ulrich Steinwandel$^{1}$,\\
Matthew C.~Smith$^{1}$,
Lars Hernquist$^{4}$,
Greg L.~Bryan$^{5, 6}$, 
Rachel S.~Somerville$^{6}$ and
Alon Gurman$^{7}$\\
}
\\%
$^{1}$Max-Planck-Institut f\"ur Astrophysik, Karl-Schwarzschild-Str. 1, D-85748, Garching, Germany\\%
$^{2}$Department of Astrophysical Sciences, Princeton University, 4 Ivy Lane, Princeton, NJ 08544, USA\\%
$^{3}$Institute for Advanced Study, 1 Einstein Drive, Princeton, NJ 08540, USA\\%
$^{4}$Center for Astrophysics | Harvard \& Smithsonian, 60 Garden Street, Cambridge, MA 02138, USA\\%
$^{5}$Department of Astronomy, Columbia University, 550 West 120th Street, New York, NY 10027, USA\\%
$^{6}$Center for Computational Astrophysics, Flatiron Institute, 162 5th Ave, New York, NY 10010, USA\\%
$^{7}$School of Physics \& Astronomy, Tel Aviv University, Ramat Aviv 69978, IS\\%
}

\date{Accepted XXX. Received YYY; in original form ZZZ}

\pubyear{2026}

\begin{document}
\label{firstpage}
\pagerange{\pageref{firstpage}--\pageref{lastpage}}
\maketitle

\begin{abstract}
We introduce PRFM-vol, a new subgrid model for star formation in cosmological simulations that aims to 
increase the physical realism of cosmological simulations by leveraging results obtained with focused ISM simulations. We deploy a modified effective equation of state and calculate the star formation rate for each gas cell as a function of the ambient densities of gas, dark matter, and stars, based on the pressure-regulated feedback-modulated (PRFM) theory of star formation. Test simulations of our model in isolated galaxies show that we match PRFM predictions and TIGRESS scaling relations remarkably well, provided sufficiently high resolution is available. In particular, we are able to clearly demonstrate the impact of the stellar potential on the star formation rate, thereby retaining an important prediction of PRFM. We then apply our new model to cosmological multizoom simulations and find, compared to our previous TIGRESS/Schmidt model, a significant increase in the stellar scale heights and a slight increase in stellar mass. We demonstrate that modifying the effective equation of state significantly affects the morphology of simulated galaxies. Pronounced stellar clumps appear if the effective pressure at low hydrogen number densities is low, and disappear for higher pressure. We show that the formation of clumps is a result of Toomre instabilities, and conclude that simulated galaxy morphologies can be used to constrain effective equation of state models. Overall, our results establish PRFM-vol as a new self-consistent, physics-motivated subgrid model for star formation in high-resolution cosmological simulations. 
\end{abstract}

\begin{keywords}cosmology: theory -- large-scale structure of Universe -- dark matter -- galaxies: haloes -- methods: numerical
\end{keywords}

\section{Introduction}

Over the last decade, hydrodynamical numerical simulations have established themselves as one of the primary theoretical tools in the field of galaxy formation physics \citep[see][for a review]{Vogelsberger2020}. Across astrophysics, simulations are routinely used to model physical processes over a vast range of mass and spatial scales, from star formation and stellar evolution \citep[e.g.][]{Krumholz2014}, to black hole dynamics \citep[e.g.][]{Rantala2017}, resolved simulations of the interstellar medium (ISM) \citep[e.g.][]{Kim2017}, and the evolution of galaxies in cosmological environments. At the largest scales, cosmological hydrodynamical simulations such as Illustris \citep{Vogelsberger2014}, EAGLE \citep{Schaye2015}, IllustrisTNG \citep{Springel2018}, SIMBA \citep{Dave2019}, ASTRID \citep{Ni2022}, and COLIBRE \citep{Schaye2026COLIBRE} model galaxy formation across representative cosmological volumes. Complementary zoom-in simulations, such as FIRE \citep{Hopkins2014}, Auriga \citep{Grand2017}, LYRA \citep{Gutcke2021}, and Vintergatan \citep{Agertz2021}, instead focus computational resources on individual galaxies and their environments, achieving significantly higher resolution for a small number of objects.

No matter the scale, all of those simulations rely on the availability of computational resources, and are thus subject to the same external constraints. In practice, this means that simulations that focus on smaller scales (such as simulations of protostellar disks or individual ISM patches) can afford to model physical processes at much higher fidelity than large cosmological simulations.

Since hydrodynamical cosmological simulations of galaxy formation aim to model the evolution of the Universe -- including the formation of structure, the formation of the first stars, reionization, and the period of cosmic noon -- it is evident that a faithful description of the physics of star formation, stellar feedback, black hole growth, and black hole feedback should be included in such simulations \citep[see][for recent reviews]{Somerville2015,Naab2017}. When combined with the constraints imposed by limited computational resources, this unfortunately collides with the other key objective of cosmological simulations, which is to model the distribution of structure in a very large volume. Resolving star formation, stellar feedback, and black hole physics on the relevant scales while also simulating large cosmological volumes remains, for the time being, out of reach. Instead, physical processes that cannot be explicitly modeled at the maximum resolution achieved in cosmological simulations need to be included in a different way. Usually, this is done by modeling the average effect of unresolved physical processes below the resolution scale in the form of so-called subgrid models. As we outlined in \citet{Burger2025}, such subgrid models can either be formulated in a resolution independent way, or explicitly depend on resolution, depending also on whether one chooses to interpret them as approximate physics models or primarily as a means to achieving optimal agreement with observations. Within the Learning the Universe (LtU) collaboration, we adopt the former point of view, meaning that we develop new subgrid models in a resolution independent way and base their implementation on analytical arguments and the results of resolved small-scale simulations. 

Several new developments are already emerging from these efforts, aimed at delivering updated versions of the TNG black hole (BH) feedback model \citep{Weinberger2017} and the TNG stellar feedback and star formation physics \citep{Pillepich2018a}, whose star formation prescription is based on the original \citet{Springel2003} approach. The TNG model is a quite successful reference in the field of subgrid models for cosmological hydrodynamical simulations, as recently demonstrated by the success of the MillenniumTNG project \citep{Pakmor2023}. When aiming to develop improvements over the TNG model, it is therefore important to define in what way the new schemes should do better. For the new generation of subgrid physics models developed within the Learning the Universe collaboration (LtU), a primary objective for improving on TNG is to put the modeling of the physics on a more faithful basis. For example, the BH dynamics model by \citet{Genina2024} aims to replace the BH repositioning algorithm with a subgrid model for dynamical friction, the ARKENSTONE wind model \citep{Smith2024a,Smith2024b,Bennett2026} aims to improve on the simple stellar wind model in TNG, and with the BRAHMA seed models \citep{Bhowmick2024} combined with a new BH accretion model, we aim to improve the modeling of early BH growth. 

In the context of star formation, the development of new subgrid models has historically proceeded along two main avenues. The first one is to either keep volumes relatively small, or to focus on high redshift environments, such that resolution requirements are less of a problem. In these cases, higher resolution ISM models ($\sim10^4\, {\rm M_\odot}$ in mass resolution), such as FIRE \citep{Hopkins2014} or SMUGGLE \citep{Marinacci2019}, have successfully been employed in the context of zoom simulations \citep[e.g.][]{Hopkins2018,Kannan2025}. While these explicitly model low-temperature gas cooling, the formation of stellar population particles, and mechanical feedback from supernovae, they are still not able to capture the hot ISM phase directly. Simulating even smaller systems, it has recently become possible to model the formation of individual stars in galaxies, accompanied by star-by-star feedback calculations \citep{Smith2026}. When targeting cosmological volumes, however, resolving metal-line cooling, the formation of molecular clouds, or accurately tracking mechanical stellar feedback remains out of reach for the time being. Instead, another way to make progress is to take the results of highly resolved simulations as input for the development of a new generation of coarse-grained subgrid models. 

In \citet{Burger2025}, we developed the TIGRESS/Schmidt model for ISM physics and star formation in exactly that spirit. The model was 
motivated by 
the results of the TIGRESS simulations \citep{Kim2017,Kim2020a, Ostriker2022}, which are a suite of shearing box simulations that focus on small, resolved patches of the ISM 
for conditions applicable in $z=0$ spiral galaxies.
Similar to the \citet{Springel2003} model and the Illustris-TNG model \citep{Pillepich2018a}, the TIGRESS/Schmidt parameterization of the high-resolution TIGRESS results consists primarily of two components. These are an effective equation of state (eEoS) that describes the pressure of cool, star-forming gas in the presence of star formation, magnetic fields, and stellar feedback, and a star formation law that determines the depletion time of gas cells as a function of the local gas density. In the TIGRESS/Schmidt model, this star formation law was inspired by PRFM-theory \citep[Pressure-regulated, feedback-modulated star formation,][]{Ostriker2010,OstrikerShetty2011,Ostriker2022}, but we neglected the contributions of stars and dark matter (DM) in maintaining the equilibrium between feedback-driven pressure and gravitational weight. The eEoS that we used was taken directly from the TIGRESS simulations, where a fitting function was provided, giving midplane pressure as a function of hydrogen number density. Thus, while we followed the approach of building our subgrid model on the results of high-resolution simulations combined with sensible physical arguments, we did not at the time implement a formalism that fully incorporates the PRFM equations. 

In the PRFM theory, the star formation efficiency per dynamical time $\varepsilon_\mathrm{dyn}$ is given by the ratio between a feedback yield $\Upsilon_\mathrm{tot}$ (the ratio between total pressure and star formation rate per unit area; see equation~\ref{eq:feedback_yield_def} below) and an effective velocity dispersion $\sigma_\mathrm{eff}$. The values of $\Upsilon_\mathrm{tot}$ and $\sigma_\mathrm{eff}$ -- and therefore $\varepsilon_\mathrm{dyn}$ -- have been calibrated from high-resolution TIGRESS simulations as a function of the ISM pressure \citep{Ostriker2022,Kim2024}; independent simulations have obtained similar results \citep{AlonUli2025}. An advantage of using pressure as the calibration variable for $\varepsilon_\mathrm{dyn}$ is that the pressure may be estimated accurately from integrated quantities even at quite coarse cosmological resolution.\footnote{This estimate assumes approximate equilibrium between total pressure at the midplane and the weight of the ISM; several simulations of star-forming galaxies with feedback have validated this assumption \citep[e.g.][]{2013ApJ...776....1K,2015ApJ...815...67K,2016MNRAS.462.3053B,2020MNRAS.498.3664G,Ostriker2022,AlonUli2025}.} In \citet{2024ApJ...975..151H}, analytic formulae were provided to obtain pressure estimates. The result was then compared to measured pressure in TNG simulations at varying resolution, showing that the expected pressure is recovered well at TNG50 resolution ($\sim 10^5\,{\rm M}_\odot$), but underestimated at coarser resolution. 
In \citet{Jeffreson2026}, two versions of PRFM-based subgrid models were implemented and intercompared in isolated galaxy simulations at a range of resolutions. One of these, ``PRFM-vol,'' employs volumetric quantities (such as gas density) measured in a simulation, and may be used when the vertical scale height of a disk galaxy is sufficiently resolved \citep[see discussion of resolution criteria in][]{2024ApJ...975..151H}. The other of these, ``PRFM-int,'' employs integrated quantities (such as gas surface density), and may be used when the vertical scale height of a galaxy is unresolved. \citet{Jeffreson2026} demonstrated that star formation rates in Milky-Way like isolated galaxies obtained using PRFM-vol and PRFM-int implementations at $10^5\,{\rm M}_\odot$ resolution are in good agreement with each other. Additionally, it was demonstrated that the PRFM-int implementation is able to capture the targeted Ostriker-Kim power-law relation between $\Sigma_\mathrm{SFR}$ and $P_\mathrm{tot}$ at all resolutions between $10^5\,{\rm M}_\odot$ and $10^7\, {\rm M}_\odot$, and PRFM-int simulations at all resolutions are able to track the same global star formation history as PRFM-vol at $10^5\,{\rm M}_\odot$ resolution.

In this work, we introduce the PRFM-vol model for cosmological simulations.
As in TIGRESS/Schmidt, we use an eEoS derived from TIGRESS results to model the pressure of gas in the star-forming ISM. We combine this with PRFM theory, modeling star formation rates as a function of the local dynamical time, which depends on ambient gas, DM, and stellar densities alike. PRFM-vol is formulated in such a way that its equations depend only on local volumetric quantities.
Making this choice requires us to be able to resolve the scale height of a galaxy with at least four gas cells. If such a resolution cannot be achieved, the alternative PRFM-int formulation presented in \citet{Jeffreson2026} is a more appropriate model; PRFM-int depends, however, on quantities that are vertically integrated over the thickness of the galactic disk, which in turn requires knowledge of the gravitational centre of each galaxy at every time step in order to identify the vertical direction. 
At a technical level PRFM-vol is much easier to implement in cosmological simulations since we do not need to track the movement of each galaxy, making it an ideal model to test PRFM predictions in a cosmological context. Moreover, we expect PRFM-vol to be applicable at TNG-50 resolution, making it a promising subgrid model for future large hydrodynamical simulations of galaxy formation. The way in which we implemented the PRFM-vol model presented here allows for easy modifications in the future, enabling us to -- for example -- incorporate alternative calibrations for the feedback yield and the eEoS from the TIGRESS-NCR simulations \citep{Kim2023,Kim2024}, which include explicit UV radiation transfer and photochemistry and provide metallicity dependent calibration, or the TIGRESS++ simulations \citep{Kim2026}, which simulate how cosmic ray feedback affects the star-forming ISM. 

This article is structured as follows. We describe the implementation of the cosmological PRFM-vol model in Section~\ref{sec:theory}. Where appropriate, we highlight similarities and differences between our new cosmological PRFM-vol technique, the TIGRESS/Schmidt approach, and the PRFM-vol and PRFM-int models presented in \citet{Jeffreson2026}. In Section~\ref{sec:setups}, we briefly introduce the initial conditions that we use to test our model in simulations. In Section~\ref{sec:isolated_galaxies}, we test our new model on simulations of isolated galaxies. Focusing on each new model feature individually, and comparing against TIGRESS/Schmidt results, we pinpoint where the two models differ, and how PRFM-vol improves upon TIGRESS/Schmidt. We then move on to discuss the results of multizoom cosmological simulations in Section~\ref{sec:multizoom}. Therein, we compare galaxy properties obtained with PRFM-vol, the TIGRESS/Schmidt model, and the TNG model. We focus on galaxy sizes, masses, scale heights, and star formation rates, and demonstrate that with our new model, the simulated galaxy scale heights increase significantly compared to TIGRESS/Schmidt, alleviating one major issue that remained in our previous method. We also demonstrate that the morphological appearance of simulated multizoom galaxies depends strongly on the eEoS, and employ a modified version of PRFM-vol that addresses the clumpy appearance of galaxies in both the TIGRESS/Schmidt model and the standard PRFM-vol model. We draw our conclusions and discuss future research avenues in Section~\ref{sec:conclusions}. 

In two appendices, we dive deeper into what drives the formation of stellar clumps in models with insufficient effective pressure at the star formation threshold. We follow clump formation with redshift in Appendix~\ref{apxsec:clumps_prfm_vol} and discuss the effects of a lack of resolution, as well as a decrease in gas surface density, in Appendix~\ref{apxsec:resolution}. 

 \section{The PRFM-vol model} \label{sec:theory}

\subsection{Introduction to PRFM}\label{ssec:PRFM_intro}

PRFM, or the pressure regulated, feedback modulated model of star formation \citep{Ostriker2022}, posits that the multiphase ISM is self-regulated, and that the feedback processes that occur as a result of star formation and stellar evolution give rise to a total pressure which, on average, balances out the gravitational pull of stars, dark matter (DM), and gas itself. This balance allows galaxies to be in a quasi-stationary state, appearing to slowly evolve in a regulated state over long intervals of time. In mathematical terms, we posit that the gravitational weight of the star-forming disk is dynamically equal to the total midplane pressure:
\begin{align}
    P_{\rm tot} = \mathcal{W} \label{eq:central_equality}.
\end{align}
Here, the weight is given as the integral 
\begin{align}
    \mathcal{W} = \int_0^\infty \rho_g(z)g(z) \,{\rm d} z, \label{eq:weight_integral}
\end{align}
where $\rho_g(z)$ denotes the average gas density as a function of the vertical coordinate, and $g(z)$ gives the gravitational acceleration, also as a function of the vertical coordinate. It is possible to split $g(z)$ into three components, namely $g(z)= g_g(z) + g_\ast(z) + g_{\rm DM}(z)$, respectively, denoting contributions from gas, stars, and dark matter (DM). As a consequence, it is also possible to write the weight itself as a sum of those same three components: 
\begin{align}
    \mathcal{W} = \mathcal{W}_g + \mathcal{W}_\ast + \mathcal{W}_{\rm DM}. \label{eq:weight_sum}
\end{align}
Following the calculations presented in \cite{2024ApJ...975..151H}, we can write the individual terms as follows:
\begin{align}
  &  \mathcal{W}_g = \frac{\pi G \Sigma_g^2}{2}, \label{eq:weight_gas} \\
 &   \mathcal{W}_\ast \approx \pi G \Sigma_g \Sigma_\ast \frac{H_g}{H_g+H_\ast}, \label{eq:weight_stars} \\
  &  \mathcal{W}_{\rm DM} = \zeta \Sigma_g \Omega^2_{\rm DM} H_g.  \label{eq:weight_dm}
\end{align}
In equation~(\ref{eq:weight_gas}), $\Sigma_g$ denotes the integrated gas surface density. Similarly, in equation~(\ref{eq:weight_stars}), $\Sigma_\ast$ refers to the integrated stellar surface density. In addition, $H_g$ and $H_\ast$ refer to the gas scale height and the stellar scale height, respectively. For each individual component, the scale height $H$ is defined as 
\begin{align}
    H = \frac{\Sigma}{2\rho(0)}, \label{eq:H_definition}
\end{align}
with $\rho(0)$ referring to the midplane density of that component. Finally, in equation~(\ref{eq:weight_dm}), $\zeta$ is a numerical prefactor, typically set to $\zeta = 1/3$, and $\Omega_{\rm DM}$ denotes the angular rotation velocity associated with the dark matter. In the case of a flat rotation curve, this can be simplified to $\Omega_{\rm DM} = 4\pi G \rho_{\rm DM} $. 

Having produced expressions for all of the weight components, i.e. the rhs terms of equation~(\ref{eq:central_equality}), we now return to the left hand side, i.e. the pressure term. Since $P_{\rm tot}$ denotes the time-averaged midplane pressure (i.e. $P_{\rm tot}= P(z=0)$ if averaged over time), we can introduce the effective velocity dispersion $\sigma_{\rm eff}^2 = P/\rho$ and write the equilibrium pressure as $P_{\rm tot} = \rho(z=0) \sigma_{\rm eff}^2(z=0)$. Combining equations~(\ref{eq:central_equality}), (\ref{eq:H_definition}), (\ref{eq:weight_sum}), (\ref{eq:weight_gas}), (\ref{eq:weight_stars}), and (\ref{eq:weight_dm}), as well as suppressing the argument $z=0$, \citet{2024ApJ...975..151H} derive an implicit equation for the gas scale height (see their equation 10), namely
\begin{align}
  H_g\left[ 1 + \frac{\Sigma_\ast}{\Sigma_g}\frac{2H_g}{H_g + H_\ast}+ \frac{2\zeta\Omega_{\rm DM}^2}{\pi G \Sigma_g}H_g\right] = \frac{\sigma_{\rm eff}^2}{\pi G \Sigma_g}. \label{eq:Hgas_implicit} 
\end{align}
We can rewrite this, using volume densities instead of integrated surface densities, to obtain \citep[see][equation 21]{2024ApJ...975..151H}: 
\begin{align}
    H_g = \frac{\sigma_{\rm eff}}{\left[2\pi G \rho_g + \frac{4\pi G \rho_\ast}{1+ H_g / H_\ast} + 2\zeta\Omega_{\rm DM}^2\right]^{1/2}}. \label{eq:Hgas_central}
\end{align}
We explicitly note here that technically all of the volume densities appearing in equation~(\ref{eq:Hgas_central}) are to be understood as midplane densities, i.e.~$\rho_g \equiv \rho_g(z=0)$ and so on. This is because all quantities were derived assuming a plane-parallel geometry, and the theory is formulated using projected (two-dimensional) quantities, such as integrated surface densities. Midplane densities can then be calculated using equation~(\ref{eq:H_definition}). We will refer to this subtlety later on, as it becomes important when constructing a subgrid model that bases its calculations on volumetric densities from the PRFM theory.

In plane-parallel geometry, we can furthermore define the (vertical) dynamical time of the system as the time it takes a particle that moves at a speed given by the effective velocity dispersion to traverse two vertical scale heights: 
\begin{align}
    t_{\rm dyn} = \frac{2H_g}{\sigma_{\rm eff}}. \label{eq:2d_dynamical_time}
\end{align}
Substituting equation~(\ref{eq:Hgas_central}) into equation~(\ref{eq:2d_dynamical_time}) and assuming a flat DM rotation curve, as well as $\zeta = 1/3$, then gives us an approximation for the dynamical time as a function of the midplane densities:
\begin{align}
    t_{\rm dyn} = \frac{2}{\left[2\pi G \rho_g + \frac{4\pi G \rho_\ast}{1+H_g/H_\ast}+\frac{8\pi G \rho_{\rm DM}}{3}\right]^{1/2}}. \label{eq:dynamical_time_formula}
\end{align}

For the purpose of calculating star formation rates in line with the PRFM model, we need an equation for the depletion time, i.e.~the time it takes a sheet of gas to turn into stars, given the current star formation rate. In the plane-parallel TIGRESS geometry, this depletion time is given by 
\begin{align}
    t_{\rm dep} = \frac{\Sigma_g}{\Sigma_{\rm SFR}}, 
    \label{eq:tdep_definition}
\end{align}
with $\Sigma_{\rm SFR}$ denoting the star formation rate surface density. PRFM further relates the star formation rate surface density to the midplane pressure, defining the so-called feedback yield $\Upsilon_{\rm tot}$ in the process: 
\begin{align}
    {P_{\rm tot}} = \Upsilon_{\rm tot} \Sigma_{\rm SFR}. \label{eq:feedback_yield_def}
\end{align}
In equation~(\ref{eq:feedback_yield_def}), $\Upsilon_{\rm tot}$ denotes the yield from three different feedback channels, the thermal yield, the turbulent yield, and the magnetic yield, which are associated with corresponding pressure components.
Assuming the total feedback yield is known, we can re-write equation~(\ref{eq:tdep_definition}) as
\begin{align}
    t_{\rm dep} = \Upsilon_{\rm tot}\frac{\Sigma_g}{P_{\rm tot}}, 
\end{align}
and, using equations~(\ref{eq:H_definition}), (\ref{eq:2d_dynamical_time}), and the definition of $\sigma_{\rm eff}$, we derive an equation for the two-dimensional (vertically averaged) depletion time in terms of the two-dimensional dynamical time \citep[][equation 29]{2024ApJ...975..151H}: 
\begin{align}
    t_{\rm dep} = \frac{\Upsilon_{\rm tot}}{\sigma_{\rm eff}}t_{\rm dyn}. \label{eq:tdep_fin}
\end{align}

Equation~(\ref{eq:tdep_fin}), combined with equation~(\ref{eq:dynamical_time_formula}), is the foundation of the PRFM-vol model we introduce here. Two important remarks should be made here at this point: 
\begin{itemize}
    \item Equation~(\ref{eq:dynamical_time_formula}) has been derived under the premise that equation~(\ref{eq:H_definition}) holds. This is generally the case in nature, but not necessarily in simulations in which the gas scale height is not resolved properly. In such cases, the proper form of the expression for the dynamical time has to be derived based on solutions of equation~(\ref{eq:Hgas_implicit}) (see equation 30 in \citealt{Jeffreson2026}). The resulting PRFM-int model formulation requires measurements of $\Sigma_\mathrm{g}$ and $\Sigma_*$, which are straightforward to obtain for an isolated galaxy whose orientation is known, but more challenging for galaxies in cosmological simulations.  The implementation of PRFM-int in a cosmological context will be valuable for large-box, low-resolution simulations, and represents a future goal for LtU.    
    \item Equation~(\ref{eq:tdep_fin}) refers to a two-dimensional (vertically averaged) depletion time. However, for a star formation model in cosmological simulations, we would ideally want a three-dimensional (volumetric) depletion time, giving the star formation rate of each gas cell as 
    \begin{align}
        \dot{m}_\ast = \frac{m_g}{\tilde{t}_{\rm dep}}\label{eq:3d_tdep_definition}.
    \end{align}
    As we will show below, the depletion times $t_{\rm dep}$ (defined in equation~\ref{eq:tdep_fin}) and $\tilde{t}_{\rm dep}$ (defined in equation~\ref{eq:3d_tdep_definition}) are usually not the same in highly resolved simulations. However, given a few reasonable assumptions, $\tilde{t}_{\rm dep}$ can easily be related to $t_{\rm dep}$ by integrating along the vertical axis. 
\end{itemize}

\subsection{3d renormalization of PRFM-vol}\label{ssec:PRFM_renormalization}

The depletion time, as given by equation~(\ref{eq:tdep_definition}), is fundamentally defined as a two-dimensional quantity. While this is not explicitly shown in equation~(\ref{eq:tdep_fin}), it is reflected implicitly, since $\sigma_{\rm eff}$ is measured locally in the midplane, and all volume densities that enter through equation~(\ref{eq:dynamical_time_formula}) are measured at the midplane point as well. As a consequence, there is no one-to-one correspondence between the 2d depletion time $t_{\rm dep}$ and the 3d depletion time $\tilde{t}_{\rm dep}$. However, a relation can readily be derived. In the following, we generalize the presentation in Section 2.4 of \citet{Jeffreson2026} to include effects of dark matter. 

In plane-parallel geometries,
\begin{align}
    \Sigma_{\rm SFR} = \int \dot{\rho}_\ast \,{\rm d}z. \label{eq:Ssfrdef}
\end{align}
Since in the \texttt{AREPO} \citep{Springel:2009aa} code the volume of a gas cell is fixed during the star formation calculation, we can define $\dot{\rho}_\ast$ similar to equation~(\ref{eq:3d_tdep_definition}):
\begin{align}
    \dot{\rho}_\ast = \frac{\rho_g}{\tilde{t}_{\rm dep}}.
\end{align}
Equation ~(\ref{eq:Ssfrdef}) then becomes
\begin{align}
    \Sigma_{\rm SFR} = \frac{\Sigma_g}{t_{\rm dep}} = \int dz\,\frac{\rho_g}{\tilde{t}_{\rm dep}}.\label{eq:3d_tdep_relation}
\end{align}
To solve equation~(\ref{eq:3d_tdep_relation}) for $\tilde{t}_{\rm dep}$, we proceed as in \citet{Jeffreson2026} and assume that $\tilde{t}_{\rm dep}$ has the same functional dependencies as $t_{\rm dep}$, but is modified by a constant, i.e. $\tilde{t}_{\rm dep} = U_f\times \tau$, where $U_f = R_f^{-1}$, with $R_f$ being the constant used in \citet{Jeffreson2026} and $\tau$ is the same in functional form and dependence (on densities and pressures) as the product of terms on the right-hand side of equation~(\ref{eq:tdep_fin}), except that it is evaluated for general $z$ rather than at $z=0$. 
To derive an analytic expression for $U_f$, we assume that the gas disk in our simulation is fully resolved, and that the vertical density profile is reasonably described by an exponential function: 
\begin{align}
    \rho_g = \rho_{\rm m} \exp{(-z/H_g)}, \label{eq:gas_density}
\end{align}
with the gas scale height $H_g$ as the vertical scale length.  We can then re-write equation~(\ref{eq:3d_tdep_relation}) as follows: 
\begin{align}
    t_{\rm dep} &= \frac{\int {\rm d}z\,\rho_g}{\int {\rm d}z\,\frac{\rho_g}{\tilde{t}_{\rm dep}}} = U_f\times \frac{\int {\rm d}z\,\rho_g}{\int {\rm d}z\,\frac{\rho_g}{\tau}}\\
    &\therefore  U_f = t_{\rm dep} \times \frac{\int {\rm d}z\,\frac{\rho_g}{\tau}}{\int {\rm d}z\,\rho_g}. \label{constant definition}
\end{align}

In analogy to the derivation of equation~(21) in \citet{Jeffreson2026}, we obtain an integral equation for $U_f$, which reads 
\begin{align}
            U_f &=  \frac{\int {\rm d}z\,\tilde{\rho}_g^\frac{\alpha+\beta}{1-2\beta}\tilde{\rho}_g\left[2\pi G \tilde{\rho}_g + \frac{4\pi G \tilde{\rho}_\ast}{1+ H_g/H_\ast}+\frac{8\pi G \tilde{\rho}_{\rm DM}}{3}\right]^{1/2}}{\left[2\pi G \tilde{\rho}_g + \frac{4\pi G \tilde{\rho}_\ast}{1+ H_g/H_\ast}+\frac{8\pi G \tilde{\rho}_{\rm DM}}{3}\right]^{1/2}\times \int {\rm d}z\, \tilde{\rho}_g}  \label{eq:complicated_integral_three},
\end{align}
where we have introduced scaled densities, i.e~densities divided by the midplane gas density: 
\begin{align}
    \tilde{\rho}_g = \rho_g/\rho_{\rm m}, \qquad \tilde{\rho}_\ast = \rho_\ast/\rho_{\rm m}, \qquad
    \tilde{\rho}_{\rm DM} = \rho_{\rm DM}/\rho_{\rm m}. 
\end{align}
The numerical constants $\alpha$ and $\beta$ are the exponents of scaling relations for the total feedback yield and the effective velocity dispersion, i.e.:
\begin{align}
    &\Upsilon_{\rm tot} = \Upsilon_0 \left(\frac{P}{P_0}\right)^{-\alpha}\\
    &\sigma_{\rm eff} = \sigma_0 \left(\frac{P}{P_0}\right)^{\beta}.\label{eq:betadef}
\end{align}
Equation~(\ref{eq:complicated_integral_three}) can be solved analytically for two instructive limiting cases. In the first case, we assume that the DM density is negligible. Moreover, we make the assumption that the vertical scaling of the stellar density is the same as that of the gas density. Under these simplifying assumptions, we derive (as in equation~(22)-(25) of \citealt{Jeffreson2026})
\begin{align}
    U_f = \frac{\int {\rm d}z\,\tilde{\rho}_g^\frac{\alpha+\beta}{1-2\beta}\tilde{\rho}_g^{\frac{1}{2}}}{\int {\rm d}z\,\tilde{\rho}_g} = \frac{\int_0^1 {\rm d}x\, x^{\frac{3}{2}+\frac{\alpha+\beta}{1-2\beta}}}{\int_0^1 {\rm d}x\,} = \frac{1}{\frac{3}{2}+\frac{\alpha+\beta}{1-2\beta}}, \label{eq:baryon_dominance}
\end{align}
where we have used the exponential dependence of the gas density on the vertical coordinate $z$ (equation~\ref{eq:gas_density}) to transform the coordinates and the integral boundaries. It is worth noting that this integral transformation implies that we can follow the vertical density profile down to very low densities\footnote{Technically, it also implies that all gas is star-forming. A more rigorous calculation would take the numerical star formation threshold into account. For simplicity -- and for the sake of deriving an analytic expression for $U_f$ -- we do not do that here, and opt to instead verify that the final expression we derive works as intended (see Section~\ref{ssec:sfr_isolated}). }. In the context of cosmological simulations, this corresponds to the limit of infinite resolution. \citet{Jeffreson2026} consider also the limit of very low resolution, which they show (see their equation~(24)) results in $U_f=1$. 
This low-resolution case is applicable in the PRFM-int implementation, which is explicitly designed for unresolved disks. 
However, given that PRFM-vol is only applicable for resolved disks, equation~(\ref{eq:baryon_dominance}) should be appropriate for the model presented here. 

For now, let us turn our attention to the case in which DM dominates the gravitational potential. If in equation~(\ref{eq:complicated_integral_three}) we neglect baryon terms in the sums, we obtain, after scaling out the gas density: 
\begin{align}
    U_f = \frac{\int {\rm d}z\, \tilde{\rho}_g^{\frac{\alpha + \beta}{1-2\beta}+1} \frac{\rho_{\rm DM}}{\rho_{\rm DM, m}}}{\int {\rm d}z\, \tilde{\rho}_g}.
\end{align}
We can then assume that the DM density does not vary appreciably throughout the disk, implying $\rho_{\rm DM}/\rho_{\rm DM,m} \approx 1$. Using this assumption, and proceeding as in the case in which baryons dominate the potential, we obtain
\begin{align}
    U_f = \frac{1}{1+\frac{\alpha+\beta}{1-2\beta}}\label{eq:DM_domination}
\end{align}
for the case in which the potential is dominated by DM. For general cases, equation~(\ref{eq:complicated_integral_three}) cannot be solved analytically, which means a fully accurate solution would have to be calculated numerically and on-the-fly, a procedure that is too tedious to employ as part of a subgrid model for hydrodynamical simulations. Instead, we assume that the integral in equation~(\ref{eq:complicated_integral_three}) is reasonably well-behaved, and that we can approximate the true solution through a suitable interpolation between the two limiting cases. One way of doing this is to fix the 3d renormalization constant to 
\begin{align}
U_f = \left(\frac{3}{2} -\frac{1}{2\left[1+\frac{\rho_g + \rho_\ast}{\rho_{\rm DM}}\right]} + \frac{\alpha+\beta}{1-2\beta}\right)^{-1}. \label{eq:renormalization_factor}  
\end{align}
Equation~(\ref{eq:renormalization_factor}) is the functional form we employ in PRFM-vol, with the indices $\alpha$ and $\beta$ fixed by power law fits to the TIGRESS simulations, as we shall discuss below. 

\subsection{Velocity kicks at birth}\label{ssec:vkicks}

From the development and application of the TIGRESS/Schmidt model \citep{Burger2025}, we know that the combination of the \citet{Ostriker2022} TIGRESS equation of state and the adopted, PRFM inspired star formation law ($\dot{\rho}_*\propto \rho_g^2$) leads to stellar disks that are substantially thinner than in the TNG model.

This did not come as a surprise, as the combination of a lower pressure in the ISM with a star formation law that favours dense regions more strongly than in TNG will result in star formation occurring closer to the midplane. On top of that, old stars are expected to stay close to the midplane, as there is less pressure to support outward migration of gas, resulting in a thinner gas disk and a deeper potential well. 

When presenting the TIGRESS/Schmidt model we verified that stellar disks are thinner, using both isolated and cosmological simulations. In PRFM-vol, we now aim to remedy this issue. In order to do so, we observe that for star-forming gas, random motion is encoded through the effective equation of state, via 
\begin{align}
    \sigma_{\rm eff} = \sqrt{\frac{P}{\rho}}. \label{eq:seff}
\end{align}
As mentioned above, the pressure measured in TIGRESS is a sum of three distinct components; turbulent pressure, thermal pressure, and magnetic pressure. However, since the relevant scale for the turbulence that sets the turbulent pressure component is unresolved in cosmological simulations, all of the three components are essentially included as thermal pressure when using an eEoS model. To compensate for that, in PRFM-vol we add velocity kicks to newly-born stars when initializing their velocities. We sample the magnitude of those velocity kicks from a Gaussian distribution with zero mean and velocity dispersion $\sigma_{\rm eff}$. 
We then kick the star along a direction randomly drawn from the unit sphere. We note that an alternative approach would be to add velocity kicks strictly perpendicular to the disk. This would increase the expected dynamical heating effect, but require knowledge of the vertical direction at each timestep\footnote{In some sense, we compensate for not doing that by using the total pressure in setting $\sigma_{\rm eff}$. In simulations where the vertical direction is known, the fully correct thing to do would be to direct the velocity kicks along the vertical coordinate, and to take only the turbulent pressure into account.}. Within the context of PRFM-vol this is not practicable in cosmological simulations, since it would add computational overhead that we attempt to avoid by not using PRFM-int. 

\subsection{Estimating the scale height ratio}\label{ssec:heightratio}

Having derived an analytic expression for $U_f$ in Section~\ref{ssec:PRFM_renormalization}, we are now close to obtaining a closed set of equations that define a completely local formulation of a PRFM-based subgrid model, i.e. PRFM-vol.
The only remaining explicit dependence on quantities that require knowledge of the orientation of the galactic disc is contained in equation~(\ref{eq:dynamical_time_formula}), namely the dependence on the ratio of the two scale heights. To circumvent having to keep track of the moment of inertia tensor of each galaxy at all times, we implement two ways to estimate this ratio by means of quantities that can be calculated locally. In a first approximation, we set $H_g/H_\ast = 1$, which means that stars and gas contribute equally to the dynamical time. In a different version of the model, we approximate the scale height ratio by means of the squared ratio of velocity dispersions, i.e.  
\begin{align}
    \frac{H_g}{H_\ast} \approx \frac{3\sigma_{\rm eff}^2}{\sigma_\ast^2}\label{eq:scale_height_ratio}. 
\end{align}
This is an approximation of the ratio that would be predicted for the equilibrium vertical distributions in response to a fixed vertical gravitational field.\footnote{\citet{Jeffreson2026} assessed an alternative estimator, $H_g/H_\ast = \sigma_\mathrm{eff}/\sigma_{\ast,z}$, which represents the relative equilibrium responses to a fixed total midplane density.} We here use the scaled 3d stellar velocity dispersion, since in cosmological simulations we will in general not know the vertical axis of any given disk galaxy. 
Later on, we shall assess the goodness of approximation~(\ref{eq:scale_height_ratio}) using isolated galaxy simulations. There, we will also compare against a different estimator based on the vertical component of the stellar velocity dispersion only, i.e.  
\begin{align}
      \frac{H_g}{H_\ast} \approx \frac{\sigma_{\rm eff}^2}{\sigma_{\ast,z}^2}\label{eq:scale_height_ratio_two}
\end{align}
which serves as a useful sanity check that can be done when the disk orientation is known. 

\subsection{Implementation of PRFM-vol} \label{ssec:imp_prfmvol}
Here we shall summarize the main steps taken to implement PRFM-vol into the moving mesh code \texttt{AREPO}. The name PRFM-vol is in line with the nomenclature in \citet{Jeffreson2026} 
but we emphasize that the specifics of the two models differ slightly (as we outline below), and that we developed this version to be applied in cosmological simulations. 

The main part of the implementation consists of calculating the local gas densities, stellar densities, and stellar velocity dispersions around each gas cell. To that end, we perform iterative neighbour searches for each gas cell in order to determine the radius of the sphere that contains a fixed number of nearest neighbours. This is done independently for DM and star particle neighbours. Within the corresponding spheres of nearest neighbours, we then calculate local densities and velocity dispersions that are stored with the gas cell at the centre of the sphere. We use a standard cubic spline kernel to weigh the contributions of individual particles (both DM and stars). In order to perform these neighbour searches, we rely on the presence of the full gravity tree. Since \texttt{AREPO} uses hierarchical timestepping, this full tree is only present on global timesteps, which means that we update the local densities and velocity dispersions at these times. In our simulations, the time between these calculations is typically of the order of a few Myrs. In the lowest resolution cosmological simulations that we run for this project, the maximum time interval between updates of local DM density and stellar density is $\sim 20$ Myrs at the end of the simulation, and much smaller at earlier times. While not insignificant, we still do not expect the structure of established $z\sim0$ disk galaxies to change drastically on such time scales. 

\begin{figure}
    \centering
    \includegraphics[width=\linewidth]{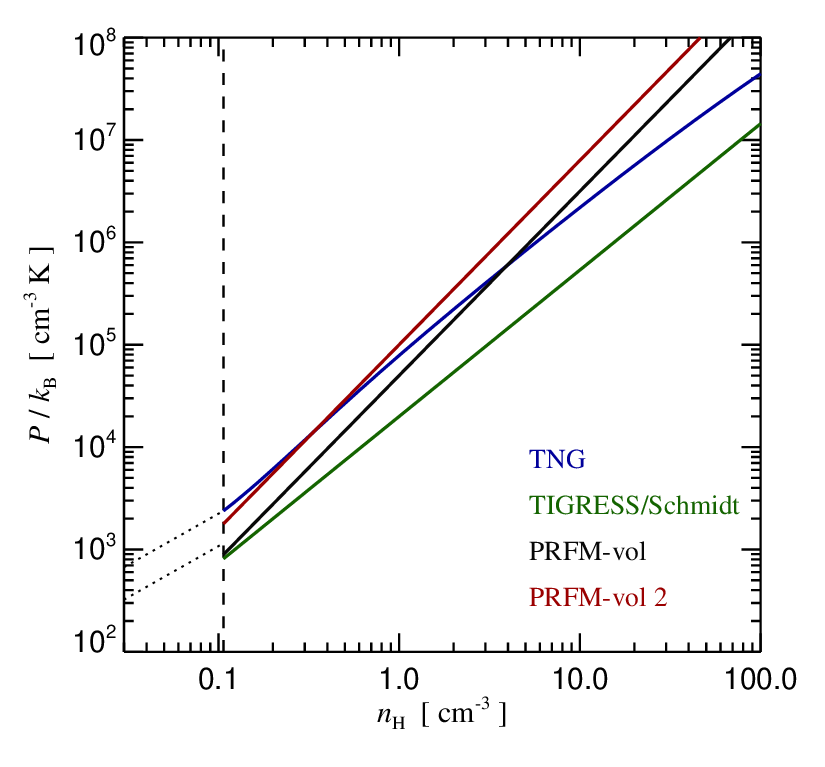}\\
    \includegraphics[width=\linewidth]{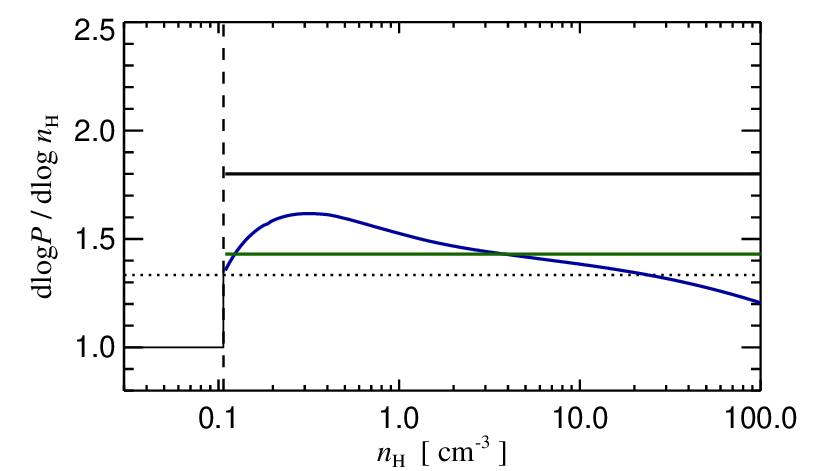}\\
    \includegraphics[width=\linewidth]{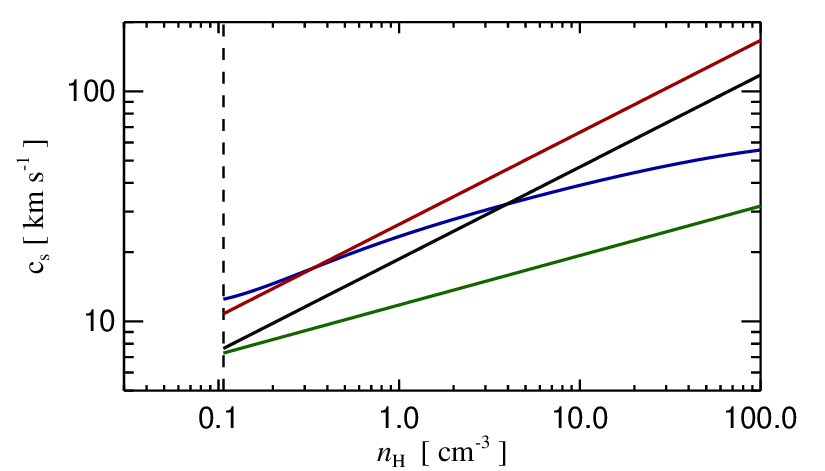}
    \caption{Comparison of the effective equations of state used in PRFM-vol (black lines), TNG (blue lines) and TIGRESS/Schmidt (green lines). In the upper panel, we show pressure as a function of hydrogen number density, while the logarithmic slope of the relations is displayed in the middle panel, and the resulting isothermal sound speeds are shown in the bottom panel. In both the upper and the bottom panel, we additionally include a line corresponding to two times the PRFM-vol eEoS pressure (red lines). The dashed vertical lines mark the adopted density threshold for star formation above which the effective equation of state is applied and the horizontal dotted line in the middle panel marks the slope below which the Jeans mass decreases with increasing density. The PRFM-vol equation of state is much harder than the other two. However, note that PRFM-vol pressure and sound speed around the star formation density threshold are significantly lower than in TNG and roughly comparable to TIGRESS/Schmidt.}
    \label{fig:eos_comp}
\end{figure}

From the local volume densities of the gas, stars, and DM, we can then calculate the depletion time of star-forming gas cells
using equations~(\ref{eq:tdep_fin}), (\ref{eq:2d_dynamical_time}), (\ref{eq:renormalization_factor}), and (\ref{eq:seff}). As is common in eEoS models, we define star-forming gas as gas whose density exceeds a numerical star formation density threshold. 

As we outlined in \citet{Burger2025}, the model we introduce here is not attempting to explicitly capture physical processes in the ISM. Instead, we use an effective equation of state to determine the pressure in the star-forming ISM. As an equation of state, we use the fit to the TIGRESS simulations presented in \cite{2024ApJ...975..151H}: 
\begin{align}
    \log(P/k_{\rm B}) &= 1.8 \times \log(n_{\rm H}) +4.7, 
\label{eq:TIGRESS_EOS}
\end{align}
where $k_{\rm B}$ is the Boltzmann constant, $n_{\rm H}=x_{\rm H}\rho_g /m_{\rm p}$ is the hydrogen number density, $x_{\rm H}$ is the hydrogen mass fraction, and $m_{\rm p}$ is the proton mass. 
This expression is derived from the mass-weighted mean value of $\sigma_\mathrm{eff}$ fit across a set of models at $P/k_{\rm B}>10^4 \ {\rm cm}^{-3}\ K$ \citep{Ostriker2022}. 

As noted in \citet{2024ApJ...975..151H}, a single power law such as equation~(\ref{eq:TIGRESS_EOS}) cannot continue to arbitrarily low density since there is a floor on $\sigma_\mathrm{eff}$ from the thermal velocity dispersion of warm gas, and indeed a flattening of the relation is seen in the TIGRESS simulation results at low density and pressure.
In \citet{Jeffreson2026}), this issue is circumvented by adopting equation~(\ref{eq:TIGRESS_EOS}) across the star-forming ISM, but imposing an effective velocity dispersion floor of $\sigma_\mathrm{eff} = 12\, \mathrm{km\ s^{-1}}$ when calculating the depletion time. This does prevent artificially low velocity dispersions at the cost of introducing a small degree of numerical inconsistency. In our default model, we here instead chose not to impose the velocity dispersion floor. However, in the context of our multizoom simulations (see Section~\ref{sec:multizoom}), we did test a version of the model that both imposes the velocity dispersion floor and alters the eEoS consistently, effectively introducing a broken power form with a logarithmic slope of 1 at low densities. We report on differences between this model and our default model in Section~\ref{ssec:modeeos}. 


In Fig.~\ref{fig:eos_comp}, we show a comparison of our effective equation of state to the ones used in the TIGRESS/Schmidt or the TNG simulations presented in \cite{Burger2025}\footnote{We note that the TIGRESS/Schmidt eEoS was derived from the same suite of TIGRESS simulations. As outlined in \citet{2024ApJ...975..151H}, the TIGRESS/Schmidt eEoS parametrizes the midplane pressure measured in TIGRESS, while our new PRFM-vol eEoS parametrizes the pressure derived from the mass weighted average velocity dispersion, which may be more appropriate to use in a subgrid model for galaxy formation simulations.}. We also show a line marking twice our new PRFM-vol equation of state (labeled PRFM-vol 2), which will be relevant in the context of our multizoom simulation analysis presented in Section~\ref{sec:multizoom}.

With the effective equations of state, we additionally show -- to the left of the numerical star formation threshold -- two dotted lines that mark two different pressure curves for gas at $10^4{\rm K}$ that approaches the threshold density from below. The upper line marks the pressure of warm ionized gas, while the lower line would be expected for cold neutral gas. In \citet{Burger2025}, we tested a way to fade into the eEoS, ensuring that the pressure is continuous at the density threshold. While numerically simple, we also found this to not make a big difference in practice. In the present work, where we test a variety of equations of state, we do not fade in, and instead note that for all of our models, the pressure at the star formation threshold remains in the range spanned by the two dotted lines. 

As highlighted by the logarithmic slopes of the equations of state shown in the middle panel of Fig.~\ref{fig:eos_comp}, the PRFM-vol eEoS is much harder than the other two, even intersecting the TNG eEoS at hydrogen number densities of $\sim 50\,{\rm cm}^{-3}$.
In principle, with the harder eEoS, 
runaway collapse of nonlinear, very dense structures could be prevented.  However, in practice the normalization of the eEoS at low density is more important because much of the gas is at densities only modestly above the star formation threshold. The lower pressure and corresponding velocity dispersion of both the TIGRESS/Schmidt eEoS and the PRFM-vol eEoS at low density makes these models more susceptible to gravitational instability than TNG. 

An alternative way of looking at this is through the lens of the bottom panel of Fig.~\ref{fig:eos_comp}, where we show the isothermal sound speeds corresponding to each eEoS. At the numerical star formation density threshold, PRFM-vol and TIGRESS/Schmidt sound speeds are far below the values achieved in TNG and PRFM-vol~2 -- the version of our model in which we multiply the eEoS with a factor of 2. As the effective sound speed directly enters the equation for the Toomre $Q$ parameter (see equation~\ref{eq:toomre}), it is easy to see that eEoS models with lower normalization will be more prone to disk instabilities. 
We will focus more on this point when discussing our multizoom simulation results.

For the total feedback yield, we use a fit to the TIGRESS simulations presented in \citet{Ostriker2022}: 
\begin{align}
    \log(\Upsilon_{\rm tot}) =  -0.212 \times \log(P/k_{\rm B}) + 3.86.\label{eq:feedback_yield_calib}
\end{align}
Note that equation~(\ref{eq:feedback_yield_calib}) immediately fixes $\alpha = 0.212$ in equation~(\ref{eq:renormalization_factor}). Moreover, equation~(\ref{eq:TIGRESS_EOS}) in combination with equation~(\ref{eq:betadef}) implies that 
\begin{align}
    \beta = \frac{1}{2}\left(1 - \frac{1}{1.8}\right)  \approx 0.22.
\end{align}
We therefore use those values to calculate the correction factor to the dynamical time, equation~(\ref{eq:renormalization_factor}). Having calculated the depletion time, the star formation rate is simply $m_{\rm g}/\tilde{t}_{\rm dep}$. At each time step $\Delta t$, a probability for star formation is calculated for each gas cell whose density is above the threshold for star formation, $p = 1 - \exp(-\Delta t/t_{\rm dep})$, which will approach 1 for very short depletion times and 0 for excessively long depletion times. Whether the gas cell is converted into a star is then decided using rejection sampling, and if a star is indeed formed, we perform the initial velocity kick described in Section~\ref{ssec:vkicks}. In case an effective model for winds is used as well, as in, e.g., Illustris-TNG \citep{Springel2018, Pillepich2018, Marinacci2018, Naiman2018, Nelson2018}, the decision procedure is slightly more complicated, since the probability for the gas cell to be converted into a wind particle also has to be considered. If that is the case, this is done analogous to how it has been described in \citet{Pillepich2018a}. We emphasize here that the above description covers just the new star formation model PRFM-vol, but we stress that PRFM-vol can seamlessly be interfaced with the galaxy formation module used in the Illustris-TNG and Millennium-TNG \citep{Pakmor2023} simulations. Below, we describe the simulation setups used to test PRFM-vol. 

\section{The simulation setups}\label{sec:setups}

In this section, we briefly describe the setup of our series of numerical simulations, which we conduct in order to test PRFM-vol against both the TNG and TIGRESS/Schmidt models. In large part, the simulation setups are identical to the ones we used in \cite{Burger2025}, and we refer the reader to that article for further details about how the initial conditions are created. Nonetheless, we summarize here the most salient points and note any specific changes that have been made for the purpose of testing PRFM-vol. 

\subsection{Isolated galaxies} \label{ssec:setups_isolated}
As in \citet{Burger2025}, we use the method outlined in \cite{SpringelHQ2005} to generate initial conditions of isolated galaxy models in dynamical equilibrium. The original models consist of four different pure gas disks (D1--D4, with disk masses given in Table~\ref{tab:isolateddisks}) in a static \cite{Hernquist:1990be} DM halo with total mass $1.65\times10^{12}\,{\rm M}_{\odot}$. 

In contrast to what was done in \cite{Burger2025}, we do not use a static DM potential here. Instead, we realize a live DM halo, assigning particle velocities by approximating the distribution function with a triaxial Gaussian and velocity dispersions calculated using the Jeans equation. We keep the relevant parameter choices for the DM halo, $v_{200}=169\,{\rm km\, s^{-1}}$ and $c=12$, and we set a DM streaming velocity through the spin parameter, which we set to $\lambda = 0.04$. The scale lengths of the galactic disks are set by demanding that the disk angular momentum be a fixed fraction of the halo spin, of the order of the disk mass fraction. The resulting implicit equation for the scale length is then solved iteratively for each setup, yielding scale lengths of $\sim 5.5\,{\rm kpc}$ for all of them. 

Including a live DM halo significantly increases the runtime of our isolated simulations. However, it is necessary to test the default version of PRFM-vol, since the calculation of the star formation rate relies on local kernel estimates of the DM density. In line with \cite{Burger2025}, we sample the gas disk in each of our initial setups with $8\times 10^5$ particles. Recall that the different disk masses of our setups then lead to corresponding differences in mass resolution \citep[see Table 1 of][]{Burger2025}. Here, we carry this difference over to the DM mass resolution as well, in order to keep the difference in the default mass resolutions fixed between the DM particles and the gas cells, which corresponds to the case typically encountered in practice. To keep the computational cost reasonable, the default DM mass resolution is set to $10\%$ of the gas mass resolution. 

In addition to the inclusion of a live DM halo, we also run setups where we aim to test the dependence of the star formation rate on the stellar density, which mainly enters through the dynamical time as given in equation~(\ref{eq:dynamical_time_formula}), as well as the renormalization factor (equation~ \ref{eq:renormalization_factor}). To that end, we generate additional sets of initial conditions in which we include a disk composed of old stars. For each setup (D1 - D4), we consider versions in which the disk contains $50\%$ stars and $90\%$ stars, respectively. These stars are included as collisionless particles, and we split the $8\times 10^5$ particles in the disk between gas cells and collisionless old star particles in proportion with the disk mass fraction of either component. While the gas disk has to be in hydrostatic equilibrium, the vertical thickness of the disk of old stars is instead supported by velocity dispersion. We set the scale lengths of both disks to be equal, and then set the stellar scale height as $0.1$ times the scale length with a corresponding vertical velocity dispersion. We assume both the gaseous and the stellar disk to be exponential and the velocities of the stars are -- as in the case of the DM particles -- calculated using the (now cylindrical) Jeans equation.
Our setups are evolved for a total of $1.5\,{\rm Gyrs}$ and snapshots taken every $37.5\,{\rm Myrs}$. We allow stars to form at the rate predicted by our model but -- as in \citet{Burger2025} -- TNG winds are turned off in the isolated galaxy simulations. An explanatory overview of the isolated disk setups is given in Table~\ref{tab:isolateddisks}. 

\begin{table*}
\centering
\begin{tabular}{ccccc}
\hline
model name & gas disk mass & baryon mass resolution & DM mass resolution  & \% old stars \\
 & $M_{\rm disk}\;[{\rm M}_\odot]$   & $m_{\rm gas}\;[{\rm M}_\odot]$ & $m_{\rm DM}\;[{\rm M}_\odot]$ & \\
 \hline
     &  & & & 0 \\
 D1  & $6.09 \times 10^9$ & $7.62 \times 10^{3}$ & $7.7\times 10^{4}$ & 50 \\
      &  &  &  & 90 \\
\hline 
   & & & & 0 \\
 D2  & $1.22\times 10^{10}$ & $1.53\times 10^{4}$ & $1.53 \times 10^{5}$ & 50\\
  & & & & 90 \\
\hline 
      & & & & 0 \\
 D3  & $2.44\times 10^{10}$ & $3.05\times 10^{4}$ & $3.29 \times 10^{5}$ & 50 \\
  & & & & 90 \\
  \hline 
  & & & & 0 \\
 D4  & $4.88\times 10^{10}$ & $6.09\times 10^{4}$ & $6.76 \times 10^{5}$ & 50\\
  & & & & 90 \\
 \hline
 \end{tabular}
\caption{Initial parameters of our isolated galaxy models in haloes of total mass $1.65\times 10^{12}\, {\rm M}_\odot$. At our default resolution, the baryonic disks are resolved with $8 \times 10^5$ resolution elements. They are comprised of gas cells and old star particles initially of the same mass. Thus, the number of gas cells and old stars used in a specific realization is determined by the old star percentage of the particular model.}
\label{tab:isolateddisks}
\end{table*}

\subsection{Multizoom simulations} \label{sse:multizoom_ics}

We use the same set of initial conditions that we introduced in \cite{Burger2025} (see Table~2 therein), where a full description of the pipeline to create multizoom initial conditions is provided. In short, multizoom initial conditions can be created using the \texttt{NGENIC} code, which is part of the simulation code \texttt{GADGET-4}. To do so, one has to select a set of haloes of interest from the group catalog of a previously run parent box simulation, generate a list of those haloes, and then pass this list to \texttt{NGENIC}, along with the corresponding snapshot of the parent box simulation\footnote{In our case, we use the $740\,{\rm Mpc}$ run from the MillenniumTNG project \citep{HernandezAguayo2023,Pakmor2023}}. With multizoom ICs, we can define several disjoint high resolution regions in a single parent box, encompassing the Lagrangian coordinates of the particles that belong to the selected haloes. This effectively allows us to zoom onto multiple objects in a single simulation. For the setup to work, it is crucial that the ICs of the parent box have also been created with \texttt{NGENIC} and that the code is used with the same random number seed. Upon receiving the list of target haloes that define the high-resolution region, the code first identifies the Lagrangian coordinates of the particles in that region. Using those, the spatial region containing those same particles is identified as the high-resolution region, and enlarged by a suitable factor $f_{\rm enlarge}$ to protect against contamination by low resolution particles. A preliminary unperturbed particle is then created such that the grid-structure of the low-resolution region is maintained, while the high-resolution region is sub-sampled with $N_{\rm zf}^3$ high-resolution particles, where ${\rm zf}$ is called the zoom factor. The high-resolution particles are arranged such that the centre-of-mass of the grid cells that host them is the same as in the parent box. Following this step, separate displacement fields are calculated for the low resolution region and the high resolution region. In a final step, the resolution outside of the high-resolution region is downgraded. This is done in a way that keeps the gravitational forces that act in the high-resolution region unchanged, at least at the initial time. For the downgrading process, one has to choose an opening angle $\theta$. \texttt{GADGET-4} then generates the gravitational oct-tree from the preliminary particle distribution that we have generated. It walks the tree for all high-resolution particles and opens the nodes that are seen under the angle $\theta$. The content of nodes that have not been opened by any high-resolution particle is replaced by a single, more massive particle at the centre-of-mass, with the node's centre-of-mass velocity as its initial velocity. 

The multizoom ICs used here \citep[and presented in detail in][]{Burger2025} use the parameters $f_{\rm enlarge} = 2.5$ and $\theta = 0.5$. Each initial condition contains 20 high-resolution haloes in a different mass bin, $\log(M_{\rm vir}/{\rm M}_{\odot}) = 11.097\pm 0.05$, $\log(M_{\rm vir}/{\rm M}_{\odot}) = 12\pm 0.05$, and $\log(M_{\rm vir}/{\rm M}_{\odot}) = 12.903\pm 0.05$. They are available with zoom factors 2, 4, and 8, with zoom factor 4 roughly corresponding to TNG-50 resolution in the high-resolution region \citep[see Table 2 of][]{Burger2025}.

While we carry out simulations starting from all the initial conditions presented in \citet{Burger2025}, we note that in the previous work, all major results that we presented have been derived from the setup with $\log({M}_{\rm vir} / {\rm M}_\odot) = 12$ and zoom factor 4. For comparison, we therefore chose this as our benchmark setup in the present work as well. Results presented in Section~\ref{sec:multizoom} are derived from this setup unless explicitly stated otherwise.

Some interesting points about resolution can be made by comparing our benchmark results to the setup with zoom factor 2, as well as the zoom factor 2, 4, and 8 setups with $\log({M}_{\rm vir} / {\rm M}_\odot) = 11.097$. We briefly discuss this in Appendix~\ref{apxsec:resolution}. We find that the more massive setups (with $\log({M}_{\rm vir} / {\rm M}_\odot) = 12.903$) display similar galaxy phenomenology as our benchmark setup, and we thus refrain from presenting those here.

\section{Isolated galaxies}\label{sec:isolated_galaxies}

Here we report on the results of our isolated galaxy runs. The main focus in this section is to highlight the impact of each individual modification introduced in PRFM-vol in comparison to TIGRESS/Schmidt. First, we look at how the star formation rate changes when including the gravitational contributions of stars and DM in the calculation of the depletion time (see equation~\ref{eq:tdep_fin}). We also investigate the role of the normalization factor (equation~\ref{eq:renormalization_factor}). Then, we compare the emergent stellar scale heights to the ones obtained when using the TIGRESS/Schmidt model, in order to assess the effect of giving velocity kicks to newly born stars. Finally, we look at individual runs with PRFM-vol and compute the ratio between the stellar scale height and the gas disk scale height, comparing it to the estimate obtained through the two different ratios of velocity dispersions discussed in Section~\ref{ssec:heightratio}.  

\subsection{Star formation rates} \label{ssec:sfr_isolated}

We examine the star formation rates through the lens of the TIGRESS scaling relation for pressure as a function of star formation rate surface density \citep{Ostriker2022}. In \cite{Burger2025}, we derived a formula for the star formation rate in the TIGRESS/Schmidt model, using this scaling relation as a constraint. Consequently, we were able to match the relation perfectly, as long as we considered pure gas disks -- meaning without pre-existing stars in the initial conditions -- and disabled the actual formation of stars in the simulations. 
However, we also demonstrated that once we allow stars to form, the simulated $P - \Sigma_{\rm SFR}$ relation no longer matched the TIGRESS prediction. Therefore, we here revisit this point, with a focus on the performance of our new subgrid model, PRFM-vol.

\begin{figure*}
    \centering
    \includegraphics[width=0.33\linewidth]{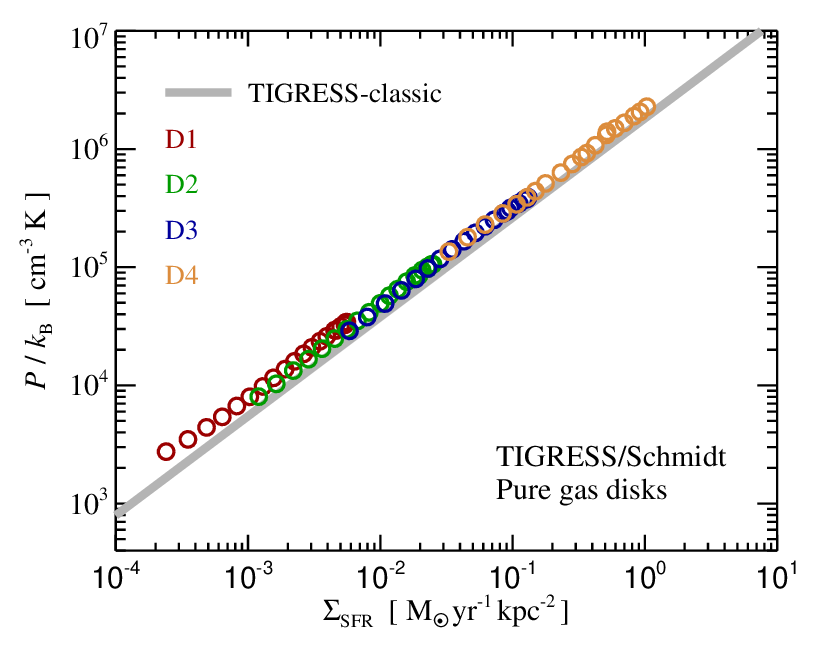}
    \includegraphics[width=0.33\linewidth]{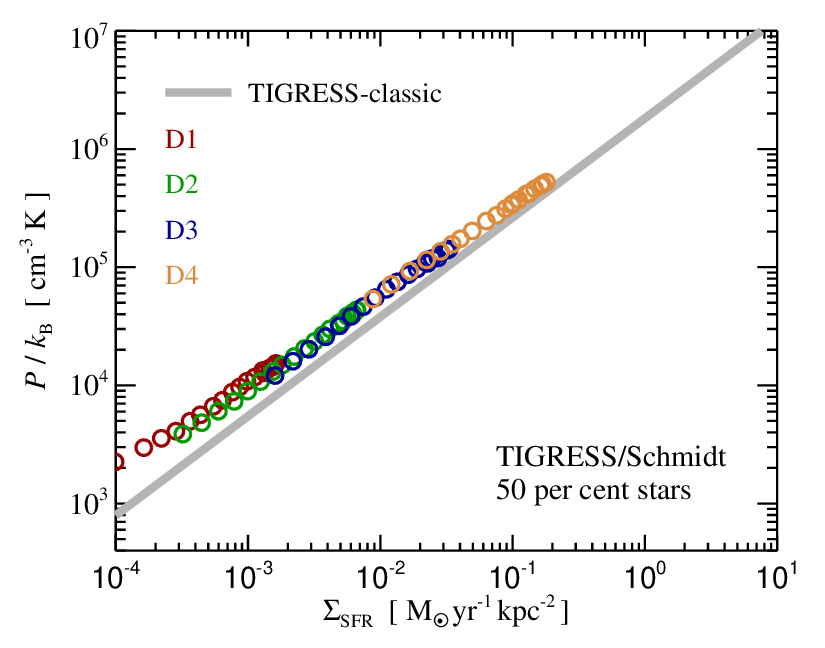}
    \includegraphics[width=0.33\linewidth]{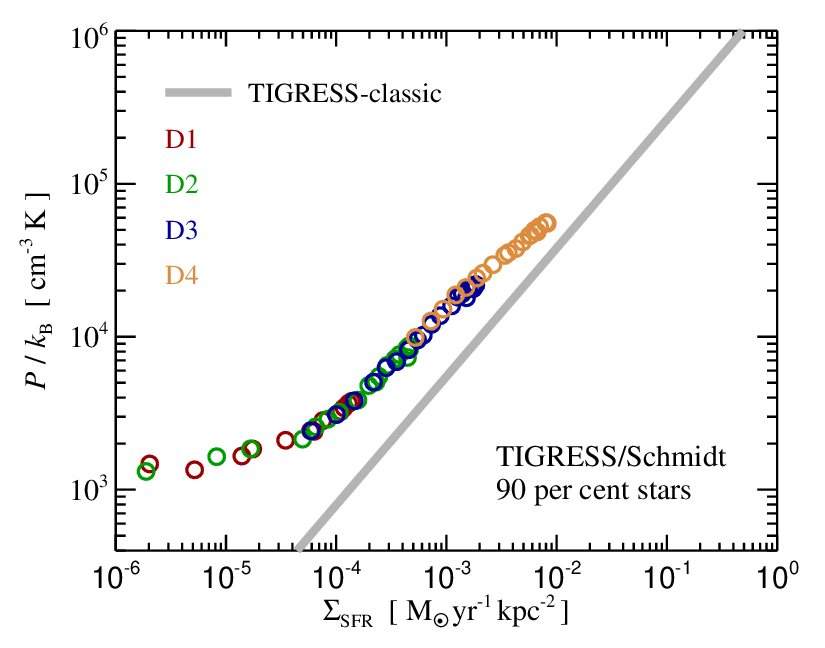}\\
    \includegraphics[width=0.33\linewidth]{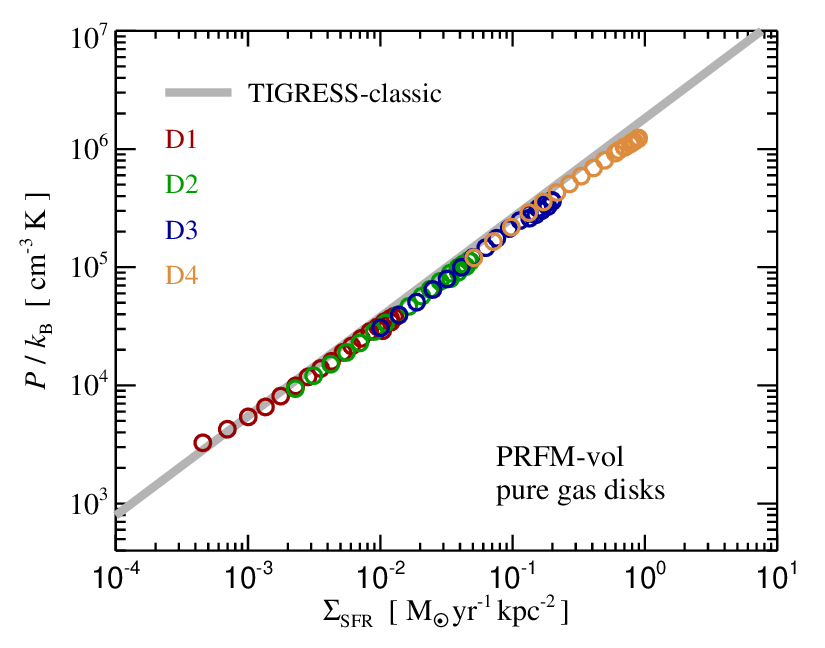}
    \includegraphics[width=0.33\linewidth]{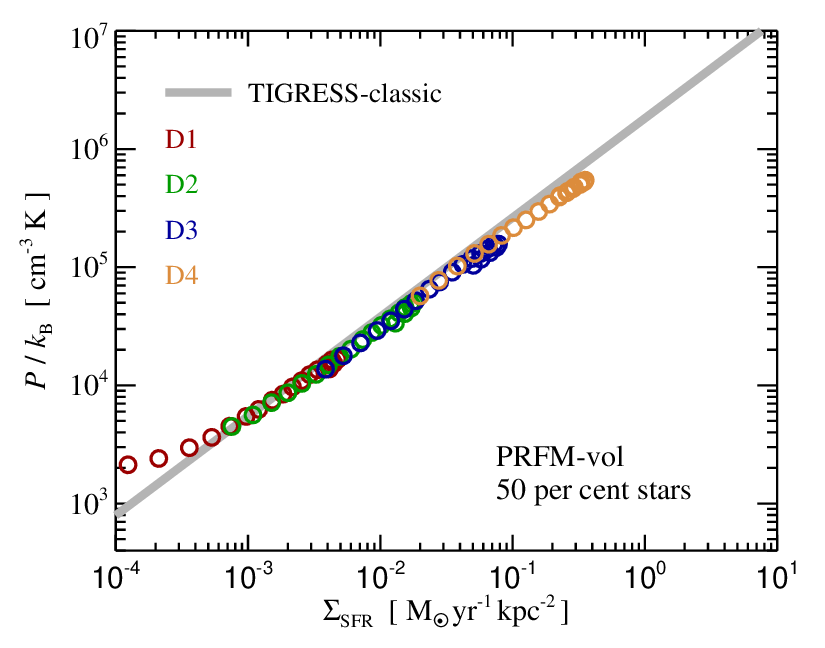}
    \includegraphics[width=0.33\linewidth]{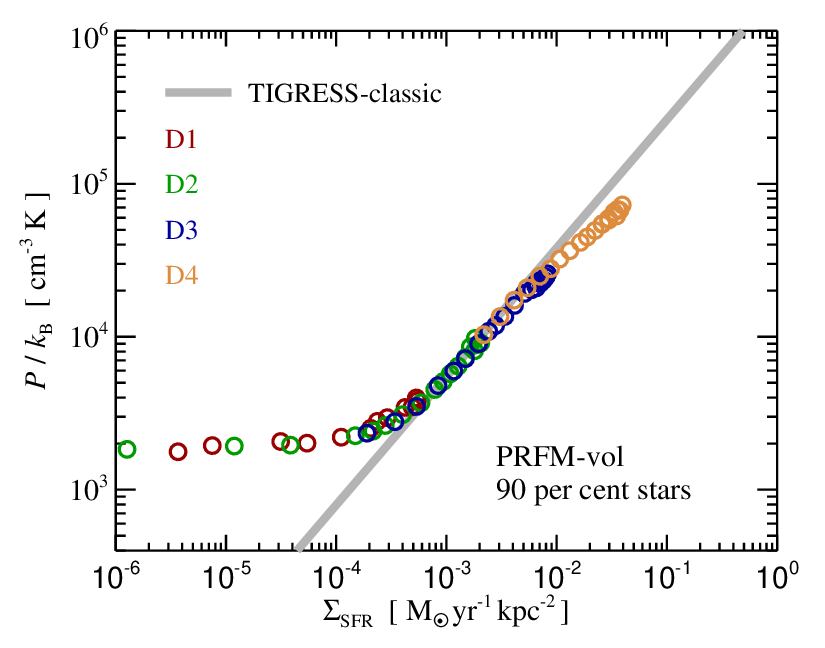}\\
    \caption{Comparison of the pressure vs. star formation rate surface density scaling relation in different model realizations. In the top row, we show results obtained using the TIGRESS/Schmidt model, while PRFM-vol results are displayed in the bottom row. The gray shaded line is the original TIGRESS scaling relation, while coloured circles show the simulation results obtained after 37.5 Myr of simulation time. Different colours denote the different simulation setups (see Table~\ref{tab:isolateddisks}). Each circle represents the relation calculated in a single bin in cylindrical radius (see text). From left to right, we increase the fraction of old disk stars that we include in the initial conditions, as indicated in each panel. Comparing the top row to the bottom row, it is evident that PRFM-vol matches the TIGRESS scaling relation better than the TIGRESS/Schmidt model, especially as the amount of stars in the galactic disk increases.}
    \label{fig:scaling_relations_modelcomp}
\end{figure*}

In Fig.~\ref{fig:scaling_relations_modelcomp}, we show a comparison between the TIGRESS/Schmidt model and the full PRFM-vol approach we introduce here. TIGRESS/Schmidt results are shown in the top row and compared against PRFM-vol results in the bottom row. In each panel, we present results from all of our four base setups (D1 -- D4), with the respective fraction of old disk stars indicated in the panel. From left to right, we increase the initial fraction of stars in the disk. 
In all cases, results are shown after 37.5 Myrs of simulation time, and star formation is enabled, which means that stars have also formed in the simulations that start out with pure gas disks. For that reason, the TIGRESS/Schmidt results shown in the upper left panel are not in perfect agreement with the TIGRESS line, as they were in \cite{Burger2025}. The colours are associated with specific setups as indicated in the legend of each panel, and each circle represents a single measurement obtained in one of 20 bins, which are logarithmically spaced in cylindrical radius, between a minimum radius of $1\,{\rm kpc}$ and a maximum radius of $6\,{\rm kpc}$. For each bin, we calculate the star formation rate surface density and the star formation rate weighted average midplane pressure. 

Some observations can be readily made from Fig.~\ref{fig:scaling_relations_modelcomp}. It is evident that both models perform better when the disk is composed mostly of gas, and their match with the TIGRESS scaling relation deteriorates as the fraction of stars in the disk increases. Notably, however, this is much less dramatic in the case of PRFM-vol, where the two more massive disks (D3 and D4) stay nearly perfectly on the relation, even for an initial stellar fraction of 90 per cent. Our least massive disk, D1, exhibits almost no star formation in the case of the rightmost panels, although even here PRFM-vol performs marginally better than TIGRESS/Schmidt. Overall, we find that the full PRFM-vol implementation yields significant improvement over the TIGRESS/Schmidt model, which underestimates star formation rates for high stellar fractions.

The remaining mismatch between the less massive disk models and the TIGRESS scaling relation is readily explained through close observation of the data. Comparison of face-on projections of star formation rates in the D1 galaxy with pure gas initial conditions to its relative with a 90 per cent fraction of stars in the initial disk reveals that in the presence of stars, there is simply insufficient dense gas remaining for the galaxy to efficiently form stars. This indicates the onset of quenching in this galaxy, as well as to some extent in D2. The mismatch with the TIGRESS relation is then simply a result of our method of calculation. When calculating $\Sigma_{\rm SFR}$, we average over all regions in space. If some regions are not star forming, gas cells there are formally still taken into account, which lowers the calculated star formation rate density. On the other hand, pressure is calculated as a star formation rate weighted average in each radial bin, so cells without star formation do not contribute. For that reason, sparsely star forming galaxies cannot be expected to lie on the $P-\Sigma_{\rm SFR}$ relation. In light of the fact that a big part of the remaining mismatch between PRFM-vol and the TIGRESS scaling can be traced back to our averaging procedure, we find the general match of PRFM-vol to the TIGRESS scaling even more impressive, and find it to clearly demonstrate that we accurately account for stellar and DM contributions to the dynamical time within our new subgrid model.   

Having demonstrated that PRFM-vol yields significant improvement over TIGRESS/Schmidt when it comes to the calculation of star formation rates, we now look to briefly discuss the impact of the renormalization factor, equation~(\ref{eq:renormalization_factor}). We argued that this factor is necessary to port the inherently two-dimensional PRFM theory into a subgrid model that depends on volumetric quantities.  

\begin{figure}
    \centering
    \includegraphics[width=\linewidth]{rev_figures/PvSfr_PRFMA_0.eps}\\
    \includegraphics[width=\linewidth]{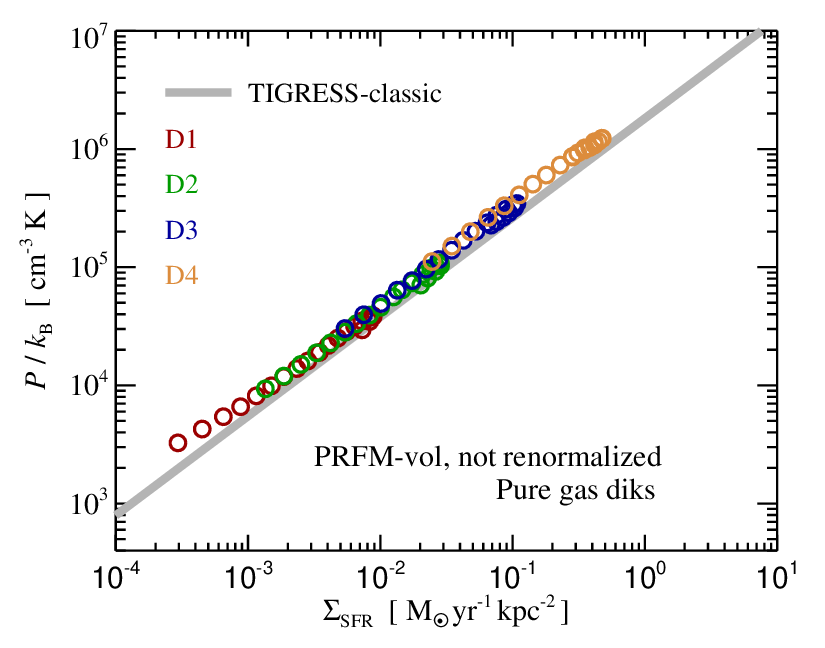}
    \caption{Comparison of the ${P} - \Sigma_{\rm SFR}$ relation measured in the full PRFM-vol model (top panel) to the relation obtained when not including the renormalization factor (bottom panel). As expected, we find overall lower star formation rates in the non-renormalized theory.}
    \label{fig:no_renorm_comp}
\end{figure}

In Fig.~\ref{fig:no_renorm_comp}, we compare the PRFM-vol setups with pure gas disks in the initial conditions to a variant of our model in which we do not renormalize the star formation rates (i.e. we set $U_f = 1$ in equation~\ref{eq:renormalization_factor}). We find, unsurprisingly, an overall decrease in the star formation rate compared to our fiducial PRFM-vol setup. As a result, we now slightly undershoot the star formation rate surface density in each radial bin. 

\begin{figure}
    \centering
    \includegraphics[width=\linewidth]{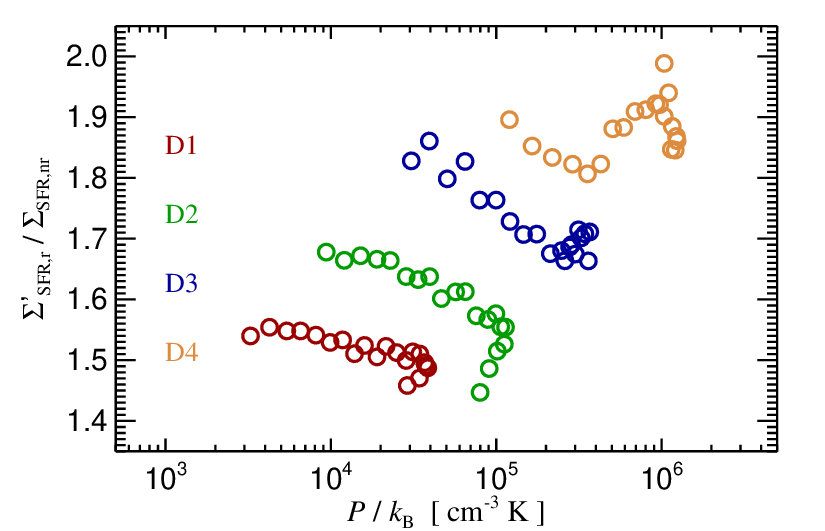}
    \caption{We show the ratio of star formation rate surface densities between benchmark PRFM-vol and the non-renormalized version of PRFM-vol, calculated in the same radial bins as in Fig.~\ref{fig:no_renorm_comp}, in order to quantify the difference between the two models. Note that here we show pressure on the horizontal axis, in order to highlight that the star formation rate is the  quantity that changes as a result of applying the renormalization. A clear separation between our simulation setups emerges, indicating that the renormalization has a stronger effect for heavier disks.}
    \label{fig:sfr_ratio}
\end{figure}

While this is not a very significant discrepancy in most bins, we find that it matters throughout, and we explicitly show the difference between the two models in Fig.~\ref{fig:sfr_ratio}. Since we calculate both quantities in radial bins for the two sets of simulations, and our calculation is carried out after some time has passed in the simulation, both $\Sigma_{\rm SFR}$ and $P$ (albeit modestly) change in each radial bin. However, given that $U_f$ specifically modifies the star formation rate, we are primarily interested in the ratio $\Sigma_{\rm SFR, r}/\Sigma_{\rm SFR, nr}$, i.e. the ratio between the star formation rate surface density in benchmark PRFM-vol and the star formation rate surface density in the non-renormalized version. For a fair comparison, we re-scale $\Sigma_{\rm SFR}$ of the benchmark model to the value it would have if the pressure were the same as in the non-renormalized model.
Concretely, this means that we show $\Sigma_{\rm SFR,r}'/\Sigma_{\rm SFR, nr}$ where
\begin{align}
\Sigma_{\rm SFR, r}' = \Sigma_{\rm SFR, r}\times \left(\frac{P_{\rm nr}}{P_{\rm  r}}\right)^{1.212};
\end{align}
the exponent is $1+\alpha$ (see \autoref{eq:feedback_yield_calib}).  
Fig.~\ref{fig:sfr_ratio} reveals a clear separation between our individual models, showing that the impact of the renormalization factor increases with the mass of the gas disk, and hence the local ratio of baryon to DM density, which is in line with our expectations from equation~(\ref{eq:renormalization_factor}). However, the overall difference between the renormalized and the non-renormalized PRFM-vol model is significantly smaller (within a factor of 2) than the difference between PRFM-vol and TIGRESS/Schmidt (see Fig.~\ref{fig:scaling_relations_modelcomp}), which is especially true once the disk contains a significant fraction of stars.    

In summary, we find that both the change in how the star formation rate is calculated and the inclusion of the renormalization factor lead to improvements of PRFM-vol over TIGRESS/Schmidt. The match to the TIGRESS scaling relation is clearly improved for all our simulation setups. The fact that the gravitational pull of DM and stars can self-consistently be taken into account when calculating the star formation rates is an important success for our subgrid model, and reassuring for the PRFM theory.   

\subsection{Velocity kicks and scale heights}\label{ssec:scaleheights}

From a physical point of view, since stars are born in turbulent clouds, they will inherit a random velocity relative to galactic rotation at the time of their birth.  We incorporate this in our subgrid model by applying velocity kicks. Beyond this underlying physical motivation, there is additional experimental evidence suggesting that adding velocity kicks may be required.
As described in Section~\ref{ssec:vkicks}, this is based on the results presented in Fig.~14 of \cite{Burger2025}, where we showed that the disk scale heights that emerge in the isolated simulations using the TIGRESS/Schmidt model are significantly smaller than the ones obtained using either the \citet{Springel2003} or the TNG model. We later confirmed this trend in cosmological multizoom simulations, but found there not to be as much of a discrepancy as in the isolated setups, likely due to a number of numerical heating mechanisms that are at play in cosmological simulations.  

While there is no clear consensus on what stellar scale heights should be for disk galaxies with a given mass, recent results using the JWST COSMOS-Web survey \citep{Yu2026} suggest scale heights of $0.67\pm 0.06\,{\rm kpc}$ for disk galaxies with $M_{\ast} > 10^{10}\,{\rm M}_{\odot}$ around redshift $z\simeq 0$. Despite the potential numerical heating effects discussed in \citet{Burger2025}, these numbers still clearly favour the TNG model over TIGRESS/Schmidt, even when considering the results of our previous suite of multizoom simulations, where the differences were more moderate than in the isolated galaxies. 
As the addition of velocity kicks to newly born stars is easily motivated as a natural part of our model, due to the presence of a (subgrid) turbulent pressure component, we have decided to include them in PRFM-vol model as described in Section~\ref{ssec:vkicks}. 

It is worth noting that in addition to implementing the velocity kicks, we have also changed the eEoS (see Fig.~\ref{fig:eos_comp}). The new, significantly harder eEoS is less likely to favour the accumulation of star forming gas in very thin disks, and should therefore also contribute to an increase in stellar scale heights in PRFM-vol compared to TIGRESS/Schmidt. 


\begin{figure}
    \centering
    \includegraphics[width=\linewidth]{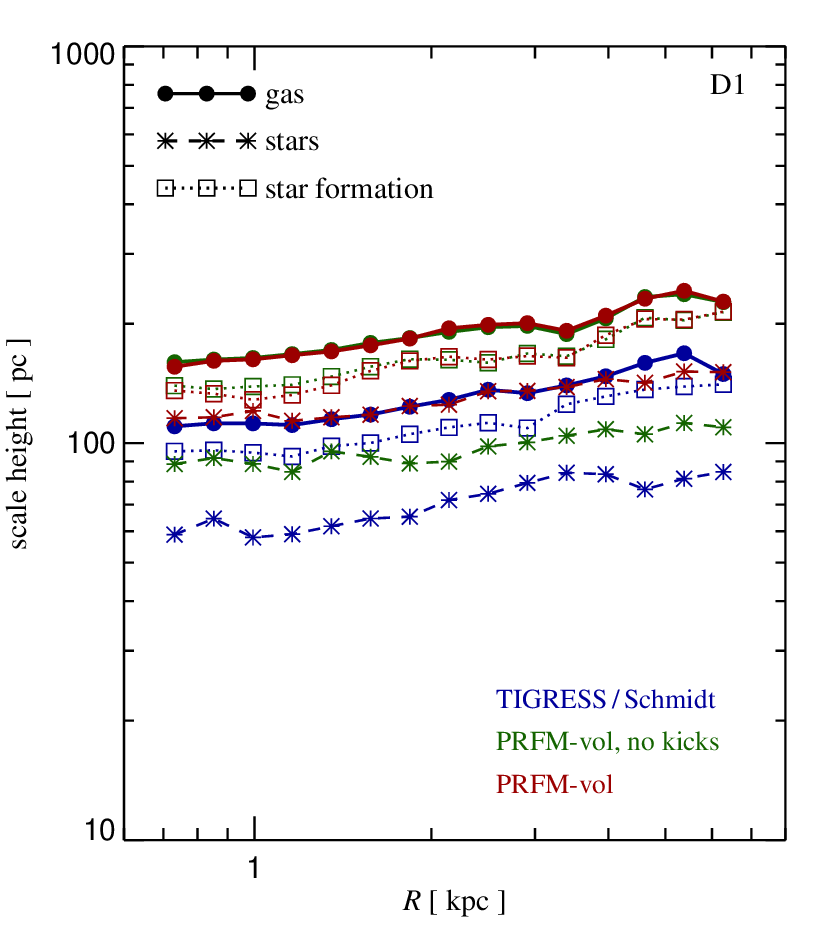}\\
    \includegraphics[width=\linewidth]{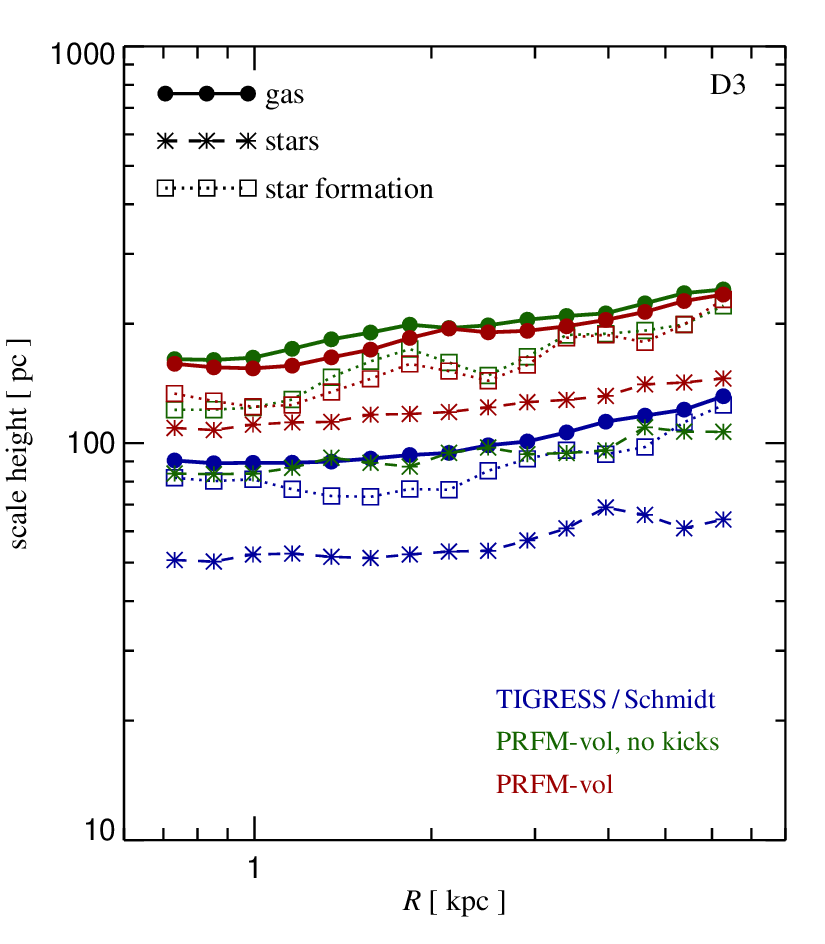}
    \caption{Scale heights of gas (filled points), star forming gas (squares), and stars (stars), for the TIGRESS/Schmidt model (blue) as well as for two versions of PRFM-vol: the full PRFM-vol implementation (red), and a version in which no velocity kicks are added to newly born stars (green). Both panels show results obtained from an average over three snapshots taken after $\sim$ 400 Myrs of simulation time. In the top and bottom panels, we show results of the D1 and D3 setups, respectively, both calculated using initial conditions with no old stars in the initial setup. Both the change in equation of state and the inclusion of velocity kicks lead to an overall increase in the stellar scale height.}
    \label{fig:scaleheights_isolated}
\end{figure}

The effects of these two changes are shown in Fig.~\ref{fig:scaleheights_isolated}. There we show, for two of our four isolated setups, and after $\sim$ 400 Myrs of simulation time, the scale heights of gas, star forming gas, and stars, for three different models: TIGRESS/Schmidt, PRFM-vol, and a version of PRFM-vol in which no velocity kicks are added to newly born stars. Scale heights are calculated as the $z$-coordinates that separately enclose half the gas, star forming gas, and stars, respectively. They are shown in 15 bins, logarithmically spaced in cylindrical radius. As in the analysis in Section~\ref{ssec:sfr_isolated}, we remove the centre of the galaxy from our analysis to exclude the region that is affected by an initial starburst, which is common in isolated galaxy simulations starting from idealized initial conditions. The scale heights shown in Fig.~\ref{fig:scaleheights_isolated} are measured from an average over three snapshots taken after $\sim 400$ Myrs of simulation time.  

In the top and bottom panels, we show results from the pure gas D1 and D3 setups, respectively.
We focus on the pure gas runs in order not to contaminate our results for the stellar scale height through the inclusion of `old' disk stars, which were added in an ad-hoc manner. In this way, we obtain an accurate measure for the scale heights that emerge self-consistently in PRFM-vol.  

Our results are consistent across setups. The gas scale height is evidently set by the effective equation of state. In both setups shown here, the scale heights of the two PRFM-vol versions are virtually identical, while the gas scale height of TIGRESS/Schmidt is much smaller. The same holds true for the scale height of star forming gas, though with lower values since star-forming gas is denser. Given the discussion in \cite{Burger2025}, this result is both desirable and expected, considering that the effective equation of state that we use in PRFM-vol is closer to the TNG effective equation of state than the one we use for TIGRESS/Schmidt. 

Comparing the stellar scale heights, we find a clear additional effect of the velocity kicks. While both versions of PRFM-vol create larger stellar scale heights than TIGRESS/Schmidt simply by virtue of the altered effective equation of state, the stellar scale heights attained in the PRFM-vol simulations with velocity kicks are significantly larger than those without velocity kicks. 
This effect is fairly consistent across the radial range we show, with the addition of the velocity kicks resulting in an increase in stellar scale height by a factor of about $1.5$. The relative difference decreases towards larger radii, where scale heights are larger to begin with. It is worth pointing out that the relative difference between the scale height of star forming gas and stars that is common to TIGRESS/Schmidt and PRFM-vol without velocity kicks is a direct consequence of the applied star formation law, since the star formation rate is proportional to $\rho_g^2$ in pure gas disks.


Overall, we find that with our new implementation of PRFM-vol, 
scale heights are larger (and presumably closer to what they physically shoud be) compared to the TIGRESS/Schmidt simulations, at least in isolated simulations. Both the change in effective equation of state, and the inclusion of velocity kicks significantly contribute to that. Therefore, we decide to include the velocity kicks in the benchmark version of PRFM-vol. 

\subsection{Scale height ratio estimates}\label{ssec:height_ratio_test}

Using our isolated setups, we are also able to assess whether estimating the scale height by means of the squared ratio of the velocity dispersions (i.e. following either equation~\ref{eq:scale_height_ratio} or equation~\ref{eq:scale_height_ratio_two}) provides considerable improvement over setting the scale height ratio to unity in equation~(\ref{eq:dynamical_time_formula}). Note that small deviations from unity would only mildly modify the stellar contribution to the dynamical time.  
\begin{figure*}
    \centering
    \includegraphics[width=0.48\linewidth]{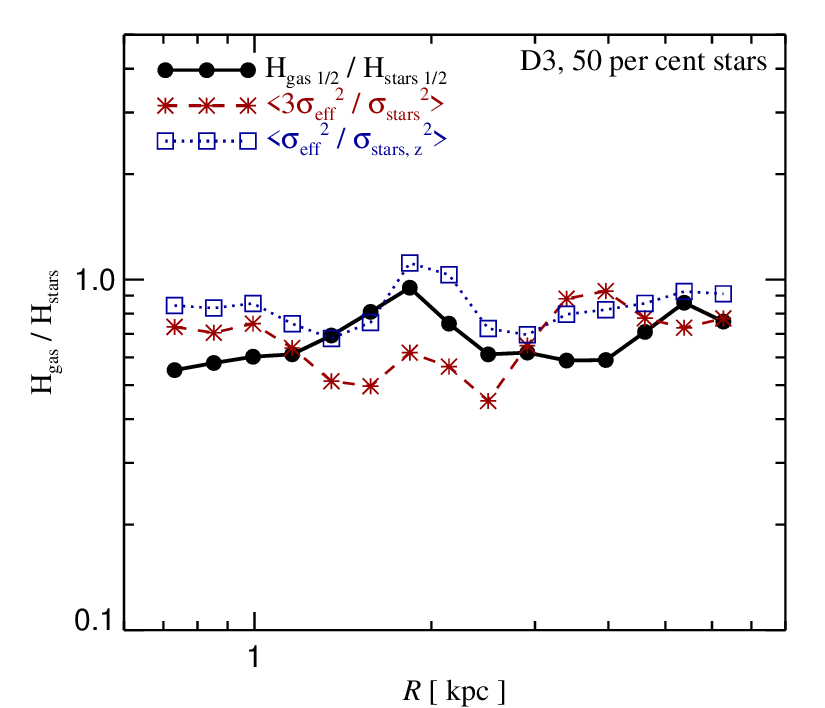}
    \includegraphics[width=0.48\linewidth]{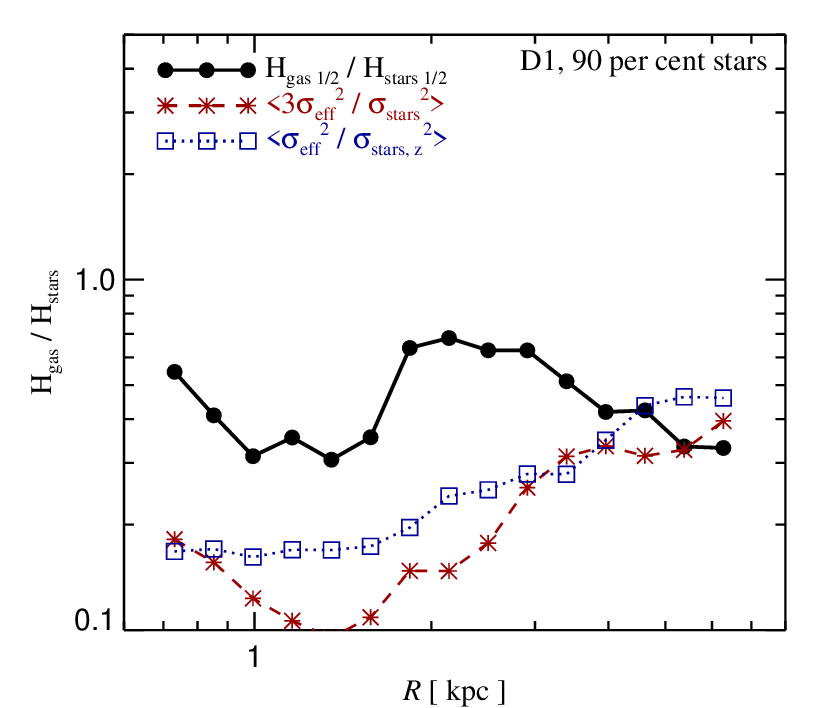}
    \caption{Scale height ratio as a function of cylindrical radius for two of our isolated setups. In the left (right) panel we show our D3 (D1) setup with an initial old star disk fraction of 50\% (90\%). For both of them, we show one measurement and two estimates of the scale height ratio, all calculated in 15 radial bins. For the measurement, we show the ratio of the half mass heights as black dots. Our two estimators correspond to equation~(\ref{eq:scale_height_ratio}) and equation~(\ref{eq:scale_height_ratio_two}): We show star formation rate weighted averages of   
    $\langle3\sigma_{\rm eff}^2/\sigma_{\ast}^2\rangle$ as red stars, and star formation rate weighted averages of  $\langle\sigma_{\rm eff}^2/\sigma_{\ast,\,z}^2\rangle$ as blue boxes. 
    For the D3 setup, both estimators perform well, with the 3d velocity dispersion performing slightly better. For the D1 setup, in which the star forming region has become discontinuous, neither estimator performs well. Here, we expect pressure support from non star-forming gas to be significant.}
    \label{fig:shre}
\end{figure*}

To assess the goodness of the scale height ratio estimates (equations~\ref{eq:scale_height_ratio} and ~\ref{eq:scale_height_ratio_two}), in Fig.~\ref{fig:shre} we show results from two of our isolated simulations -- the D3 setup with 50\% of old stars in the initial disk in the left panel, and the D1 setup with 90\% of stars in the initial disk in the right panel.

We deliberately chose setups with old stars in the initial conditions to assess how the estimators perform in the presence of massive stellar disks. Note that in contrast to the pure gas setups shown in Fig.~\ref{fig:scaleheights_isolated}, where only stars that formed during the simulation were considered, it is now possible for stellar scale heights to be larger than gas scale heights. In fact, we find this to be the case in both setups shown in Fig.~\ref{fig:shre}. Since we set up the gaseous disk and the stellar disk with the same scale height initially, the mismatch after 375 Myrs of simulation time is related to the collapse of the gas disk, which occurs as a result of the onset of gas cooling and star formation shortly after starting the simulation.  

As a measurement for the scale height ratio, we show the ratio between the half mass height of the gas disk and the half mass height of the stellar disk, calculated in the same 15 radial bins as in Fig.~\ref{fig:scaleheights_isolated}. Again, we limit our analysis to regions that are sufficiently far from the disk centre in order to avoid being affected by an initial central starburst. For each bin, we further calculated star formation rate weighted averages of the two estimators (equation~\ref{eq:scale_height_ratio} and~\ref{eq:scale_height_ratio_two}), where we use kernel estimates of the stellar velocity dispersions that were calculated on the fly for each star forming gas cell. 

For the D3 setup with an old star fraction of 50\%, we find remarkable agreement between the measurement of the scale height ratio and the predictions by the two estimators. The quality of the match between the measured scale height ratio and the ratio estimated using equation~(\ref{eq:scale_height_ratio}) is perhaps a bit surprising, since only the vertical stellar velocity dispersion should be dynamically related to the scale height. Yet, for the majority of our isolated setups, we find that it performs as well as, if not better than, equation~(\ref{eq:scale_height_ratio_two}), showing that in general, stellar velocity dispersions can reasonably be approximated as isotropic. At face value, this result is very encouraging and suggests that using equation~(\ref{eq:scale_height_ratio}) as an estimator in cosmological simulations could be a good way to increase the accuracy with which we track the stellar contribution to the dynamical time without requiring us to keep track of galaxy orientations. However, we note that the high resolution of our isolated simulations might bias this result. While we do not check for this specifically, it is possible that using a fixed number of nearest neighbours to calculate the velocity dispersion might affect the performance of the estimator at low resolution. This would occur if the kernel spans a significant in-plane area, at which point the galaxy rotation could artificially contribute to the 3d velocity dispersion calculated in Cartesian coordinates.

For now, we do not investigate this issue further due to the results shown in the right panel of Fig.~\ref{fig:scaleheights_isolated}. For the D1 setup with an initial old star fraction of 90\%, we find that neither of our estimators gives good agreement, as they significantly underestimate the scale height ratio for most of the radial bins. We checked the results of all of our 12 initial setups (see Table~\ref{tab:isolateddisks}) and found that for most of them, both estimators showed good agreement with the scale height ratio measurement. The only three setups for which this was not the case are the D1 setup with 50\% old stars and the D1 and the D2 setup with 90\% old stars. Comparison with Fig.~\ref{fig:scaling_relations_modelcomp} reveals that these are the exact three setups that deviated from the TIGRESS $P-\Sigma_{\rm sfr}$ relation, with their star formation rate surface densities being too low. In Section~\ref{ssec:sfr_isolated}, we had confirmed through visual inspection of those setups that the star forming region in their disks had become discontinuous, with a significant fraction of the gas disk being non-star forming. Thus, the vertical equilibrium in this non-star forming region is also no longer maintained by the star-forming gas, but by the non-star forming gas instead. For that reason, it makes sense that the estimators no longer track the scale height ratio properly.

Our takeaway message from this test is twofold. On the one hand, we have shown that both estimators provide good estimates of the scale height ratio in most cases, potentially resulting in an improvement over setting the scale height ratio to unity in equation~(\ref{eq:dynamical_time_formula}). On the other hand, we find that neither of them performs well in galaxies that are undergoing quenching, a common occurrence in cosmological simulations. Moreover, we suspect that the performance of the 3d velocity dispersion based estimator might decrease with resolution. 
Since the measured ratio is also fairly close to unity in most of our setups, we choose not to use 
equation~(\ref{eq:scale_height_ratio}) in our multizoom cosmological simulations for the moment, and instead opt to set $H_{\rm gas} / H_\star = 1$.
We present the results of the multizoom simulations below, and note here that in some specific cases (such as when focusing on high-redshift, star-forming objects in highly resolved zoom simulations), using 
equation~(\ref{eq:scale_height_ratio}) to estimate the scale height ratio may be appropriate.  

\section{Multizoom cosmological simulations}\label{sec:multizoom}

Below, we present results of the multizoom simulations with $\log({M}_{\rm vir}/{\rm M}_\odot) = 12$ and zoom factor 4 (i.e., baryon mass resolution $m_{\rm gas}=4.82\times10^{5}\,{\rm M}_\odot$ and DM mass resolution $m_{\rm DM}=3.1\times 10^6\,{\rm M}_\odot$; see Section~\ref{sse:multizoom_ics}). Differences seen in our other multizoom runs, specifically the ones with $\log({M}_{\rm vir}/{\rm M}_\odot) = 11.097$ and the ones with $\log({M}_{\rm vir}/{\rm M}_\odot) = 12$ and zoom factor 2 are briefly discussed in Appendix~\ref{apxsec:resolution}. We focus on galaxy morphology first. Then, we introduce an alternative effective equation of state (PRFM-vol~2) and discuss how galaxy morphology is affected, and quantify the morphological differences that arise when changing the eEoS. In doing so, we put a strong emphasis on the role of the pressure normalization when it comes to determining whether or not star forming disks become Toomre unstable. Afterwards, we analyze scaling relations, stellar masses, star formation rates, and galaxy scale heights, and compare against observations where appropriate. 
\subsection{Galaxy morphologies}\label{ssec:morphs}
\begin{figure*}
    \centering
    \includegraphics[width=\linewidth]{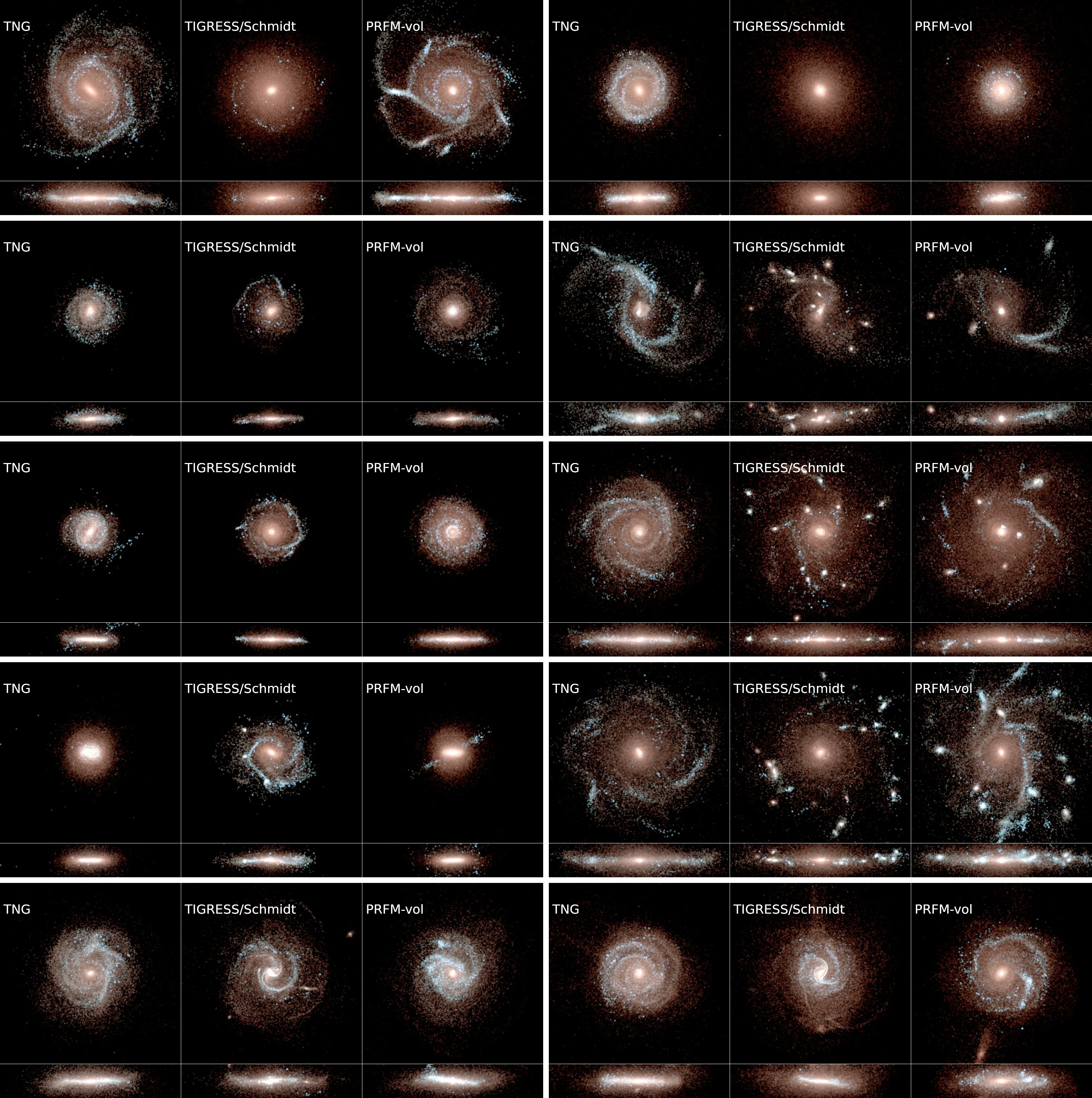}
    \caption{Galaxy stellar light projections for different star formation models. We show 10 side-by-side comparisons for galaxies in matched host haloes
    from cosmological multizoom simulations using the TNG, TIGRESS/Schmidt, and the PRFM-vol models. The images are created by mapping luminosities in the K-, B-, and U-bands to RGB colour space. For each galaxy, we show face-on and edge-on projections with a panel side-length of 60 kpc. The logarithmic colour mapping is the same for all galaxies.}
    \label{fig:galaxy_morphologies_one}
\end{figure*}

To assess the morphologies of the galaxies simulated with the benchmark PRFM-vol model, we show side by side comparisons with results from equivalent TNG and TIGRESS/Schmidt runs in Fig.~\ref{fig:galaxy_morphologies_one}. We logarithmically map K-, B-, and U-luminosities of the star particles at redshift zero to RGB space and then show edge-on and face-on projections of the observed stellar light. The colour mapping is the same for all galaxies, and the side length of each panel is $60\,{\rm kpc}$. Galaxies are matched by comparing the positions of their host haloes to the group catalog of the parent box. 

Fig.~\ref{fig:galaxy_morphologies_one} allows for a few general statements. First, the observed sizes and shapes of the galaxies are fairly consistent between all three models, with no major deviations appearing due to the new PRFM-vol model. A second, more important point can be made through comparison of the PRFM-vol galaxies against the TIGRESS/Schmidt galaxies. We find that the stellar clumps that we identified in the TIGRESS/Schmidt galaxies in \cite{Burger2025} do not disappear when changing to the PRFM-vol model. In fact, clumps are found in many of the same galaxies as before. In light of that, we have to comment on two conjectures that we made in \cite{Burger2025}. We suggested that the occurrence of stellar clumps may be reduced by the inclusion of stellar velocity kicks, which would disperse the clumps. Evidently, given that PRFM-vol includes physically motivated velocity kicks to newly born stars, we find this not to be the case. This does not mean that including the velocity kicks is futile, however. As one can already see from the edge-on projections -- and as we will quantify later -- the inclusion of the velocity kicks leads to increased scale heights in PRFM-vol compared to TIGRESS/Schmidt, which is a favourable outcome for our new model. A second potential reason that we gave for the formation of stellar clumps was the Toomre disk instabilities \citep{Toomre1964} that can arise as a consequence of the lower pressure in the star forming gas as compared to TNG, a point that has already been made in the literature, for example in Figure 6 of
\cite{SpringelHQ2005}.

We follow the formation of clumps seen in Fig.~\ref{fig:galaxy_morphologies_one} in Appendix~\ref{apxsec:clumps_prfm_vol}. In short, we find that stellar disk formation occurs around redshift $z=1$, and that clumps form after that, when disks are already present. Combined with the observation that stellar clumps coincide with gas density peaks, as well as peaks in star formation rate, this lends further credibility to our theory that gas disk fragmentation due to Toomre instabilities causes the formation of stellar clumps. 

In Section~\ref{ssec:toomre}, we verify that this is indeed the case, and take a closer look at the Toomre $Q$ parameter in clumpy galaxies.


\subsection{A modified effective equation of state} \label{ssec:modeeos}

What is curious about the result presented in Section~\ref{ssec:morphs} is that the eEoS that we use in PRFM-vol deviates significantly from the TIGRESS/Schmidt eEoS, with the pressure at a given hydrogen number density generally being higher (see Fig.~\ref{fig:eos_comp}). Yet, at low hydrogen number densities, we find that the pressure in PRFM-vol roughly equals the TIGRESS/Schmidt pressure. It is due to the harder eEoS that the PRFM-vol pressure eventually surpasses the TNG pressure. However, if star formation were to occur predominantly at low densities, Toomre instabilities can feasibly explain the clumpy morphology of some PRFM-vol galaxies compared to their TNG counterparts. 
Recall that the Toomre~$Q$ parameter is mainly determined by the gas sound speed and the gas surface density:
\begin{equation}
Q= \frac{\kappa\, c_{\rm s}}{\pi  G \Sigma_{\rm gas}}. \label{eq:toomre}
\end{equation}

\begin{figure*}
    \centering
    \includegraphics[width=\linewidth]{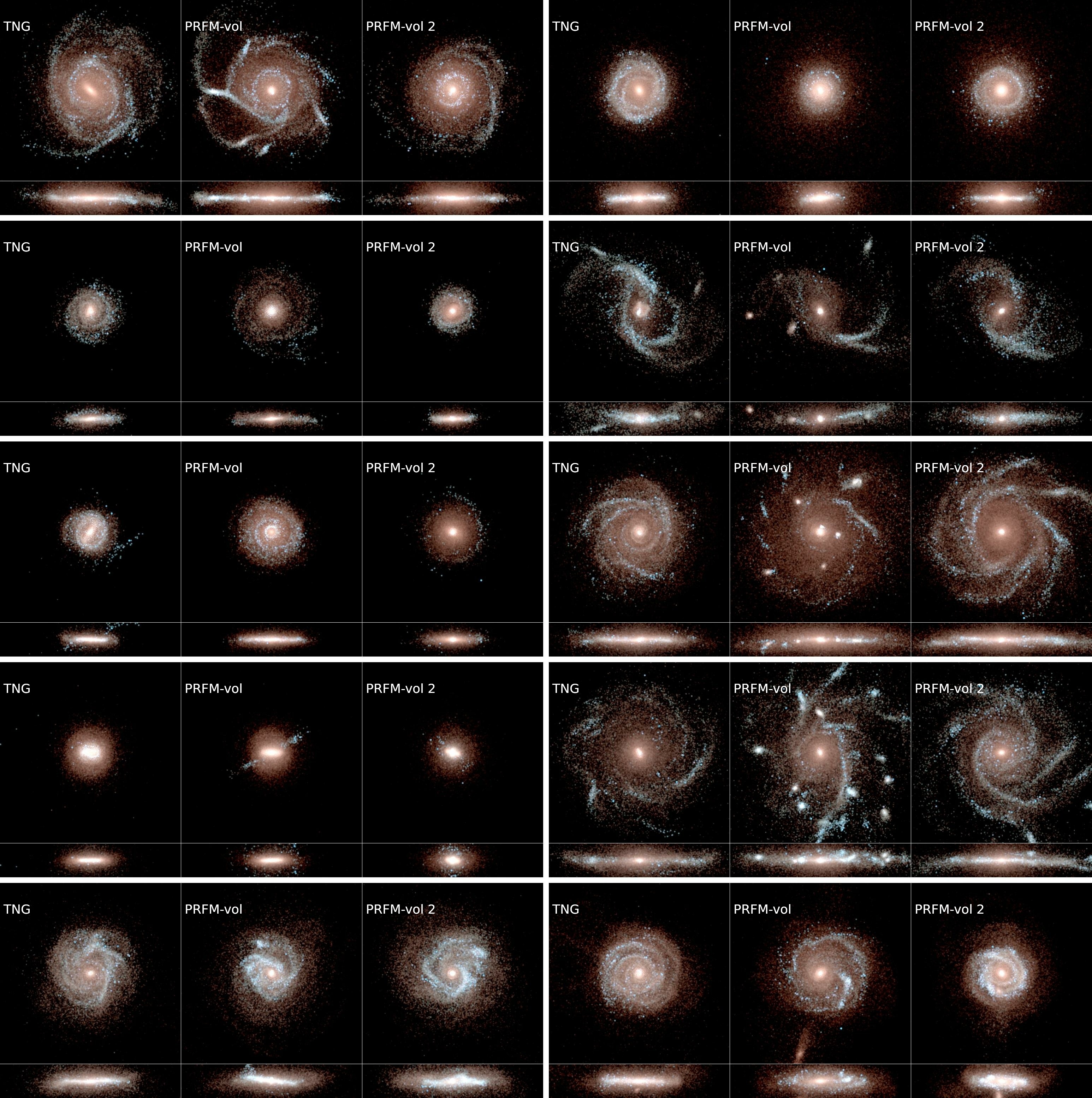}
    \caption{Same as Fig.~\ref{fig:galaxy_morphologies_one}, but this time showing a comparison between TNG, PRFM-vol, and PRFM-vol with a modified eEoS in which the pressure is multiplied by two. Stellar clumps no longer appear in the latter ``PRFM-vol~2'' run.}
    \label{fig:galaxy_morphologies_two}
\end{figure*}

Assuming that the issue is that the pressure at low hydrogen number densities is significantly lower than in TNG, we can attempt to eliminate the clumps by modifying the eEoS. In a first attempt to do so, we tested a model in which we implemented the velocity dispersion floor of $12\,{\rm km\,s}^{-1}$ that was introduced in \citet{Jeffreson2026} and then assigned a consistent pressure at low hydrogen number densities. This increases the pressure at the density threshold to about the TNG value, but then assigns a logarithmic slope of one to the eEoS at low hydrogen number densities, up to densities at which the effective velocity dispersions in  PRFM-vol exceed the floor value (see Fig.~\ref{fig:eos_comp}), at which point the two equations of state become identical. Unfortunately, we found that this physically motivated way of altering the model did not make much difference. Clumpy galaxies remained clumpy, for the most part. Moreover, we found that the very soft low density branch of this eEoS resulted in a further decrease of the scale heights of simulated galaxies, something we actively tried to counteract by introducing the velocity kicks. For that reason, we did not pursue this model further. Instead, we looked for other ways to alter the eEoS. 

Noting that the main issue appears to be the low density regime, and that TNG galaxies are generally not clumpy, we multiply the PRFM-vol eEoS by a factor of two. This effectively sets the eEoS to similar values as in TNG in the low density regime, while preserving the logarithmic slope and the steep rise of the implied isothermal sound speed with hydrogen number density. With this altered model, we carry out an additional multizoom simulation, starting from the $\log({M_{vir}}/{\rm M}_\odot) = 12$ initial conditions with zoom factor 4. 


We show results of this simulation in Fig.~\ref{fig:galaxy_morphologies_two}. Here, we present stellar light projections that we derive in the same way as for Fig.~\ref{fig:galaxy_morphologies_one}. However, we now replace the TIGRESS/Schmidt panels with panels that contain the PRFM-vol galaxies with a modified eEoS, which we dub ``PRFM-vol~2''. The matching of the galaxies is done in the same way as before. We find that the sizes and shapes of the galaxies again do not change substantially with respect to either TNG or the original PRFM-vol. However, stellar clumps are no longer present in the PRFM-vol~2 galaxies. This implies that increasing the pressure by a factor of 2
has a strong effect on galaxy morphology. To investigate why this makes such a substantial difference, we look at the cumulative mass of star forming gas as a function of hydrogen number density at different redshifts, both in the PRFM-vol and the PRFM-vol~2 simulations. 

\begin{figure}
    \centering
    \includegraphics[width=\linewidth,trim = 0.5cm 0.5cm 0.5cm 0.5cm, clip]{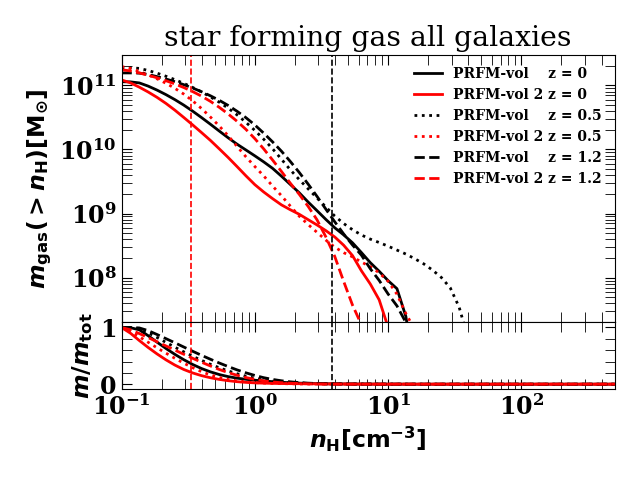}
    \caption{Complementary cumulative star-forming gas mass as a function of hydrogen number density, shown for the PRFM-vol (black lines) and PRFM-vol~2 (red lines) simulations, at redshifts $z=0$ (solid), 0.5 (dotted), and 1.2 (dashed). Star-forming gas cells are identified  within a $30\,{\rm kpc}$ sphere around the host galaxy's centre of potential, and with instantaneous star formation rate larger than zero. In the small bottom panel, we show the cumulative mass of star forming gas normalized by the total mass $m_{\rm tot} = m_{\rm gas}(>n_{\rm th})$. The dotted vertical lines indicate the intersection points of the TNG eEoS with the PRFM-vol (black) and the PRFM-vol-2 (red) effective equations of state.}
    \label{fig:cum_density_functions}
\end{figure}

In Fig.~\ref{fig:cum_density_functions}, we show the cumulative mass of the star forming gas as a function of hydrogen number density in both of these simulations, and at three different redshifts, $z=0$, $0.5$, and $1.2$. We also show where the PRFM-vol eEoS and the TNG eEoS intersect, indicated by a vertical dashed black line (see also Fig.~\ref{fig:eos_comp}). At higher densities, the PRFM-vol pressure is larger than the TNG pressure, and vice versa at lower densities. In addition, we show the density at which the TNG eEoS intersects two times the PRFM-vol eEoS (aka the PRFM-vol~2 eEoS) as a red dotted vertical line. In a small bottom panel, we show the cumulative mass functions normalized by their respective maximum values. 

From Fig.~\ref{fig:cum_density_functions} it is apparent that at all redshifts the fractional amount of star forming gas for which the PRFM-vol pressure is larger than the TNG pressure at the same hydrogen number density is negligible. In turn, this implies that almost all of the gas is at lower pressures than it would be in TNG, and thus, it makes sense that PRFM-vol galaxies are prone to disk instabilities and clump formation, just as the TIGRESS/Schmidt galaxies. Analyzing the placement of the red vertical line reveals, however, that for PRFM-vol-2, the fraction of gas cells whose pressure is higher than in TNG is around $50$ \% -- slightly higher than that at redshift $z=1.2$ and slightly lower than that at redshift $z=0$. In addition to this, the pressure of gas at densities lower than the intersection point marked by the red vertical line is also much closer to the corresponding TNG value. From the star forming gas mass function, we can thus deduce why the stellar clumps do not appear in PRFM-vol~2.
The pressure of the star forming gas is similar to TNG, where clumps are not present. This increase in pressure implies a corresponding increase in isothermal sound speed, which would lead to higher Toomre $Q$ values at similar gas surface densities. 

We can further observe from Fig.~\ref{fig:cum_density_functions} that for both PRFM-vol and PRFM-vol~2 -- but even more so for PRFM-vol~2 -- gas at higher densities is progressively depleted as stars form, with the redshift $z=0$ distributions more strongly favouring densities close to the star formation threshold than the redshift $z=1.2$ mass functions. This is no surprise, as the effective velocity dispersion modifies the depletion time (see equation~\ref{eq:tdep_fin}). It also suggests that stars may predominantly be born at densities that are a bit larger than the numerical star formation threshold. 


\begin{figure}
    \centering
    \includegraphics[width=\linewidth,trim = 0.5cm 0.5cm 0.5cm 0.5cm, clip]{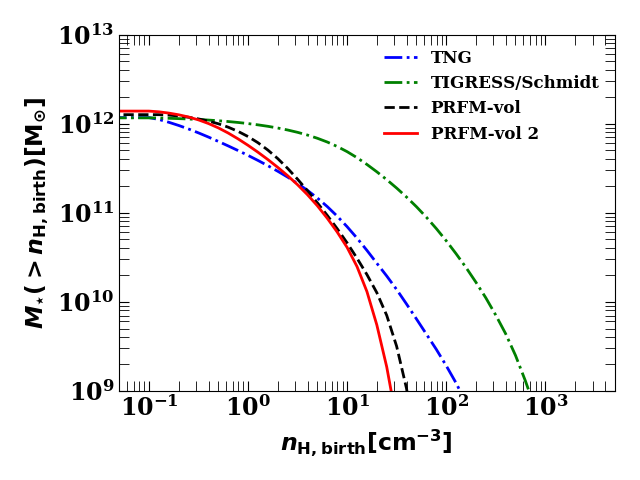}
    \caption{Complementary cumulative stellar mass as a function of birth density -- here defined as the physical density of the parent gas cell at the time the star was formed. We show the cumulative mass of all stars formed until redshift $z=0$, stacked over the twenty simulated high-resolution galaxies, for the multizoom simulations with $\log(M_{\rm vir}/{\rm M}_\odot) = 12$ and zoom factor 4, using the TNG (blue line), TIGRESS/Schmidt (green line), PRFM-vol (black line), and PRFM-vol-2 models (red line), respectively. The softer the model's eEoS, the longer the tail towards higher birth densities.}
    \label{fig:birth_densities_mass_function}
\end{figure}

Fig.~\ref{fig:birth_densities_mass_function} further illustrates this point. We show the cumulative mass of stars that are within 30 kpc of the centre of one of the 20 high-resolution multizoom galaxies and were formed in gas cells with densities exceeding a given birth density (mass-weighted cumulative birth density functions), calculated at redshift zero for multizoom simulations run with the TNG, TIGRESS/Schmidt, PRFM-vol, and PRFM-vol-2 models. Observing both the PRFM-vol curve and the PRFM-vol~2 curve confirms the results presented in Fig.~\ref{fig:cum_density_functions}. In the PRFM-vol models, stars are typically not born at densities $\sim 0.1{\rm cm}^{-3}$, but rather at densities of $\sim 1\,{\rm cm}^{-3}$. From Fig.~\ref{fig:cum_density_functions}, we know that at those densities, the PRFM-vol~2 pressure is already of the order of the TNG pressure. The PRFM-vol pressure, on the other hand, is still significantly lower, resulting in less stable gas disks.  
We also find that the mass-weighted cumulative stellar birth density functions have tails that extend to larger densities than we have seen in Fig.~\ref{fig:cum_density_functions}, which confirms our previous analysis: Stars born at densities larger than $10\,{\rm cm}^{-3}$ must have formed before redshift $z=1.2$, since such densities were not reached afterwards, further indicating that gas at high densities is depleted faster, due to its shorter depletion time.

Comparing the two PRFM-vol density functions to the TNG and the TIGRESS/Schmidt density function provides further insight. The effective equations of state employed in these two models are softer than the PRFM-vol eEoS (see Fig.~\ref{fig:eos_comp}), with the TIGRESS/Schmidt eEoS being the softest across a large range of densities. Fig.~\ref{fig:birth_densities_mass_function} demonstrates that the difference in logarithmic slope is reflected not only in how fast high-density star forming gas is depleted, but also in whether such high densities are reached to begin with. The tail of the TNG cumulative birth density functions extends beyond $n_{\rm H, birth} = 100\,{\rm cm}^{-3}$, about a factor of 5 further than the PRFM-vol curves, although the bulk of the stellar mass is formed at about the same densities in the two PRFM-vol models and in TNG. The tail of the TIGRESS/Schmidt curves extends substantially further than any of the other models, reaching birth densities of almost $1000\,{\rm cm}^{-3}$. This highlights the complex role of the eEoS in regulating the star forming ISM. Softer equations of state enable gas to reach higher densities, while the pressure at such densities regulates how fast this gas is depleted. These effects are closely intertwined with effects associated with the assumed star formation law, and therefore, choosing a suitable eEoS is a crucial part of designing a subgrid star formation model for cosmological simulations.  

We comment on the physical implications of adopting the PRFM-vol (or PRFM-vol~2) eEoS in Section~\ref{sec:conclusions}.  

\subsection{Toomre $Q$ in clumpy galaxies}\label{ssec:toomre}

\begin{figure*}
    \centering
    \includegraphics[width=\linewidth, trim = 3cm 0cm 3cm 0cm, clip]{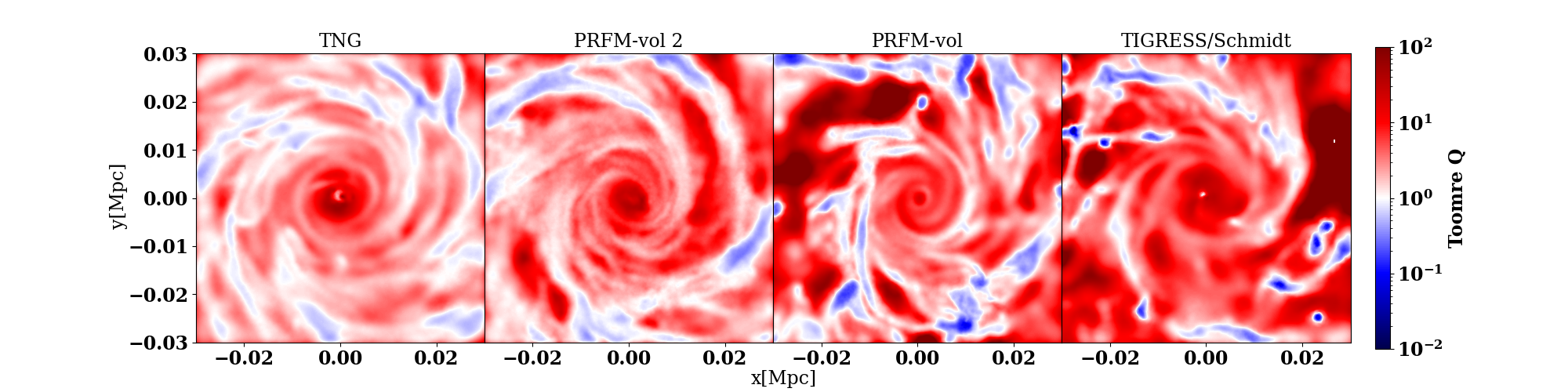}\\
    \includegraphics[width=\linewidth, trim = 3cm 0cm 3cm 0cm, clip]{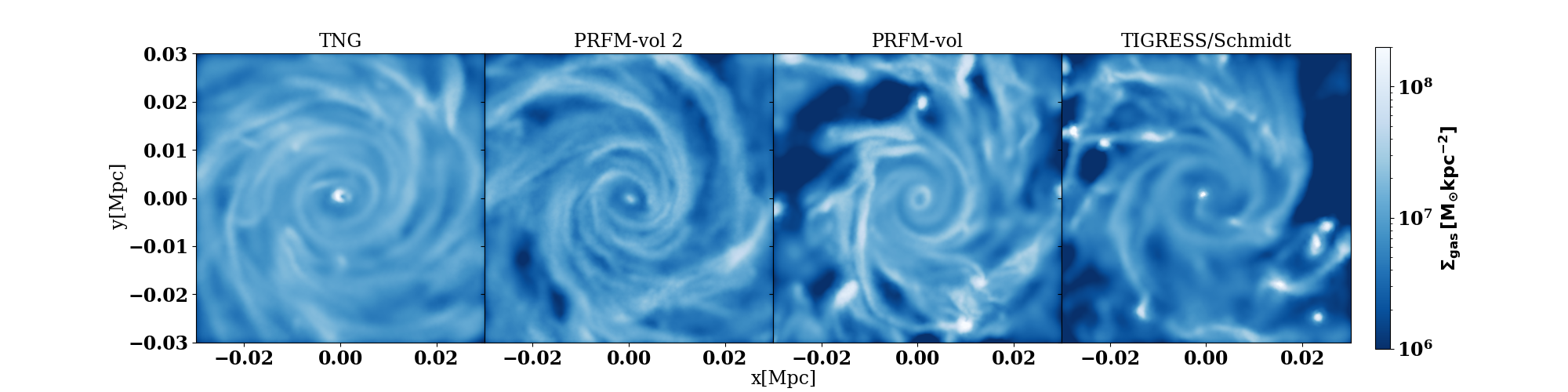}\\
    \includegraphics[width=\linewidth, trim = 3cm 0cm 3cm 0cm, clip]{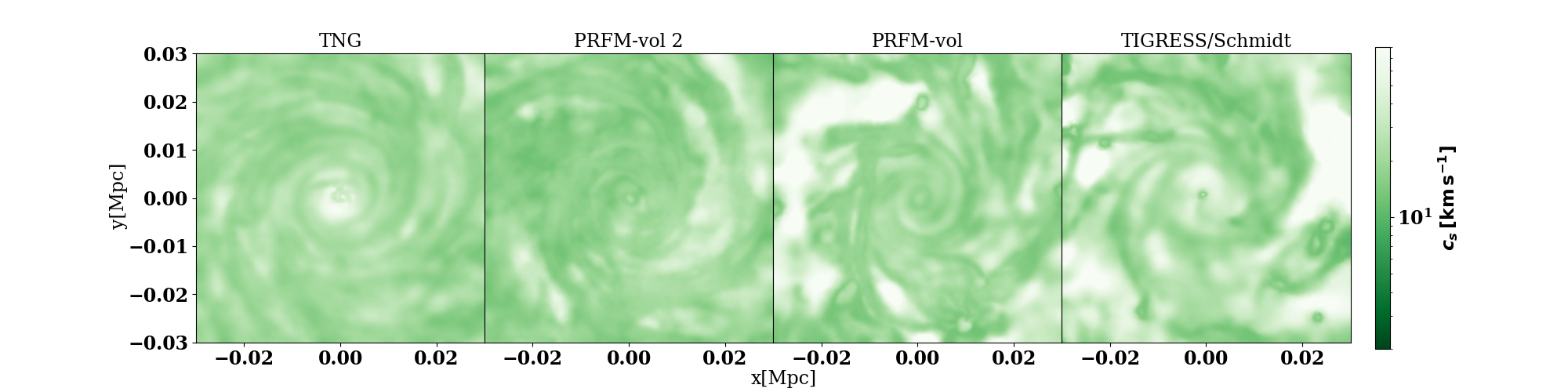}
    \includegraphics[width=\linewidth, trim = 3cm 0cm 3cm 0cm, clip]{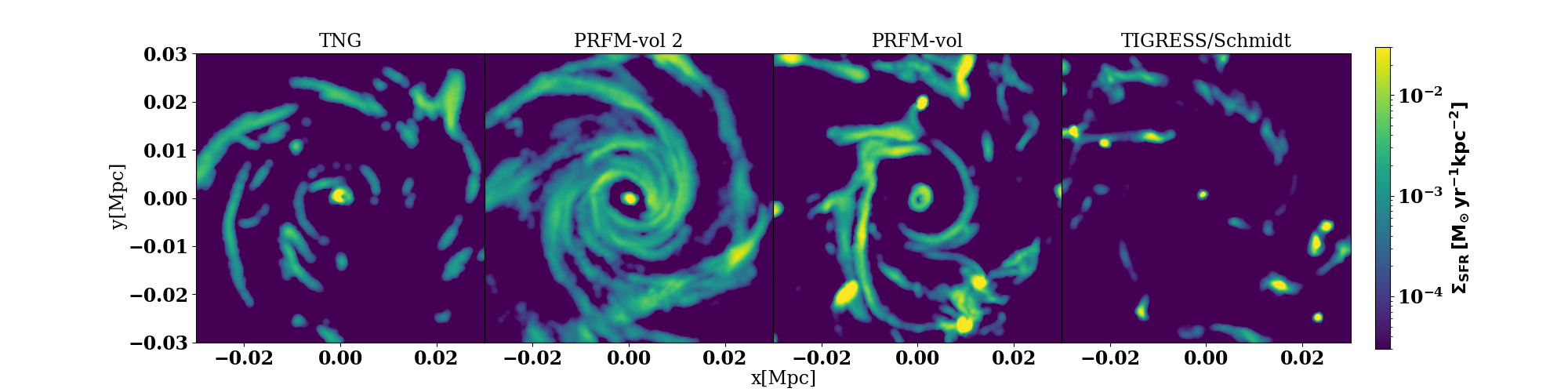}
    \caption{From top to bottom, we show maps of  Toomre $Q$, gas density, isothermal sound speed, and star formation rate surface density in a single galaxy picked from Fig.~ \ref{fig:galaxy_morphologies_one}. Concretely, we chose the fourth galaxy from the top in the right column of Fig.~\ref{fig:galaxy_morphologies_one}.  We compare across models, showing, from left to right, results of the TNG, PRFM-vol-2, PRFM-vol, and TIGRESS/Schmidt simulations. Note that we arrange the models in such a way that from left to right, the pressure in the star-forming ISM decreases. In the lower pressure models, this leads to the gravitational collapse of gas into dense clumps, with an increasing number of such clumps for lower overall pressure. Clumpy regions are associated with low values of Q and high star formation rate surface densities. No strong correlation between clumps and gas sound speed is found at redshift $z=0$. }
    \label{fig:toomreq_face_on}
\end{figure*}

\begin{figure*}
\includegraphics[width=0.48\linewidth, trim = 0.5cm 0.5cm 0.5cm 0.5cm, clip]{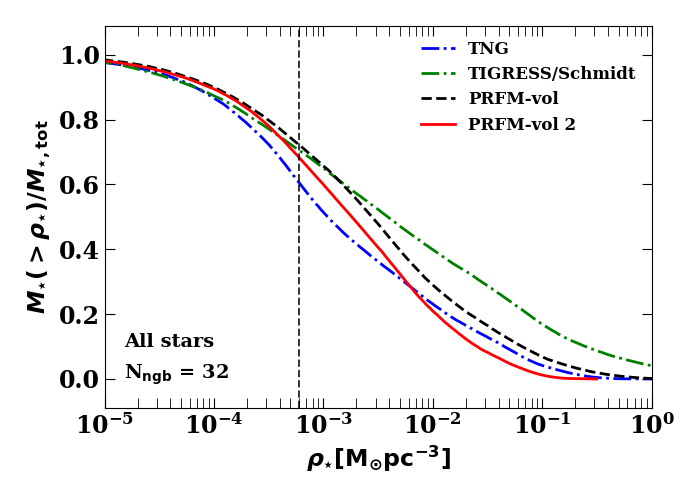}
    \includegraphics[width=0.48\linewidth, trim = 0.5cm 0.5cm 0.5cm 0.5cm, clip]{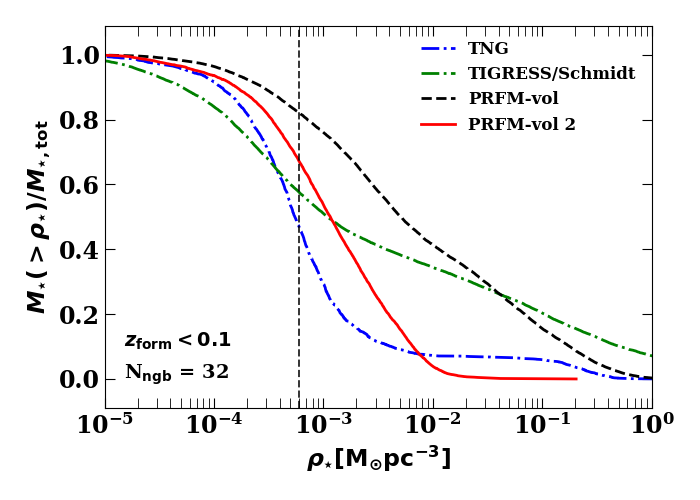}
    \caption{Complementary cumulative distribution of stellar mass as a function of ambient stellar density for the fourth galaxy in the right column of Figs~\ref{fig:galaxy_morphologies_one} and~\ref{fig:galaxy_morphologies_two}. In both panels, we compare between the distributions measured in the TNG (blue), TIGRESS/Schmidt
    (green), PRFM-vol (black), and PRFM-vol~2 (red) models. The left panel shows the cumulative distribution inferred from all stars in the galaxies, while in the right panel, we limit the analysis to stars with a formation redshift below  0.1. In both cases, ambient stellar densities are estimated using a cubic spline kernel with $N_{\rm ngb} = 32$ nearest neighbours. The vertical lines mark a numerical reference density: the gas mass resolution divided by the third power of the gravitational softening length. In both panels, we have removed the central $2\,{\rm kpc}$ from our analysis.}\label{fig:clumps_cumulative}
\end{figure*}

So far, we have claimed that the formation of stellar clumps occurs as a result of Toomre instabilities in star-forming disks. In Appendix~\ref{apxsec:clumps_prfm_vol}, we show explicitly that clumpy star-forming regions appear past redshift $\sim 1$, and after gas has settled into star-forming disks. With Fig.~\ref{fig:toomreq_face_on}, we take a closer look at the gas disk of one of our simulated multizoom galaxies at redshift $z=0$. Specifically, we show face-on projections of four quantities: the Toomre $Q$ parameter, gas surface density, isothermal sound speed, and star formation rate surface density -- from top to bottom. The chosen galaxy is the one shown in the fourth row from top, in the right columns of both Fig.~\ref{fig:galaxy_morphologies_one} and~\ref{fig:galaxy_morphologies_two}. We show each of the quantities for every single one of our models; TNG, PRFM-vol, PRFM-vol~2, and TIGRESS/Schmidt. Following our discussion in Section~\ref{ssec:modeeos}, we assume that the pressure in the star-forming ISM determines how prone galaxies are to forming stellar clumps. To test that assumption, in Fig.~\ref{fig:toomreq_face_on}, we arrange the maps in a specific way. On the left, we show the results of the TNG run, our model with the highest average pressure in the star-forming ISM. To the right of the TNG results, we display results of the PRFM-vol~2 run, our model with the second highest average pressure. We continue ordering the models by average pressure, showing next results of the PRFM-vol simulation and concluding with the TIGRESS/Schmidt run, whose eEOS assigns the lowest effective pressure to gas at all densities.     

The quantities shown in Fig.~\ref{fig:toomreq_face_on} reveal a few clear trends. With decreasing pressure (at equivalent densities), regions with lower values of $Q$ emerge. These regions become more localized as the mean effective pressure decreases. Their positions correlate strongly with the positions of overdense regions in the associated gas surface density maps.
For the two examples in which clumps have formed, we find no strong correlation between the positions of clumps and regions of very low sound speed -- at least not nearly as strong as between Toomre $Q$ and gas surface density. But, we find that in PRFM-vol~2 and TNG, where no clumps have yet formed, regions with $Q<1$ do correlate with regions with low isothermal sound speeds. When looking at the galaxies at redshift~0.5, we find that the same is true at that time for PRFM-vol and TIGRESS/Schmidt. This highlights that the physical process responsible for the formation of stellar clumps is the gravitational collapse of the star forming disk. Since pressure counteracts gravitational collapse, clump formation is enhanced in models in which the star-forming ISM is less pressurized. Collapsing regions have low isothermal sound speeds, and once they are collapsed, they become self-gravitating, forming high-density, highly star-forming gas clumps. In our two models with lower pressure, we are looking at the end result of this process. The gas disks of PRFM-vol and TIGRESS/Schmidt contain very dense clumps with high star formation rate surface densities. Since collapse has already happened, the correlation between $Q$ and sound speed is less pronounced. In TNG and PRFM-vol~2, regions with $Q<1$ correlate with high surface density and low sound speed, indicating that disk collapse is ongoing. 

In all cases, we find that the star formation rate surface density peaks correlate with low $Q$ values. This highlights that in our simulations some gravitational instability is required to trigger star formation. This regime appears to be rather fine-tuned, given that changing the pressure by a factor of $\sim 2$ can lead to drastically different morphologies, at least when looking at the stellar light emitted by young stars\footnote{We note that for a fully conclusive comparison of the galaxy morphologies with observational data, one should generate mock observations using radiative transfer. While beyond the scope of the current work, we aim to investigate this further in the future.}.

As a final note, we point out that in our benchmark model setup, explicit stellar feedback is modeled using hydrodynamically decoupled wind particles, meaning that there is no local and self-consistent coupling of the feedback into the ISM. Such local feedback, when taken directly into account, could potentially disperse dense, star forming clouds. The numerical need for higher effective pressure in the star forming ISM can thus be interpreted as a consequence of our choice of an effective model. In other words, while the physical mechanism of Toomre instability that we present here is correct, it is likely not exactly what actually happens in nature. However, in eEoS-based ISM models commonly used in large cosmological simulations, simulated galaxy morphologies can put numerical -- but not truly physical -- constraints on the normalization of the eEoS.

\subsection{Stellar densities in clumpy and non-clumpy galaxies}\label{ssec:stellar_clumpy_densities}

So far, we have only identified stellar clumps as a visual feature that is characteristic of galaxies that emerge in cosmological simulations when using an eEoS that provides comparatively little pressure in the star forming ISM. Here, we aim to quantify how different the PRFM-vol and TIGRESS/Schmidt galaxies are from their TNG or PRFM-vol~2 counterparts. To that end, we continue to focus on the same galaxy that we analyzed in Fig.~\ref{fig:toomreq_face_on}. 

In Fig.~\ref{fig:clumps_cumulative}, we investigate how stars in this galaxy are distributed. To that end, we show their cumulative mass distributions as a function of ambient stellar density, and compare them between our four models. In the left panel, we show the distributions calculated from all stars in the galaxy. The first thing to note is that, on average, the densities that we can reliably resolve and maintain long-term are not excessively high. It is therefore not possible to associate the visually apparent clumps with observed overdense structures, such as nuclear star clusters. In fact, we find that for all of our models, a large fraction of the stars in the simulated galaxy live in regions that are denser than our numerical reference density of $\sim 6\times 10^{-4}\,{\rm M}_\odot{\rm pc}^{-3}$. At higher densities, gravitational forces between stars are softened, and, as a consequence, overdensities are numerically washed out with time. 

Nonetheless, we find significant differences between our models at densities larger than the reference density, even when considering all of the stars\footnote{As we show in Appendix~\ref{apxsec:clumps_prfm_vol}, clumps do not appear before redshift 1, which is past cosmic noon. So the fact that a significant difference remains between the models highlights that the difference seen in younger stars has to be quite significant.}.
In the left panel, we find that TNG has the lowest fraction of stars that live in environments with $\rho_\ast > 10^{-3}\,{\rm M}_\odot {\rm pc}^{-3}$. The PRFM-vol and PRFM-vol~2 curves look fairly similar, but the fraction of stellar mass above a given density is consistently higher in PRFM-vol than in PRFM-vol~2 at densities larger than the reference density, indicating that PRFM-vol stars tend to live in denser environments, consistent with the presence of clumps in PRFM-vol -- and their absence in PRFM-vol~2. The TIGRESS/Schmidt curve again looks very different from the other models, with an extended tail towards high densities, highlighting the fact that the soft eEoS supports the collapse of gas into very dense clumps. 

Above the reference density, and when considering all stars, the spread between models is of order $20\%$, which is significant, but not dramatic. This changes drastically when limiting our analysis to young stars, which intuitively makes sense, given the blue appearance of the stellar clumps in Figs~\ref{fig:galaxy_morphologies_one} and~\ref{fig:galaxy_morphologies_two}. In the right panel of Fig.~\ref{fig:clumps_cumulative}, we see a clear separation between models in which the galaxy in question is clumpy and models in which it is not. Given how similar the PRFM-vol and the PRFM-vol~2 curves are when considering all stars, their difference when focusing on young stars is particularly striking. In PRFM-vol, young stars lie in environments with much higher densities than in PRFM-vol~2 or TNG. For example, close to 50\% of the young stellar mass in PRFM-vol has ambient stellar densities larger than $10^{-2}\,{\rm M}_\odot{\rm pc}^{-3}$, compared to none of the young stellar mass in PRFM-vol~2 (and almost none in TNG). The TIGRESS/Schmidt curve looks similar to the corresponding curve in the left panel, but with an even larger fraction of stars at high ambient stellar densities. 

These results strongly confirm some of our earlier conclusions. First, clump formation is predominantly a late time phenomenon. Second, it occurs primarily in models with low pressure in the star-forming ISM, which means that it can be regulated by changing the normalization of the eEoS. Finally, we note that the right panel of Fig.~\ref{fig:clumps_cumulative} suggests that the phenomenon is fairly significant, especially when considering recently formed stars. Over the whole stellar population in a given galaxy, the difference between models is less pronounced, which is both due to density peaks being numerically ``washed out'' with time, and due to clump formation being a late time phenomenon.  


\subsection{Stellar masses} \label{ssec:stellarmasses}

\begin{figure}
    \centering
    \includegraphics[width=\linewidth]{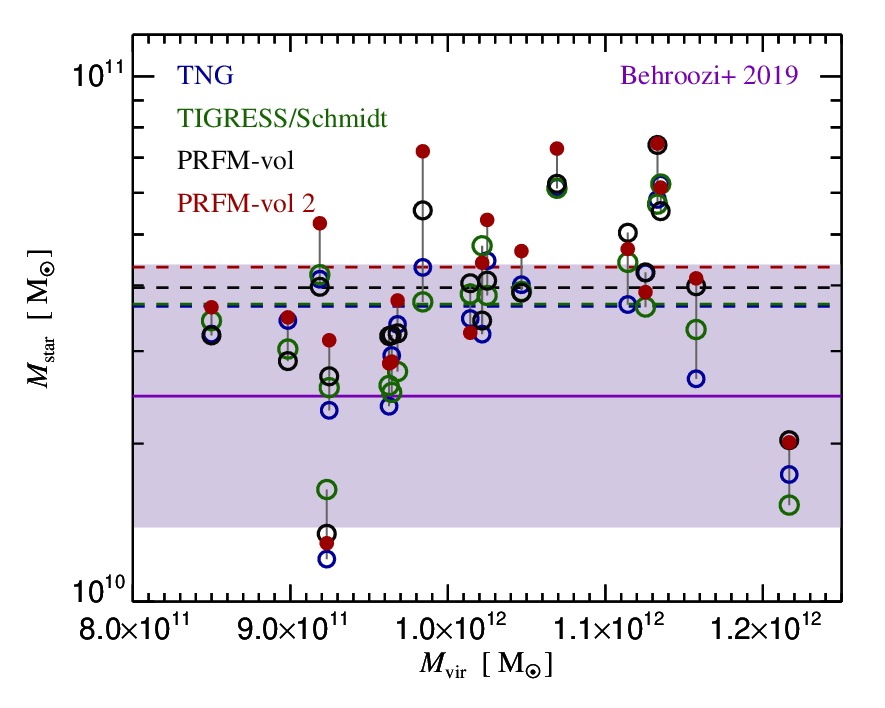}
    \caption{Comparison of the final stellar masses of the 20 simulated zoom galaxies with $\log(M_{\rm vir}/{\rm M}_\odot) = 12$ and zoom factor 4 for four different models, TNG (blue open circles), TIGRESS/Schmidt (green open circles), PRFM-vol (black open circles) and PRFM-vol 2 (red closed circles). Stellar masses are shown as a function of halo mass, and galaxies corresponding to the same halo are matched as indicated by the connecting black vertical lines. Dashed horizontal lines show averages over 20 galaxies, with colours corresponding to the respective models, as indicated. In purple, we show masses consistent with the median (solid purple line) and the galaxy-to-galaxy scatter (shaded purple band) of the stellar to halo mass relation at redshift $z=0$ reported in \citet{Behroozi:2019}}
    \label{fig:stellarmassesz0}
\end{figure}
\begin{figure}
    \centering
    \includegraphics[width=\linewidth]{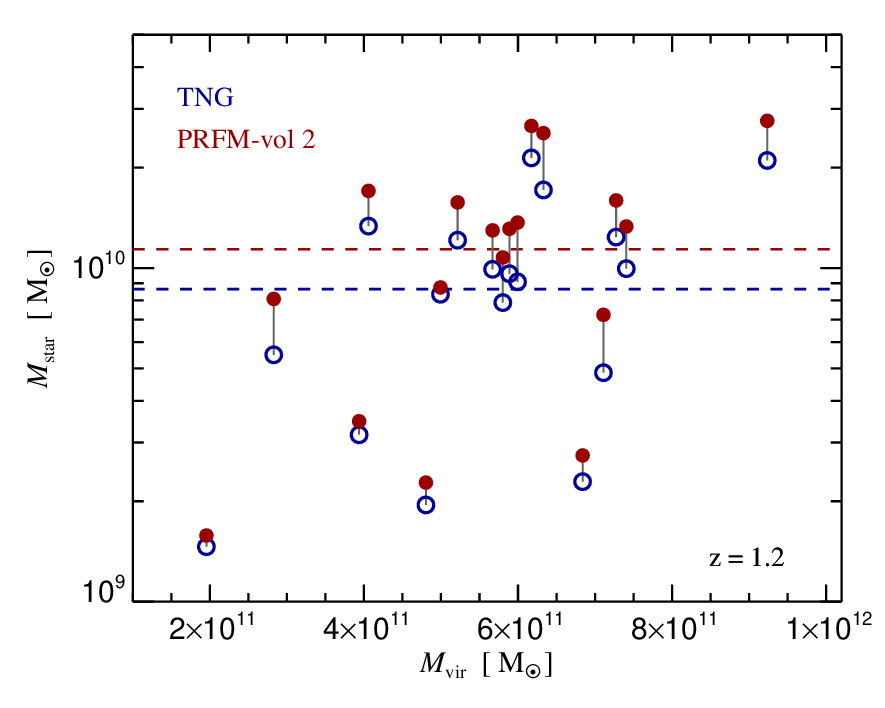}
    \caption{Comparison of the stellar masses in TNG (blue open circles) and PRFM-vol-2 (red closed circles) as a function of halo mass at redshift $z=1.2$. Galaxies are matched as in Fig.~\ref{fig:stellarmassesz0} and vertical lines indicate averages over 20 galaxies.  }
    \label{fig:stellarmassesz12}
\end{figure}

\begin{figure*}
    \centering
    \includegraphics[width=0.48\linewidth, trim = 0.5cm 0.5cm 0.5cm 0.5cm, clip]{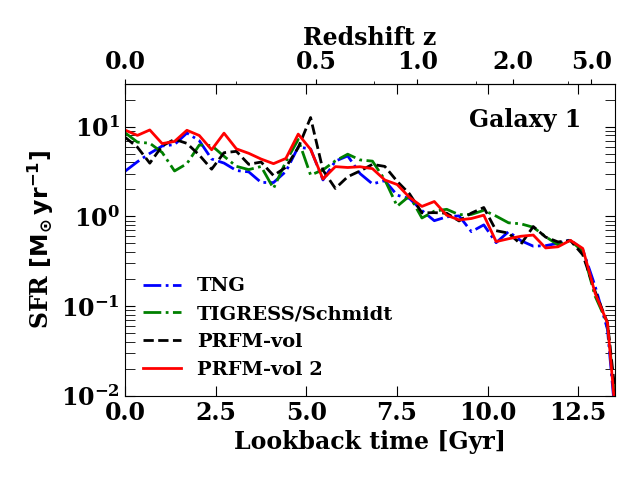}
    \includegraphics[width=0.48\linewidth,trim = 0.5cm 0.5cm 0.5cm 0.5cm, clip]{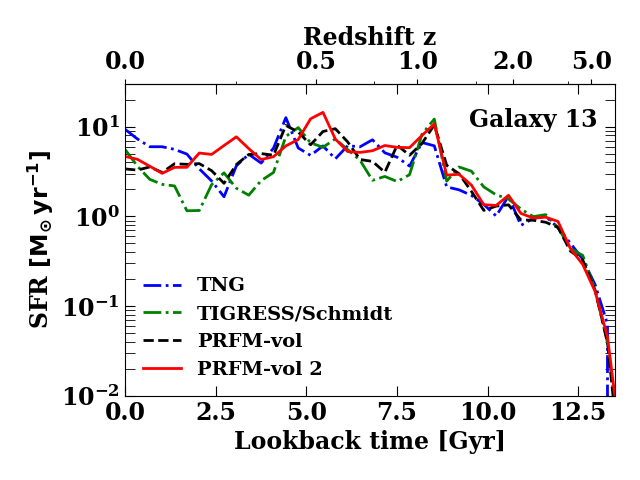}
    \caption{Star formation histories of two selected cosmological multizoom galaxies. We show star formation rate as a function of lookback time (and redshift) for the TNG (blue), TIGRESS/Schmidt (green), PRFM-vol (black) and PRFM-vol-2 (red) models. In most galaxies, differences between the star formation histories are fairly noisy. However, we find that in a few of them, the TIGRESS/Schmidt rate systematically drops below the PRFM-vol and PRFM-vol-2 rates at late times. An example of this is shown in the right panel (Galaxy 13).}
    \label{fig:sfrs_ind_gals}
\end{figure*}
We now compare the stellar masses between our models. In Fig.~\ref{fig:stellarmassesz0}, we show stellar mass as a function of halo mass at redshift $z=0$ for the TNG, TIGRESS/Schmidt, PRFM-vol, and PRFM-vol-2 multizoom simulations. Galaxies are paired by matching their host haloes, and we show the average stellar masses for each model as dashed horizontal lines. While there is substantial scatter in the galaxy stellar mass, we do find that on average, PRFM-vol -- and especially PRFM-vol-2 -- galaxies are slightly more massive than TNG and TIGRESS/Schmidt galaxies. We had previously asserted that the redshift $z=0$ stellar masses in TNG and TIGRESS/Schmidt are approximately equal  \citep[see][]{Burger2025}, and thus the fact that stellar masses in the two PRFM-vol models are slightly larger should come as no surprise. Star formation rates in PRFM-vol are in general higher than in TIGRESS/Schmidt, as we discussed in the case of isolated galaxies in Section~\ref{ssec:sfr_isolated}. This is largely a consequence of the star formation rate being modified due to the gravitational contributions from stars and DM. Evidently, this effect prevails in spite of the presence of TNG winds, which were not included in the isolated galaxy simulations.

The even further increase in stellar mass at fixed halo mass leads to some slight tension between PRFM-vol~2 and the results of the UniverseMachine \citep{Behroozi:2019}, although it is important to note that we did not recalibrate any component of our models except for the new star formation models, and that star formation, in cosmological simulations and in nature -- and especially at late times -- is largely regulated by stellar and black hole feedback. Galactic winds drive mass and energy out of the ISM and into the circumgalactic medium (CGM), where it can then cool and eventually fall back onto the galaxy. A recalibration -- or a change -- of the wind model used in cosmological simulations will therefore affect the stellar mass in a simulated galaxy at $z=0$ much more than a change in the star formation model. With that in mind, we take the tension between PRFM-vol~2 and UniverseMachine as a sign that we may need to slightly recalibrate the wind model, instead of as a significant shortcoming of our model.  

The fact that PRFM-vol~2 stellar masses are even larger than PRFM-vol stellar masses can be explained by the direct dependence of gas depletion time on the effective equation of state (see also equation~\ref{eq:tdep_fin}). Since this effect should dominate at higher redshift, we also look at a comparison between the stellar masses in TNG and in PRFM-vol-2 at redshift $z=1.2$.

Such a comparison is shown in Fig.~\ref{fig:stellarmassesz12}. Therein, stellar masses are shown as a function of halo mass at redshift $z=1.2$. As we show in Appendix~\ref{apxsec:clumps_prfm_vol}, this is before disks have fully formed in most cases. Here, the difference in stellar mass has to originate from the different star formation law dependence on gas density\footnote{At least for a fixed wind model.} \citep[see also our discussion in][]{Burger2025} or the difference in effective equation of state. As we can see, even at redshift $z=1.2$, stellar masses in PRFM-vol~2 are significantly higher than in TNG. This indicates that star formation in PRFM-vol~2 was also significantly higher than in TNG at early times, when the difference is likely to originate from the difference in eEoS.

\subsection{Star formation rates}\label{ssec:sfr_cosmo} 
We continue our discussion from Section~\ref{ssec:stellarmasses} by looking at the star formation rates of individual matched galaxies and comparing them between models.

Two examples of this are shown in Fig.~\ref{fig:sfrs_ind_gals}. We compare the star formation rates as a function of lookback time between the multizoom galaxies simulated with the TNG, TIGRESS/Schmidt, PRFM-vol, and PRFM-vol~2 star formation and ISM models. Instantaneous star formation rates are calculated by identifying member stars of the galaxies and then using their stored information on birth time and initial particle mass. This means that we do not differentiate between in-situ and ex-situ star formation.  

From the example star formation rates shown in Fig.~\ref{fig:sfrs_ind_gals}, it is not easy to identify general trends, since the star formation histories are fairly noisy. One trend that emerges in several galaxies, however, can be identified in galaxy 13 (right panel). There, we see that below a lookback time of about $4\,{\rm Gyr}$, the TIGRESS/Schmidt star formation rate drops below the star formation rates in the other models, and in particular below the PRFM-vol and the PRFM-vol-2 rates. We find this to be the case in several of our simulated galaxies. This is likely a consequence of the stellar and DM potentials contributing to the star formation rate, which, compared to TIGRESS/Schmidt\footnote{A direct statement like this cannot be made about TNG, since the star formation law is entirely different.}, leads to higher star formation rates once a substantial number of stars has formed (see also the discussion in Section~\ref{ssec:sfr_isolated}). While in other galaxies, such as the one on the left panel, we do not observe this trend, we do note that star formation rates in individual galaxies are generally quite noisy. For that reason, we next focus on an average of the star formation rates over all 20 simulated galaxies for each of the models. 

\begin{figure}
    \centering
    \includegraphics[width=\linewidth, trim = 0.5cm 0.5cm 0.5cm 0.5cm, clip]{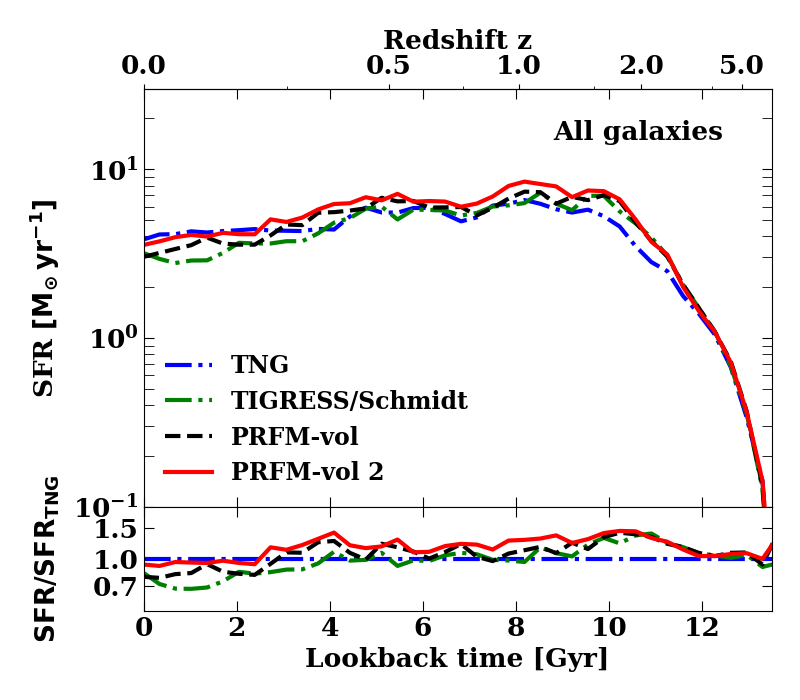}
    \caption{Average of the star formation histories of all 20 simulated galaxies, compared between the TNG (blue), TIGRESS/Schmidt (green), PRFM-vol (black), and PRFM-vol~2 (red) models. The bottom panel gives the ratio of the star formation rate in the three new models compared to TNG.}
    \label{fig:stack_sfr}
\end{figure}

 In Fig.~\ref{fig:stack_sfr}, we present a comparison of the averaged star formation rates in all four models. To that end, for each of the models, we average over the instantaneous star formation rates of each of the 20 galaxies. In the smaller bottom panel, we show the ratio of the TIGRESS/Schmidt, PRFM-vol, and PRFM-vol~2 star formation rates to the TNG star formation rate. We find that in TIGRESS/Schmidt, PRFM-vol, and PRFM-vol~2, the stacked star formation rates begin higher than in TNG, with PRFM-vol~2 having the highest overall star formation rate at lookback times above $2\,{\rm Gyr}$. PRFM-vol, and especially TIGRESS/Schmidt, exhibit faster drops towards the TNG star formation rate, with the TNG and TIGRESS/Schmidt star formation rates being roughly equal between lookback times of 4~Gyr and 8~Gyr, and PRFM-vol staying slightly higher. At lookback times below 4~Gyr, TIGRESS/Schmidt and PRFM-vol star formation rates start to drop below the TNG line, but PRFM-vol star formation rates remain fairly close to TNG. PRFM-vol~2 galaxies, on the other hand, continue to have star formation rates equal to or higher than in the TNG model all the way until lookback time 0. 

Two main conclusions can be drawn from Fig.~\ref{fig:stack_sfr}. First, we can directly relate the fact that star formation rates in TIGRESS/Schmidt drop relative to PRFM-vol and PRFM-vol~2 to our discussion in Section~\ref{ssec:sfr_isolated}. Since PRFM-vol star formation rates are higher than TIGRESS/Schmidt rates in the presence of a significant stellar component, the relative drop in star formation rate can be linked to the gradual formation of stars and the buildup of a stellar disk past $\sim$ redshift $z=1$. Second, we note that the increased star formation rates in PRFM-vol~2 are reflected in both the final distribution of stellar masses (see Section~\ref{ssec:stellarmasses}) and the final distribution of star forming gas (see Fig.~\ref{fig:cum_density_functions}). Interestingly, despite the clear drop in star-forming gas mass seen in Fig.~\ref{fig:cum_density_functions}, the star formation rates in the PRFM-vol-2 galaxies still remain at the level of TNG. Especially when compared to the late time drop seen in PRFM-vol, we interpret this as a direct consequence of the shorter depletion times in gas with higher pressure (see equation~\ref{eq:tdep_fin}). 

Overall, however, we find that in spite of these minor differences, the star formation rates and final stellar masses in all of our models are still quite similar. Beyond that, they are also strongly dependent on the wind model used in the simulations, as we mentioned in Section~\ref{ssec:stellarmasses}, which limits the effect that small scale star formation laws can have on cosmological star formation rates. It may thus be difficult to use observed star formation rates or stellar masses at $z\leq 2$ for the purpose of model selection. 

\subsection{Galaxy sizes} \label{ssec:gal_sizes_mzoom}

\begin{figure}
    \centering
    \includegraphics[width=\linewidth]{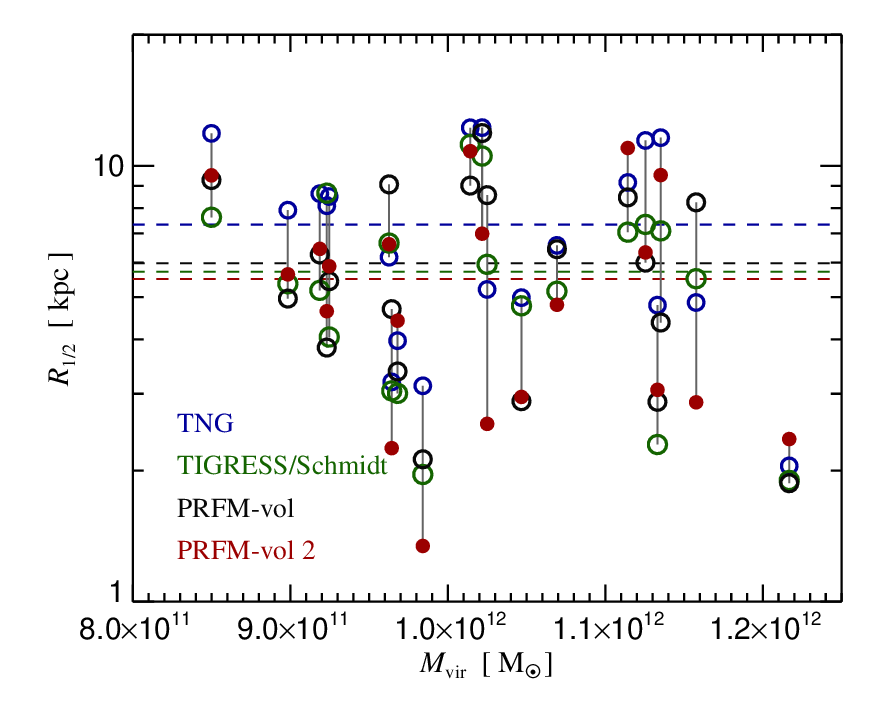}
    \caption{Half-mass radii of the simulated galaxies as a function of halo virial mass. We show a comparison between the TNG (blue open circles), TIGRESS/Schmidt (green open circles), PRFM-vol (black open circles), and PRFM-vol-2 (red closed circles) models. Horizontal dashed lines show averages over the 20 multizoom galaxies simulated for each model. Galaxies with corresponding host haloes are matched and connected by black vertical lines.}
    \label{fig:galaxy_sizes}
\end{figure}

We move on to analyze the sizes of the simulated multizoom galaxies. In Fig.~\ref{fig:galaxy_sizes}, we show galaxy half mass radii as a function of halo virial mass for simulations using the TNG, TIGRESS/Schmidt, PRFM-vol, and PRFM-vol~2 models. Galaxies are matched in the same way as in previous plots, and we calculate a galaxy's half mass radius as the spherical radius in which half of the stellar mass is contained. We find very strong scatter in the sizes of the simulated galaxies. Looking at the averages, which we show as dashed horizontal lines, we find that TNG galaxies tend to be larger than the galaxies simulated with any of the other models. While this may appear significant, we note that in \cite{Burger2025} we found that galaxies simulated using the original \citet{Springel2003} model had average sizes similar to the TIGRESS/Schmidt galaxies. Given how close the PRFM-vol and PRFM-vol-2 half mass radii are to TIGRESS/Schmidt, one may conclude that that PRFM based models may naturally prefer smaller galaxies than emerge from the TNG model. However, given how large the scatter is and considering the fact that we are still using TNG winds which have not been recalibrated, we refrain from making any definitive statement.   

\subsection{Galaxy scale heights}

Finally, we move on to compare the scale heights in our sample of 20 galaxies with host halo mass  $\log({M_{\rm vir}}/{\rm M}_\odot) = 12$ and zoom factor 4. In \cite{Burger2025}, we had found that scale heights in the TIGRESS/Schmidt model are very small compared to those that emerge in TNG. This was particularly true for young stars, and related to the fact that the scale height of the star-forming gas is set in large part by the eEoS, which, as we have seen in Fig.~\ref{fig:eos_comp}, assigns much lower pressure to the star-forming gas in TIGRESS/Schmidt than in TNG. To alleviate this issue, we changed the eEoS and introduced velocity kicks (see Sections~\ref{ssec:vkicks} and \ref{ssec:scaleheights}). We thus expect that scale heights will be larger in PRFM-vol, and especially in PRFM-vol-2, where the pressure is substantially higher than in TIGRESS/Schmidt, and the velocity kicks are larger in magnitude, since they are sampled from a Gaussian with width $\sim \sigma_{\rm eff}$. 

\begin{figure}
    \centering
    \includegraphics[width=\linewidth]{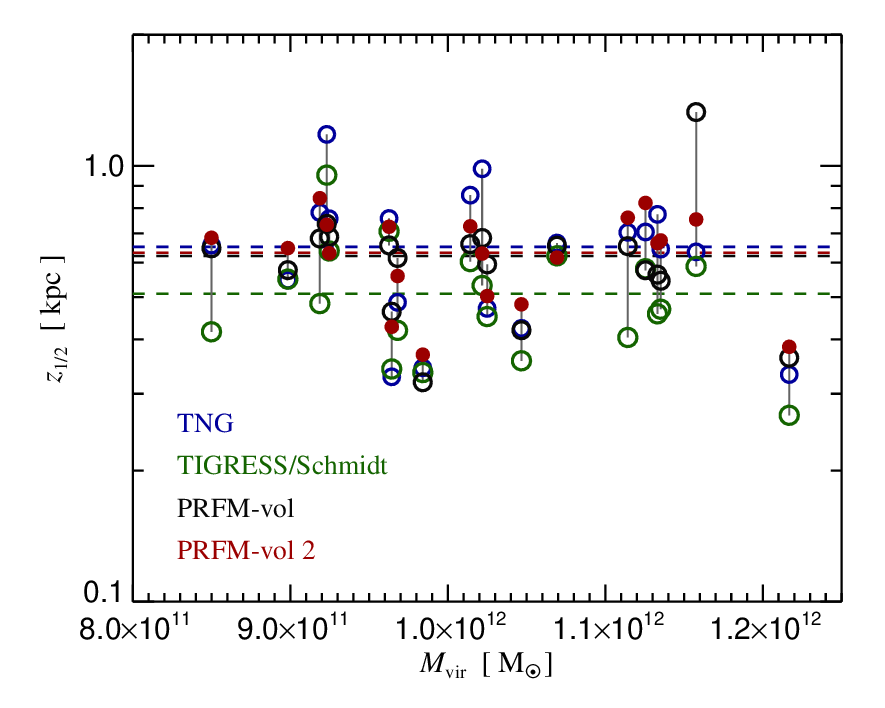}\\
    \includegraphics[width=\linewidth]{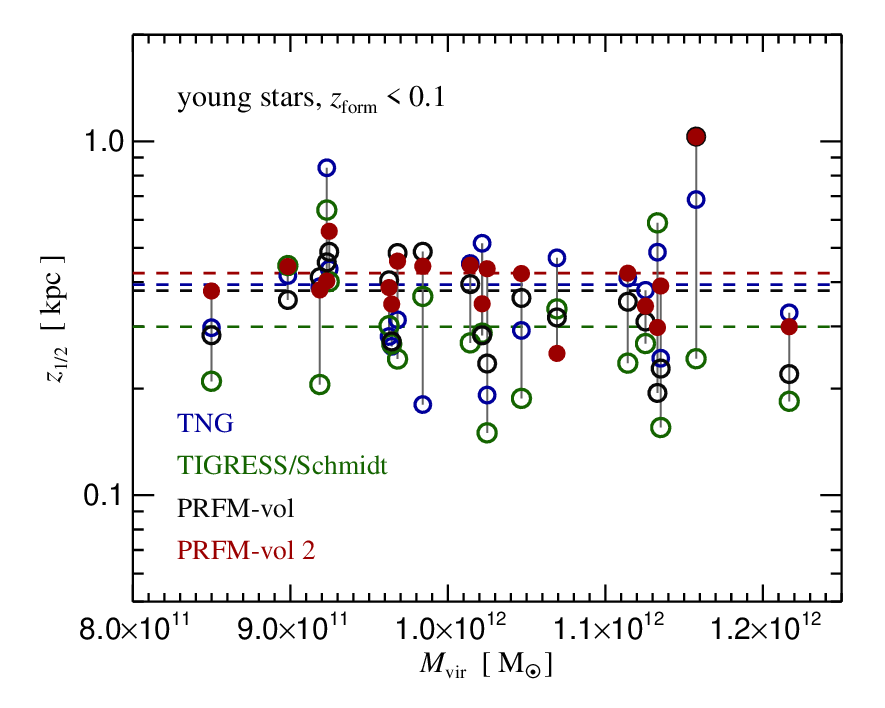}
    \caption{Comparison of the stellar scale heights of our 20 multizoom galaxies simulated using the TNG (blue open circles), TIGRESS/Schmidt (green open circles), PRFM-vol (black open circles), and PRFM-vol~2 (red filled circles) models. Scale heights are shown as a function of halo virial mass, and galaxies are matched as in previous figures, with vertical dashed lines showing averages over the 20 galaxies, and line colours corresponding to models as indicated. The upper panel shows the half mass heights calculated from all stars belonging to a galaxy, while in the lower panel we limit our analysis to stars with a formation redshift of less than $z < 0.1$.  }
    \label{fig:cosmo_scaleheights}
\end{figure}

We show a comparison of stellar scale heights in Fig.~\ref{fig:cosmo_scaleheights}. For each galaxy, we calculate the stellar scale height as the half-mass height, i.e.~the vertical coordinate that encloses half of the galaxy's stellar mass. We compare our four models, TNG, TIGRESS/Schmidt, PRFM-vol, and PRFM-vol~2. In the upper panel, we use all of each galaxy's stars to calculate the half-mass height, whereas we limit ourselves to stars with formation redshift $z_{\rm form} < 0.1$ in the lower panel in order to focus on young stars, which should trace the scale height of star forming gas and thus be more sensitive to model differences. 

Comparing the stellar scale heights calculated from all the stars, we find that switching to PRFM-vol fixes the issue of very small scale heights that we encountered in TIGRESS/Schmidt. The inclusion of velocity kicks, along with the change in eEoS, drives up the scale heights in the two PRFM-vol models.

Notably, however, we find no pronounced difference between PRFM-vol and PRFM-vol-2 when comparing the scale heights calculated from all member stars. There could be a few reasons for this. Stars at larger heights are likely older, and may have formed before the galaxy settled into a stable disk, diminishing the importance of relatively small differences in pressure. In addition, heating of orbits can drive stars outward, increasing the scale height and also potentially erasing minor differences. We note in passing that the TNG, PRFM-vol, and PRFM-vol~2 models are almost perfectly consistent with the number reported by \citet{Yu2026}, which is $z_{1/2} = 0.67\pm 0.06\,{\rm kpc}$.
It is worth pointing out, however, that not all observations agree with the \citet{Yu2026} value. For example, the majority of mass in the Milky Way's disk has a scale height half of the above \citep[e.g.][]{2008ApJ...673..864J,2013ApJ...772..108Z,2022MNRAS.511.3863E}, similar to the thin disks of galaxies with two components \citep[e.g.][]{2025MNRAS.540.3493T}.

Focusing on young stars only (i.e.~the lower panel of Fig.~\ref{fig:cosmo_scaleheights}), a clearer picture of the effects of increasing the eEoS pressure emerges. We find that, while PRFM-vol scale heights are again comparable to TNG, PRFM-vol~2 scale heights are even a little larger. The additional factor of two in the eEoS and the increase in velocity kick magnitude work together to convincingly fix the issue of small scale heights that we encountered in the TIGRESS/Schmidt model. 

In our new PRFM-vol model, the scale height of young stars (and hence the star forming gas) is clearly sensitive to the normalization of the eEoS, both due to the change in vertical dynamical equilibrium, and due to the fact that the magnitude of the added velocity kicks is modified along with it. In light of our discussion on the physical meaning of the eEoS in the absence of local feedback (or rather the lack thereof) in Section~\ref{ssec:toomre}, this makes the normalization of the eEoS a potentially promising parameter to tune when aiming to calibrate a new combination of subgrid physics models, in which PRFM-vol is included as an ISM and star formation model.    

\section{Summary and Conclusions} \label{sec:conclusions}

We have implemented, tested, and presented a new subgrid model for cosmological simulations, based on both PRFM theory and scaling relations measured from the TIGRESS simulation suite \citep{Ostriker2022}. Our present study follows up previous work introducing the TIGRESS/Schmidt model in \citet{Burger2025}, and is a subgrid model in the same vein. \citet{Jeffreson2026} previously implemented and compared two versions of star formation subgrid models, PRFM-vol and PRFM-int, in isolated galaxy simulations. Here we develop PRFM-vol for cosmological simulations, where the numerical resolution is sufficient to resolve the scale height of gas disks, but cannot directly resolve the multiphase ISM and the self-regulation of star formation and feedback.
For that reason, our model relies on an effective equation of state to prescribe the mean pressure as a function of average gas density in the star-forming ISM. As in our work on the TIGRESS/Schmidt model,
we base the implementation of such an effective equation of state on the results of highly resolved simulations of the ISM. Concretely, we here use the relation derived from the mass averaged effective velocity dispersion that was presented in \citet{Ostriker2022}, which results in a much harder eEoS compared to what we used in the TIGRESS/Schmidt model (see Fig.~\ref{fig:eos_comp}). We emphasize, however, that the bulk implementation of PRFM-vol is independent of the specific choice of eEoS and feedback yield. Testing alternative calibrations (such as those obtained from the TIGRESS-NCR simulations of \citet{Kim2024} or the GHOSDT simulations of \citet{AlonUli2025}) requires a change of no more than two lines of code.  

In addition to the different eEoS, we introduce a new star formation law, which is based entirely on PRFM star formation theory. The star formation rate is determined locally for each gas cell, and is defined through a depletion time, the calculation of which requires the calibrated feedback yield, the eEoS, and the local dynamical time (see equation~\ref{eq:tdep_fin}). In PRFM-vol, the dynamical time is calculated from the local gas, DM, and stellar densities. We compute stellar and DM densities using nearest neighbour kernel density estimates. We have demonstrated that in order to apply the results obtained in the TIGRESS shearing box simulations to estimate local star formation rates (without projecting to surface-based quantities), we need to renormalize the calculated dynamical time.  We extend the renormalization presented in \citet{Jeffreson2026} to include a factor that depends on the local ratio of DM density to baryonic density. To address some shortcomings of the TIGRESS/Schmidt model, and motivated theoretically by the presence of a subgrid turbulent pressure component, 
we added velocity kicks to newly born stars, with a magnitude consistent with the expected local velocity dispersion. We also considered the local squared ratio between the stellar velocity dispersion and the effective velocity dispersion of the gas as a proxy for the ratio between the corresponding scale heights, and tested the accuracy of that approximation. 

In large part, the PRFM-vol model introduced here is similar to the one presented in \citet{Jeffreson2026}. However, there are some key differences related to the use of PRFM-vol in cosmological simulations that should be pointed out, and so we comment on those here: 
\begin{itemize}
    \item In our implementation, we define the nearest neighbour search kernel by a fixed number of nearest neighbours instead of a fixed search length. This can be advantageous in scenarios where, for example, the stellar population in a young galaxy is still very sparse. 
    \item We add velocity kicks to newly born stars. 
    In addition to the underlying physical motivation for this, it allows us to address the small stellar scale heights that we encountered with the TIGRESS/Schmidt model \citep{Burger2025}. 
    \item In the present version of PRFM-vol, the renormalization of the dynamical time allows for the possibility that dark matter dominates the gravity. This is necessary due to the expected application of the model in widely varying conditions, as opposed to as a baseline comparison model to PRFM-int in isolated simulations only. 
\end{itemize}

We would like to stress here that the explicit implementation of the relationship between the dynamical time and the depletion time that emerges from PRFM theory
constitutes the main motivation for the development of both the PRFM-vol and PRFM-int subgrid models. 
The PRFM theory is built upon contemporary understanding of the co-regulation of star formation rates and multiphase ISM properties, as  confirmed in TIGRESS as well as other high-resolution star-forming ISM simulations, and PRFM predictions are in agreement with observations of galaxies from several surveys. 
As shown in \citet{2024ApJ...975..151H}, the TNG model does not match the predicted dynamical time vs. depletion time relation, which makes it desirable to replace the \citet{Springel2003} model with one that accurately predicts star formation rates in disk galaxies. A key new conceptual element is the inclusion of the gravitational weight of stars and DM in the calculation of the gas dynamical time.
This also offers practical advantages by expanding the range of galactic conditions where cosmological simulations can be compared to observations, e.g. to include ultra-diffuse galaxies \citep{Kado-Fong2022}. 

We first tested our model on idealized, isolated galaxies, which we set up similar to our test galaxies in \citet{Burger2025}, with some modifications to allow for specific tests of our new model features. In particular, we included setups with an old stellar component in the galactic disk, in order to verify that the PRFM-vol depletion time formula accurately captures the dependence of the dynamical time on the local stellar density. Three main conclusions can be drawn from our tests on isolated setups: 
\begin{itemize}
    \item PRFM-vol performs much better than TIGRESS/Schmidt if a substantial part of the gravitational potential is due to stars or DM. In cases where, for example,  $90\,\%$ of the disk is initially made up of stars, PRFM-vol still predicts star formation rates that are in line with the TIGRESS $P-\Sigma_{\rm SFR}$ relation, while star formation rates in the TIGRESS/Schmidt model are systematically lower, as they only depend on the local gas densities \citep{Burger2025}. 
    \item Velocity kicks contribute strongly to increasing the scale height of newly formed stars, as does the change in eEoS. We showed that gas scale heights, star formation rate scale heights, and stellar scale heights increase compared to TIGRESS/Schmidt when switching to the PRFM-vol star formation law and eEoS, and the stellar scale height further increases upon adding the velocity kicks. This change alleviates the issue of very small scale heights in the TIGRESS/Schmidt model. 
    \item The squared ratio of the effective velocity dispersion to the 3d stellar velocity dispersion (equation~\ref{eq:scale_height_ratio}) can, in most of our isolated galaxies, be used as a very good estimator for the ratio between the gas scale height and the stellar scale height, on par with using the vertical stellar velocity dispersion instead (equation~\ref{eq:scale_height_ratio_two}). However, for three of our galaxies with very low overall star formation rates (and discontinuous star forming regions in the disk), we find that neither of the two estimators perform well. Given that quenching is a common phenomenon in cosmological simulations, we decided not to use any scale height estimator for now, and instead simply adopt $H_{\rm gas} \simeq H_\star$ in our default version of PRFM-vol. We verified that this introduces only negligible errors and point out that the use of a scale height estimator may be beneficial for future work targeting star forming galaxies at high redshift.   
\end{itemize}

Having successfully tested our model in isolated galaxies, we then turned to cosmological multizoom simulations. We find that the expected changes with respect to the TIGRESS/Schmidt do indeed manifest themselves: 
\begin{itemize}
    \item Stellar scale heights in PRFM-vol are larger than in TIGRESS/Schmidt, at the level of scale heights in TNG. Comparing the simulated scale heights to recent observations \citep{Yu2026}, we find good agreement between TNG, PRFM-vol (and PRFM-vol~2) and observations. Scale heights are also increased when considering young stars only.
    \item Average star formation rates and final stellar masses are higher, due to the change in eEoS and due to increased star formation in the presence of a significant stellar component. 
\end{itemize}
Not much changes when considering the galaxy sizes. They remain comparable to the galaxy sizes found in TIGRESS/Schmidt, and slightly smaller than the galaxy sizes in TNG. 

Looking at the morphologies of our simulated galaxies at redshift $z=0$, we find that the stellar clumps that were present in the TIGRESS/Schmidt simulations shown in \citet{Burger2025} persist. We argued in the past that this is due to Toomre instabilities, owing to the fact that in PRFM-vol, the pressure at a given hydrogen number density is lower than in TNG, where no such stellar clumps appear. 
In an attempt to remedy this (without adding additional complications to the subgrid model to represent localized feedback within clumps), we rerun our benchmark multizoom setup with a modified PRFM-vol model, where we multiply the eEoS by a factor of two. Indeed, we do find  improved galaxy morphologies, with no remaining stellar clumps. Given that the PRFM-vol pressure is lower than the TNG pressure close to the numerical density threshold for star formation, but eventually surpasses the TNG pressure, the idea that the clumps form as a result of Toomre instabilities only makes sense if the majority of the star forming gas resides at densities close to the star formation threshold. We verify explicitly that this is the case, by looking at the mass weighted cumulative density functions of star forming gas at different redshifts, for both the PRFM-vol run and the run with the modified eEoS, which we name PRFM-vol~2. For PRFM-vol, virtually all of the star-forming gas has densities below the value at which the PRFM-vol pressure surpasses the TNG pressure, at any of the redshifts we analyze. In the PRFM-vol~2 run, on the other hand, about half of the star forming gas now has higher pressure than it would have in TNG. As a result, the disks remain Toomre stable, and no stellar clumps form.

Due to the relatively hard eEoS, the majority of the star forming gas in both PRFM-vol and PRFM-vol~2 simulations lives in environments with fairly low hydrogen number densities. In spite of that, we find that the bulk of the stars is formed at hydrogen number densities of $\sim 1{\rm cm}^{-3}$, a factor of 10 above the numerical star formation threshold, but still in a regime where the PRFM-vol pressure is significantly lower than the TNG pressure at the same densities (see Fig.~\ref{fig:birth_densities_mass_function}). This is in stark contrast to the TIGRESS/Schmidt model, where stars form at much higher densities than in either the PRFM-vol model or in TNG, owing to the softer eEoS. The eEoS thus affects the star forming ISM in two ways. Its normalization determines how fast gas is depleted at a given density, while its power law index determines how dense star forming gas can become.  

In addition to PRFM-vol~2, we tested a version of our model in which we modified the eEoS at low densities, consistent with the effective velocity dispersion floor imposed in \citet{Jeffreson2026}. We found that this did not result in strong changes of the galaxy morphologies at redshift $z=0$. This points toward the fact that in models with a relatively hard eEoS, such as PRFM-vol, deriving a physically motivated star formation threshold value constitutes an important future task, as we find that the eEoS at low hydrogen number densities plays a crucial role in determining the morphology of simulated galaxies.  

Having confirmed that PRFM-vol~2 significantly changes galaxy morphologies compared to PRFM-vol, we took a closer look at a galaxy with clumpy morphology, and compared face-on projections of Toomre $Q$, isothermal sound speed, gas surface density, and star formation rate surface density between our four models (see Fig.~\ref{fig:toomreq_face_on}). We found that stellar clumps are associated with Toomre unstable gas density peaks, as well as peaks in gas surface density. We could also trace back the process of disk fragmentation. In models with less pressure in the star forming disk, Toomre instabilities arise when sound speeds are too low to counteract gravity. Clumps are the end product of disk collapse. We then moved on to quantify the significance of the clumps in that same galaxy (see Fig.~\ref{fig:clumps_cumulative}), and found strong model differences in the distribution of stellar densities, especially when considering young stars, showing that the morphological differences in stellar disks can be fairly significant, depending on which eEoS is chosen. We emphasize here that while this puts numerical constraints on the eEoS, these constraints should not be considered physical, and cosmological simulations such as ours do not allow us to comment on the validity of the TIGRESS results, or the results of resolved ISM simulations in general. Likely, the reason that eEoS models require relatively high pressure in the star forming ISM is to compensate for the lack of an explicit local stellar feedback mechanism that could disperse dense clumps. 

Following this line of thought, we interpret the results obtained when renormalizing the eEoS as providing information about how far away we are from a fully small-scale informed model, rather than as a failure of PRFM-vol itself. Given these results, one approach going forward would be to expand the subgrid model to include early feedback that would disperse dense, star-forming clumps -- as occurs in real galaxies. Another approach would be to treat the eEoS coefficient as a parameter that can be adjusted, effectively 
calibrating to observed galaxy morphological properties. 
The latter approach would arguably be at odds with the goal of developing a subgrid model based purely on a fundamental theory of dynamical equilibrium and the results of resolved small-scale simulations. 
However, adopting a renormalization factor can also be seen as a pragmatic alternative to introducing a more complex subgrid model, and indeed both routes can be pursued at the same time.  
Moreover, we believe that comparing PRFM-vol predictions to TNG predictions in systems where the changed star formation law is expected to make a strong difference will showcase the full impact of our new model. For such upcoming projects, we will remove the renormalization factor from the calculation of the depletion time, in order to directly compare PRFM star formation rate predictions to their TNG counterparts.

Apart from fixing the galaxy morphologies, adopting PRFM-vol~2 also leads to a further increase in stellar scale heights, an increase in star formation rates, both early and at late times, and as a result, also an increase in total stellar mass at redshift $z=0$. All of these changes are expected, given that an increase in pressure leads to stronger velocity kicks and shorter depletion times. Overall, we find that none of the PRFM-vol~2 scaling relations are in strong tension with the TNG results, though there is some slight tension between the stellar masses at redshift $z=0$ and the results of \citet{Behroozi:2019}. Given that we have used the TNG wind model, which we have not recalibrated, this slight mismatch should not be seen as a problem of our model, but rather as a challenge to combine it with a recalibrated wind model, or potentially with ARKENSTONE \citep{Smith2024a,Smith2024b,Bennett2026}, a new model for stellar winds, also developed within LtU. Considering the strong effect galactic wind  models have on cosmic star formation rates, it is also not too surprising that the changes that arise in stellar mass and star formation rates when altering the small scale star formation model are relatively small. It is important to note, however, that the current multizoom simulations do not address high-redshift star formation of the first massive galaxies, when cosmic densities were much higher than the present.  Since the new models have stronger density dependence of the SFR than in TNG, it will be very interesting to explore this regime, which we aim to do in forthcoming work. 



In conclusion, we have developed a physics-informed alternative to the TNG ISM and star formation model, based on both PRFM theory and results of high resolution ISM simulations. Our model works well, produces reasonable galaxies at redshift $z=0$, and is ready for use in high resolution cosmological simulations. It serves as one of many new and improved subgrid models developed within the Learning the Universe collaboration, and as such is ready to play an integral part in future next generation cosmological simulation projects.   

\section*{Acknowledgements}
We thank R\"{u}diger Pakmor for helpful comments and discussion. 
The authors acknowledge support by the Simons Collaboration on
``Learning the Universe''.
GLB acknowledges support from the NSF (AST-2307419), NASA TCAN award 80NSSC21K1053, and the Simons Foundation. The work of ECO and CGK was supported by grants 888968 from  the Simons Foundation and SFI-MPS-LU-00008515-02 from Simons Foundation International to Princeton University for the Learning the Universe Collaboration.

\section*{Data availability}

The data underlying this article will be shared upon reasonable re-
quest to the corresponding authors.

\bibliographystyle{mnras}
\bibliography{manuscript}

\appendix 

\section{Clump formation in PRFM-vol} \label{apxsec:clumps_prfm_vol}

\begin{figure*}
    \centering
    \includegraphics[width=\linewidth]{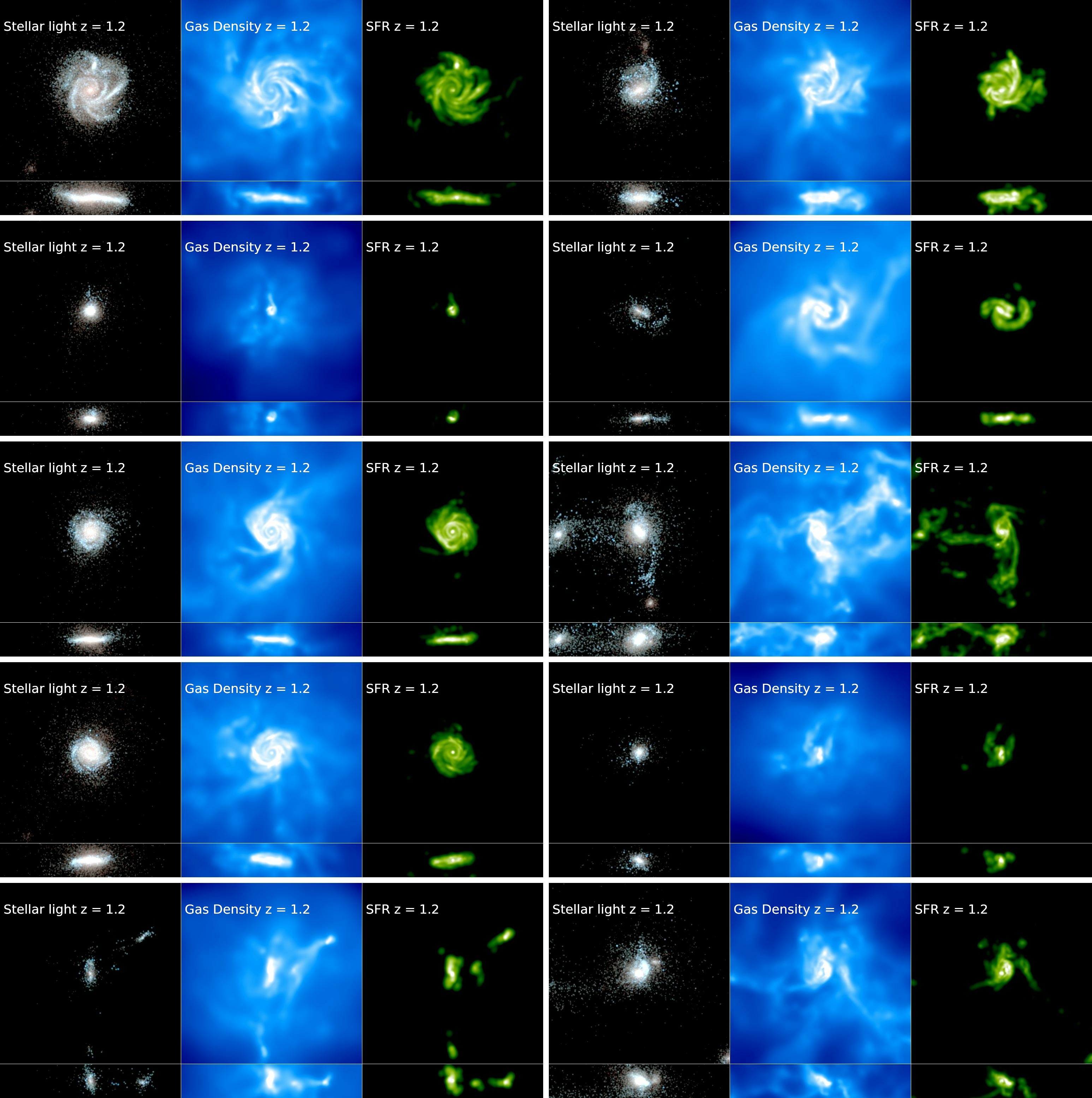}
    \caption{Different projections of the first 10 PRFM-vol galaxies at redshift $z=1.2$. For each galaxy, we show face-on and edge-on projections of stellar light (left panel), gas density (middle panel), and star formation rate (right panel). Stellar light projections are made as in Figs~\ref{fig:galaxy_morphologies_one} and \ref{fig:galaxy_morphologies_two}. The field of view for gas density and star formation rate projections is the same as for the stellar light, and in all cases we use the stellar moment of inertia tensor to define directions for the edge-on and face-on projections. An SPH smoothing kernel is applied to gas density and star formation rates. In contrast to the stellar light projections, the colour maps of gas density and star formation rate are each normalized to their own maximum values, in order to better highlight peaks in density and star formation rate. }
    \label{fig:prfm_vol_one}
\end{figure*}

\begin{figure*}
    \centering
    \includegraphics[width=\linewidth]{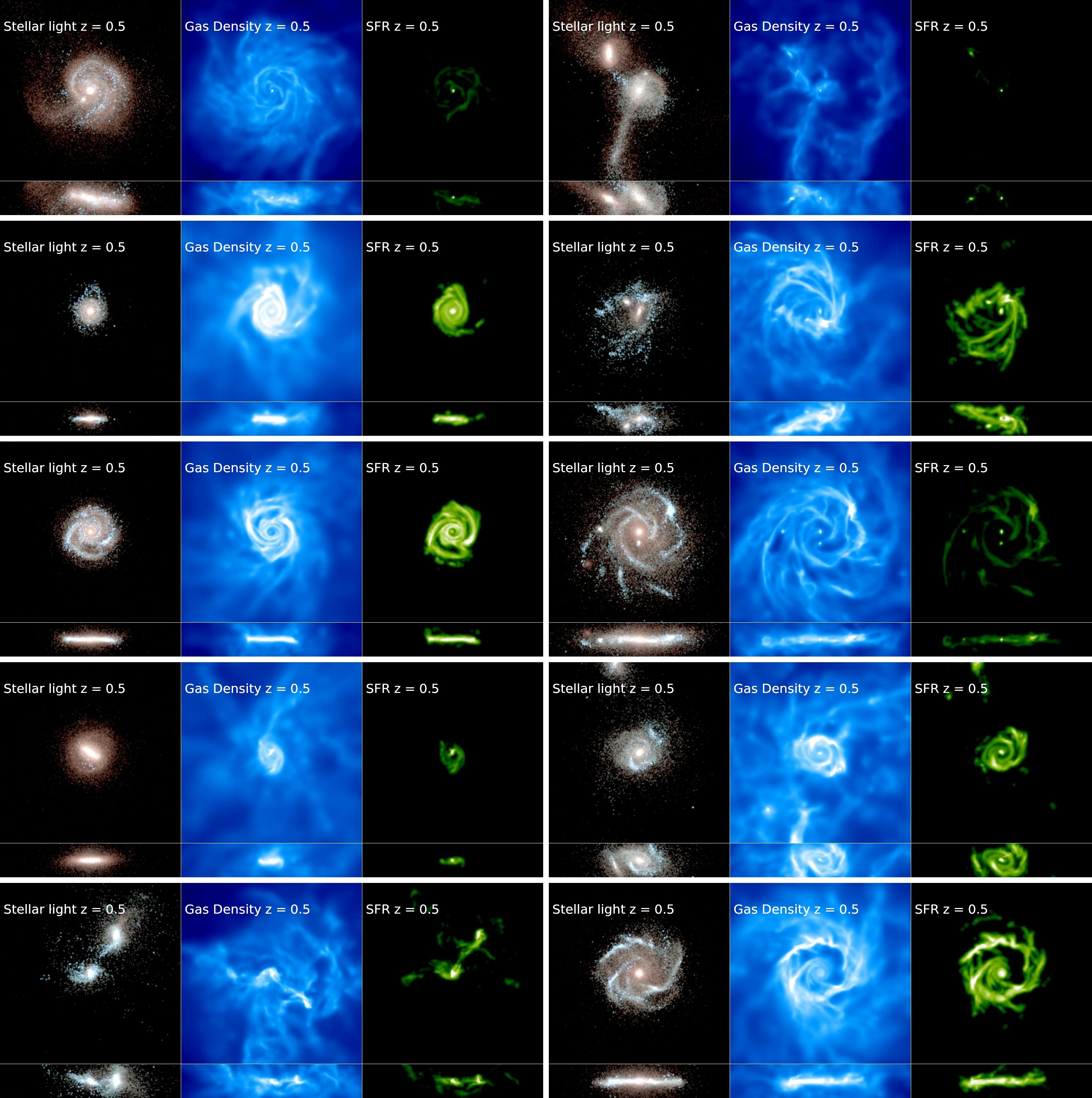}
    \caption{Same as Fig.~\ref{fig:prfm_vol_one} (also selecting the same haloes), but shown at redshift $z=0.5$. Notice how disks have mainly settled, and clumps begin to form in the third galaxy on the right hand side. These clumps coincide with peaks in the gas density and the star formation rate.}
    \label{fig:prfm_vol_two}
\end{figure*}

\begin{figure*}
    \centering
    \includegraphics[width=\linewidth]{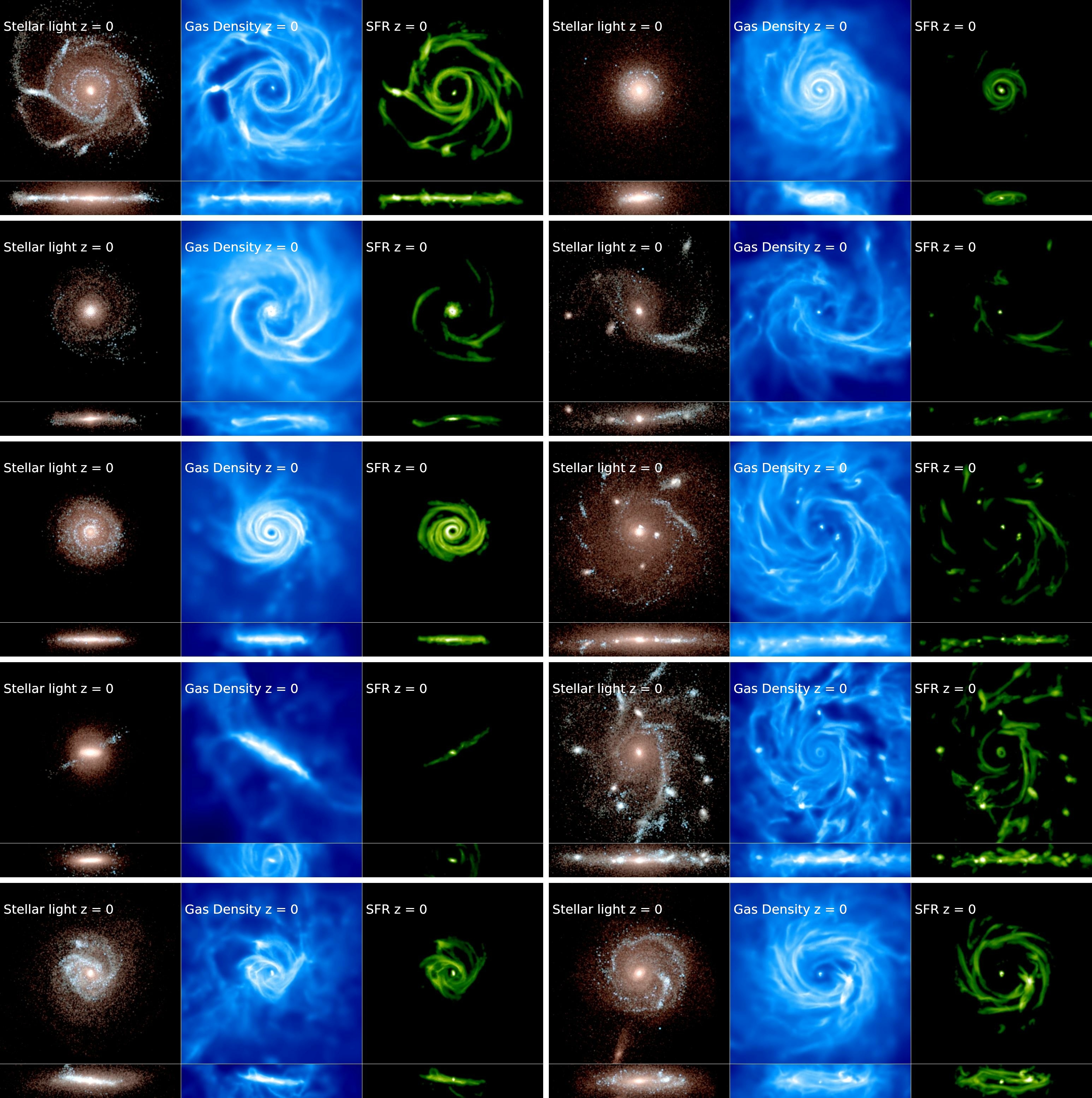}
    \caption{Same as Figs~\ref{fig:prfm_vol_one} and \ref{fig:prfm_vol_two}, but shown at redshift $z=0$. It is apparent that with decreasing redshift disks become more prone to form clumps at density peaks. Clumps do not disappear once formed, but persist while more of them can form. Notice that the correlation between the clumps and the gas density (or star formation rate) in the third galaxy on the right is visually weaker than in Fig.~\ref{fig:prfm_vol_two}. In the fourth galaxy, however, which has formed clumps more recently (it was not yet clumpy in Fig.~\ref{fig:prfm_vol_two}), the locations of the stellar clumps still correlate more strongly with peaks in gas density and star formation rate. }
    \label{fig:prfm_vol_three}
\end{figure*}

In our main article we paint a concise picture of how stellar clumps form in our multizoom setups. As we outlined throughout the article and argued in Section~\ref{ssec:morphs}, the clump formation process is closely related to disk instabilities that arise as a consequence of the fact that the PRFM-vol pressure at low hydrogen number densities is significantly lower than the TNG pressure. The actual process of clump formation should then proceed as follows. First, galaxies have to settle into disks. Once massive disks are present, gas will fragment due to Toomre instabilities. Star formation occurring in the resulting density peaks will eventually result in stellar clumps, which remain in place unless there is a localized feedback mechanism that can disperse them (which is not the case when using the TNG wind feedback).

To confirm the validity of this picture, we follow the formation of stellar clumps in the PRFM-vol model in Figs~\ref{fig:prfm_vol_one}, \ref{fig:prfm_vol_two}, and \ref{fig:prfm_vol_three}, where, for 10 of the 20 PRFM-vol galaxies we show face-on and edge-on projections of stellar light, gas density, and star formation rate at redshifts $z=1.2$, 0.5, and 0, respectively. All projections are made using the stellar moment of inertia tensor to define the coordinate system. Galaxies are matched at different redshifts, in such a way that their position (in terms of the panel they are shown in) does not change between figures. While the stellar light scale is kept fixed and consistent between all galaxies, gas density and star formation rate maps are normalized individually for each galaxy, in order to maximize contrast and clearly highlight the positions of density and star formation rate peaks. 

Fig.~\ref{fig:prfm_vol_one} shows that at redshift $z=1.2$, stellar disks have yet to settle in most cases, with only about half of the galaxies looking significantly different in face-on and edge-on projections. While none of the galaxies have formed clumps yet, galaxies appear very blue (with vigorous star formation) and sport identifiable stellar structures that correlate with peaks in gas density and star formation rate (see for example the first galaxy on the left side and the first two galaxies on the right side). 

In Fig.~\ref{fig:prfm_vol_two}, we find that at redshift $z=0.5$, almost all galaxies have settled into disks. Moreover, we can see stellar clumps starting to form in the second and third galaxy on the right hand side of the figure. The positions of the clumps correspond to peaks in gas density, as well as in star formation rate. 

Finally, we examine redshift $z=0$ in Fig.~\ref{fig:prfm_vol_three}. We find that clumps have formed in four of the 10 galaxies, the first one on the left side and the second, third, and fourth on the right side. Notably, for the second and third galaxy on the right side, the correspondence between the locations of stellar clumps and peaks in gas density and star formation rate is not as clear cut as it is for the other two galaxies, which must have formed their clumps between redshift $z=0.5$ and redshift $z=0$, as opposed to by redshift $z=0.5$. Our takeaway from this is that with the wind model used in our simulations, clumps are not dispersed once they have formed. This confirms the overall picture of clump formation that we painted in the main article. The gas disk fragments due to Toomre instabilities, leading to sharp density peaks, clustered star formation, and the subsequent buildup of stellar clumps. 

\section{Stellar clumps, simulation resolution, and galaxy size}\label{apxsec:resolution}

In this section, we take a look at some of our other multizoom simulation results. In particular, we focus on two runs in which we end up with fewer stars per galaxy compared to our fiducial, $\log(M_{\rm vir}/{\rm M_\odot})$ setup with zoom factor 4. First, we look at results obtained in another run that also focuses on galaxies with $\log(M_{\rm vir}/{\rm M_\odot}) = 12$, but this time with a zoom factor of 2, thus a factor of 8 poorer mass resolution.

\begin{figure*}
    \centering
    \includegraphics[width=\linewidth]{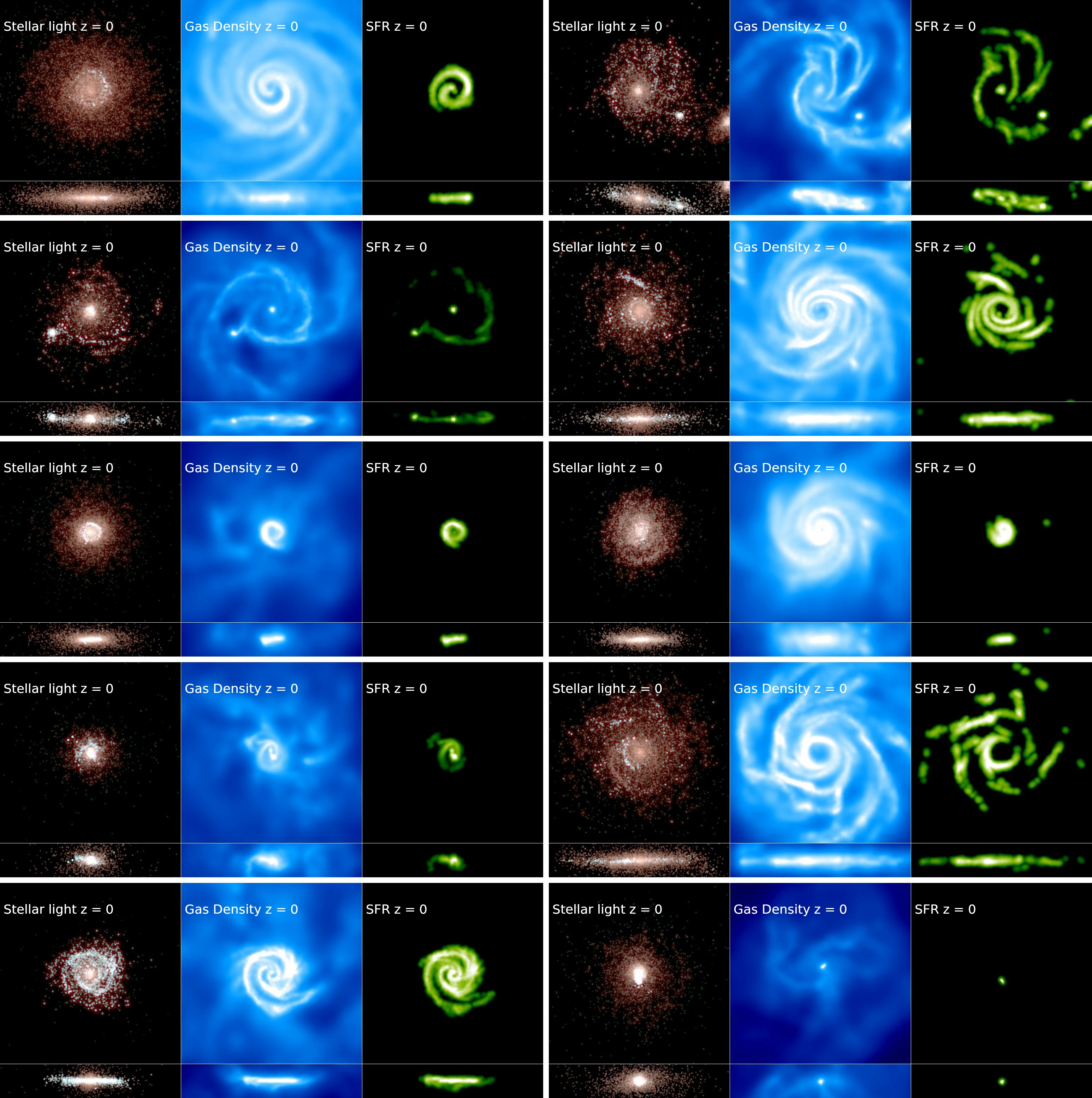}
    \caption{Different projections of 10 PRFM-vol galaxies, taken from the multizoom run with $\log(M_{\rm vir}/{\rm M_\odot}) = 12$ and zoom factor 2. For each galaxy, we show face-on and edge-on projections of stellar light (left panel), gas density (middle panel), and star formation rate (right panel).
    Projections are calculated in the same way as in Fig.~\ref{fig:prfm_vol_one}. While the galaxies surface brightness appears quite similar to our default setup at 8 times better mass resolution, we find an overall reduced occurrence of stellar clumps at this lower resolution.}
    \label{fig:reduced_resolution}
\end{figure*}

We show  redshift zero face-on and edge-on projections of stellar light, gas density, and star formation rate in Fig.~\ref{fig:reduced_resolution}. Remember that in contrast to stellar light, gas density and star formation rate are not globally normalized, just as in Fig.~\ref{fig:prfm_vol_three}. Compared to our fiducial setup, which we show in Fig.~\ref{fig:prfm_vol_three}, we find a clear reduction in the appearance of stellar clumps. Only two of the galaxies shown here exhibit any clumps at all -- the second galaxy on the left side and the first galaxy 
on the right side. They have exactly one stellar clump that, in both cases, coincides with peaks in gas density and star formation rate. In general, it appears that the decreased resolution suppresses the formation of clumps. This is best explained by comparing the gas density projections shown in Fig.~\ref{fig:reduced_resolution} to the ones shown in Fig.~\ref{fig:prfm_vol_three}. Especially in the face-on projections, it is apparent that the ISM structure is much better resolved in our fiducial setup. From Fig.~\ref{fig:prfm_vol_three}, we also see that the spatial scale of gas clumps is relatively small\footnote{In fact, it is close to the gravitational softening length in the zoom factor 2 simulations.}. Structure on that scale is poorly resolved in Fig.~\ref{fig:reduced_resolution}. This shows that the formation of clumps in galaxies which are prone to Toomre instabilities can be suppressed numerically through a lack of resolution. 

We now turn our attention to another multizoom run, the run with $\log(M_{\rm vir}/{\rm M_\odot}) = 11.097$ and a zoom factor of 4. This run has the same resolution as our fiducial setup, but focuses on haloes that are 8 times lighter on average, corresponding to smaller, less massive galaxies.  ~
\begin{figure*}
    \centering
    \includegraphics[width=\linewidth]{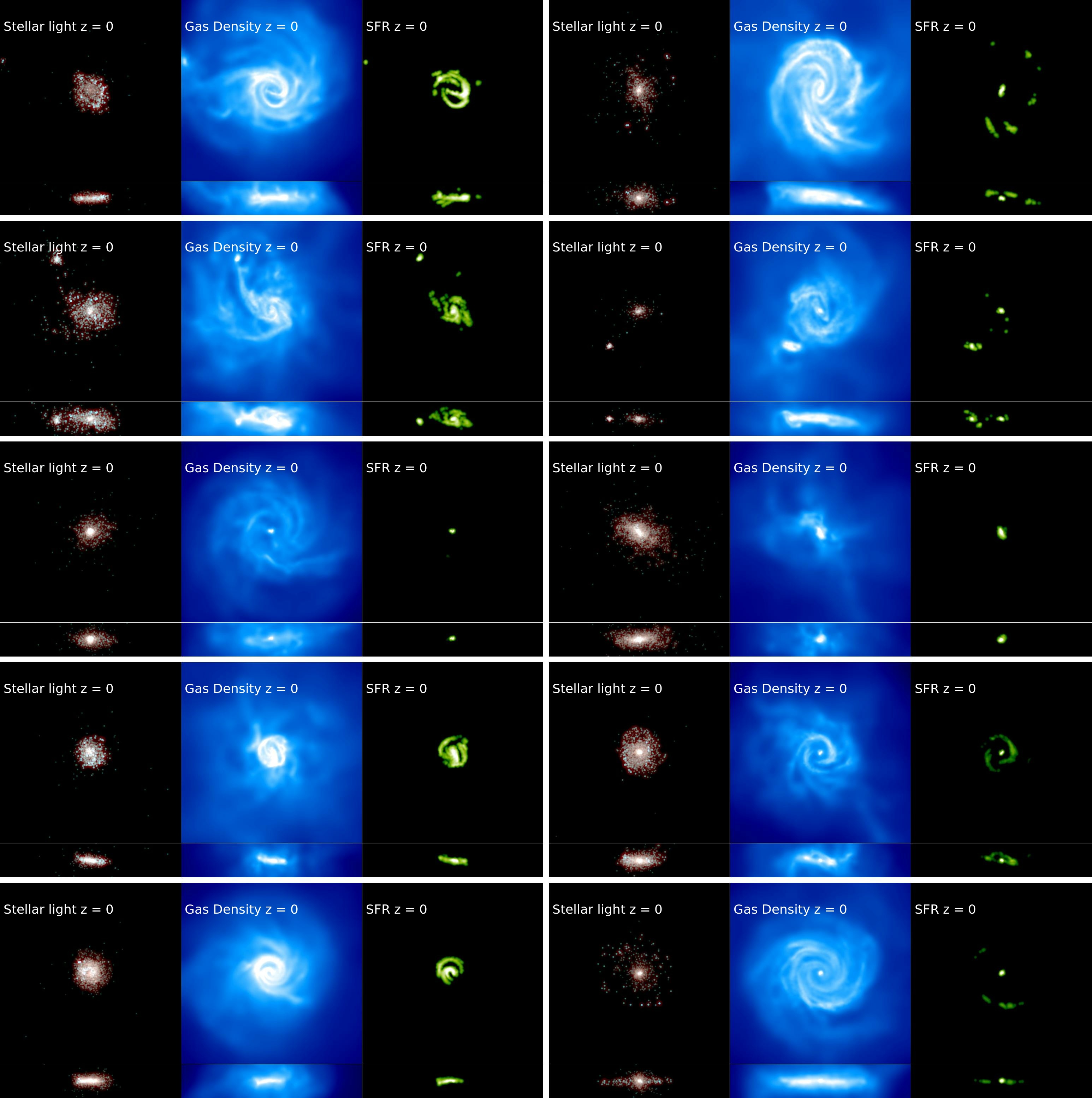}
    \caption{Same as Fig.~\ref{fig:reduced_resolution}, but for the run with $\log(M_{\rm vir}/{\rm M_\odot}) = 11.0969$ with zoom factor 4. For the stellar light projections, we have adapted how we map K, B, and U band luminosities to RGB space, amplifying the projected brightness by a factor of 10 compared to previous stellar light projections. }
    \label{fig:smaller_galaxies}
\end{figure*}

In Fig.~\ref{fig:smaller_galaxies}, we show redshift zero face-on and edge-on projections of stellar light, gas density, and star formation rate for 10 galaxies simulated in this multizoom run. Compared to similar previous figures, we change how we project K, B, and U band luminosities to RGB space, effectively increasing the projected brightness by a factor of 10. The simulated galaxies in Fig.~\ref{fig:smaller_galaxies} do not contain clearly identifiable stellar clumps. This is readily explained by equation~(\ref{eq:toomre}). Their reduced gas surface densities lead to increased $Q$-values, making their disks less prone to instabilities than the larger galaxies of our fiducial multizoom simulations.

We have thus identified scenarios in which no stellar clumps appear in the original PRFM-vol model -- one physical and one numerical: 
\begin{enumerate}
    \item When focusing on galaxies with comparatively low surface densities, clumps do not appear since these galaxies are Toomre stable. 
    \item In galaxies that should be Toomre unstable, clump formation can be suppressed through a lack of resolution. 
\end{enumerate}

\bsp	
\label{lastpage}
\end{document}